\documentclass[journal]{IEEEtran}
\usepackage{makecell}
\usepackage{tcolorbox} 
\usepackage{graphicx}
\usepackage{array}
\usepackage{longtable}
\usepackage{comment}
\usepackage{lscape}
\usepackage{mathtools}
\usepackage{booktabs}
\usepackage{hyperref}
\usepackage{float}
\usepackage{amsmath}
\usepackage{mathpazo}
\usepackage{datetime}
\usepackage{color, colortbl}
\usepackage{adjustbox}
\usepackage[belowskip=0pt,aboveskip=-1pt]{caption}

\definecolor{VeryHigh}{rgb}{0.90,0.44,0.43}
\definecolor{High}{rgb}{0.92,0.59,0.47}
\definecolor{Medium}{rgb}{0.94,0.72,0.51}
\definecolor{Low}{rgb}{0.68,0.81,0.52}
\definecolor{VeryLow}{rgb}{0.47,0.73,0.50}  

\definecolor{Black}{rgb}{0,0,0}
\definecolor{Grey}{rgb}{0.85,0.85,0.85}

\definecolor{White}{rgb}{1,1,1}

\hypersetup{
    colorlinks=true,
    linkcolor=blue,
    filecolor=magenta,      
    urlcolor=cyan,
    pdftitle={Overleaf Example},
    pdfpagemode=FullScreen,
    }

\usepackage{bookmark}
\usepackage{multirow}
\usepackage{xcolor}
\usepackage{xspace}
\usepackage{tabularx}

\usepackage{url}
\usepackage{todonotes}
\usepackage{listings}
\usepackage{authblk}
\usepackage{subcaption}

\newcommand{\thickhline}{%
    \noalign {\ifnum 0=`}\fi \hrule height 1pt
    \futurelet \reserved@a \@xhline}
\newcolumntype{"}{@{\hskip\tabcolsep\vrule width 1pt\hskip\tabcolsep}}

\newcommand{\oran}{O-RAN\xspace}
\newcommand{\nrt}{Near-RT RIC}
\newcommand{\nonrt}{Non-RT RIC}
\newcommand{\mlaa}{ML-assisted application}
\definecolor{bottleGreen}{RGB}{13, 172, 91}
\begin{document}
\date{}

\title{\Large \bf{Adversarial Machine Learning Threat Analysis and Remediation in \\ Open Radio Access Network (O-RAN)}}

%\author{}
%\begin{comment}
\author[1]{\rm Edan Habler}
\author[1]{\rm Ron Bitton}
\author[1]{\rm Dan Avraham}
\author[1]{\rm Eitan Klevansky}
\author[1]{Dudu Mimran}
\author[1]{Oleg Brodt}
\author[2]{Heiko Lehmann}
\author[1]{Yuval Elovici}
\author[1]{Asaf Shabtai}

\affil[1]{Ben-Gurion University of the Negev}
\affil[2]{Deutsche Telekom AG, T-Labs}

% \author{Edan Habler$^1$, Ron Bitton$^1$, Dan Avraham$^1$, Dudu Mimran$^1$, Eitan Klevansky$^1$, 
% Oleg Brodt$^1$, Heiko Lehmann$^2$, Yuval Elovici$^1$, Asaf Shabtai$^1$}
% \affiliation{%
%   \institution{$^1$Ben-Gurion University of the Negev}
%   \institution{$^2$Deutsche Telekom AG, T-Labs}
% }

%\author{

%    \IEEEauthorblockN{
%        Edan Habler, %\textsuperscript{1}, 
%        Ron Bitton, %\textsuperscript{1}, 
%        Dan Avraham, %\textsuperscript{1}, 
%        Eitan Klevansky, %\textsuperscript{1}, 
%        Dudu Mimran, %\textsuperscript{1},
%        Oleg Brodt %\textsuperscript{1}, 
%        \\ 
%        Heiko Lehmann, %\textsuperscript{2}, 
%        Yuval Elovici, and %\textsuperscript{1}, and 
%        Asaf Shabtai %\textsuperscript{1}
%    }
    
%    \IEEEauthorblockA{
%        Ben-Gurion University of the Negev\textsuperscript{1} \\
%        Deutsche Telekom AG, T-Labs\textsuperscript{2} 
%    }
%}
%\end{comment}

%\pagestyle{plain}
\maketitle

\thispagestyle{empty}

\subsection*{Abstract}
%\vspace{-3pt}
%Open Radio Access Network (
O-RAN is a new, \textit{open}, \textit{adaptive}, and \textit{intelligent} RAN architecture. 
Motivated by the success of artificial intelligence in other domains, O-RAN strives to leverage machine learning (ML) to automatically and efficiently manage network resources in diverse use cases such as traffic steering, quality of experience prediction, and anomaly detection. 
Unfortunately, it has been shown that ML-based systems are vulnerable to %are not free of vulnerabilities; specifically, they suffer from a special type of logical vulnerabilities that stem from the inherent limitations of the learning algorithms. 
%To exploit these vulnerabilities, an adversary can utilize
an attack technique referred to as adversarial machine learning (AML). 
This special kind of attack has already been demonstrated in recent studies and in multiple domains.
In this paper, we present a systematic AML threat analysis for O-RAN.
We start by reviewing relevant ML use cases and analyzing the different ML workflow deployment scenarios in O-RAN. 
Then, we define the threat model, identifying potential adversaries, enumerating their adversarial capabilities, and analyzing their main goals. 
Next, we explore the various AML threats associated with O-RAN and review a large number of attacks that can be performed to realize these threats and demonstrate an AML attack on a traffic steering model.
In addition, we analyze and propose various AML countermeasures for mitigating the identified threats. 
Finally, based on the identified AML threats and countermeasures, we present a methodology and a tool for performing risk assessment for AML attacks for a specific ML use case in O-RAN.

\begin{IEEEkeywords}
%\keywords{
Open radio access networks, adversarial machine learning, security and privacy, threat analysis
%}
\end{IEEEkeywords}

%\vspace{-15pt}
\section{\label{sec:into}Introduction}
%\vspace{-10pt}
In recent years, the number of cellular network users has increased dramatically.
According to Statista's 2021 report,\footnote{Statista's Forecast: \url{https://www.statista.com/statistics/245501/multiple-mobile-device-ownership-worldwide/}} in 2021 there were 15 billion unique mobile devices, a number which is expected to reach 18.22 billion by 2025. 
There has also been rapid growth in the number of connected IoT devices, with a projection of 27 billion connected IoT devices in 2025,\footnote{IoT Analytics State of IoT Summer 2021: \url{https://iot-analytics.com/product/state-of-iot-summer-2021/}} and thus, the requirements of the cellular network are not limited to just mobile devices.
Currently, cellular networks must support a diverse set of use cases~\cite{parkvall2017nr} involving various types of user equipment, including smartphones, smart watches, drones, industrial IoT devices, distributed sensors, and connected cars. 
New devices, use cases and applications pose new challenges for cellular networks.
Such challenges include the need to serve billions of devices, while maintaining low expenses, and the need to offer adaptive bandwidth and latency requirements, in real time, for different application and use cases.

% \begin{tcolorbox}
% [colback=gray!5!white,colframe=gray!75!black,fonttitle=\small,left=2pt,right=2pt,title=O-RAN’s Vision: key objectives defined by the alliance,boxrule=2pt,arc=0.6em,boxsep=0mm]
% \footnotesize
%     \textbf{Open:}
%     All design documents, interfaces, and software must be open. 
%     The openness aspect promotes multi-vendor deployments with open interfaces between all decoupled RAN components, enabling a more competitive ecosystem.
    
%     \textbf{Adaptive:}
%     RAN components must be able to adapt themselves, in real time, to support different use cases and service requirements.
%     To achieve this goal, the alliance promotes the cloudification of the RAN technology and an overall shift to cloud-native technologies, where network components are virtualized and controlled by software-defined networking. 
%     Cloudification facilitates flexible resource provisioning and enables centralization of the RAN infrastructure and a reduction of operational costs~\cite{alliance2020ran}. 
    
%     \textbf{Intelligent:}
%    RAN management must not rely on human intensive means.
%     Motivated by the success of %artificial intelligence and 
%     machine learning (ML) in other domains, the O-RAN strives to leverage ML for efficient automated network resource management.
%     This includes a large set of use cases such as: traffic steering, quality of experience prediction, network traffic prediction, and anomaly detection.
% \end{tcolorbox}

%\vspace{-13pt}
\subsection{\label{subsec:oran}The Open Radio Access Network}
In order to support new cellular network requirements, vendors have started investigating new radio access network (RAN) architectures.
A promising RAN architecture that has gained worldwide acceptance is the Open Radio Access Network (O-RAN), which was suggested by the O-RAN Alliance~\cite{oran-website}.
The O-RAN Alliance (founded in February 2018 by AT\&T, China Mobile, Deutsche Telekom, NTT DOCOMO, and Orange) is a worldwide community of mobile network operators, vendors, and academic institutes. 
The alliance's vision is to reshape the RAN industry toward the establishment of an \textit{open}, \textit{adaptive}, and \textit{intelligent} RAN.
The alliance's vision is to reshape the RAN industry toward the establishment of an \textit{open}, \textit{adaptive}, and \textit{intelligent} RAN:

    \textbf{Open:}
    All design documents, interfaces, and software must be open. 
    The openness aspect promotes multi-vendor deployments with open interfaces between all decoupled RAN components, enabling a more competitive ecosystem.
    
    \textbf{Adaptive:}
    RAN components must be able to adapt themselves, in real time, to support different use cases and service requirements.
    To achieve this goal, the alliance promotes the cloudification of the RAN technology and an overall shift to cloud-native technologies, where network components are virtualized and controlled by software-defined networking. 
    Cloudification facilitates flexible resource provisioning and enables centralization of the RAN infrastructure and a reduction of operational costs~\cite{alliance2020ran}. 
    
    \textbf{Intelligent:}
   RAN management must not rely on human intensive means.
    Motivated by the success of artificial intelligence and machine learning (ML) in other domains, the O-RAN strives to leverage ML for efficient automated network resource management.
    This includes a large set of use cases such as: traffic steering, quality of experience prediction, network traffic prediction, and anomaly detection.

%\vspace{-13pt}
\subsection{\label{subsec:oransec}Security of the O-RAN}
%\vspace{-3pt}
The introduction of new concepts and technologies into the RAN is promising in terms of meeting the new requirements~\cite{parvez2018survey}.
Unfortunately, when new technologies are introduced, they are accompanied by new cybersecurity threats; as a result, the RAN's attack surface may change dramatically when integrating new technologies.
Understanding the new attack surface is crucial for securing the new Open RAN architecture.

Recent studies have provided a throughout security analysis of the new Open RAN architecture~\cite{shen2022security,polese2022understanding,SecOpenRAN2021}, however these papers mainly focused on (a) reviewing recent attacks on the traditional RAN and evaluating their applicability to the O-RAN; (b) reviewing recent attacks on cloud environments and evaluating their applicability to the O-RAN; and (c) reviewing threats to open-source architectures and evaluating their impact on the O-RAN.
To the best of our knowledge, no recent studies have evaluated the security threats introduced by the use of ML in the O-RAN; i.e., the robustness of the O-RAN to adversarial machine learning threats.

%\vspace{-10pt}
\subsection{\label{subsec:amlintro}Adversarial Machine Learning}
%\vspace{-3pt}
ML systems are not vulnerability-free.
Specifically, ML systems suffer from a special type of logical vulnerabilities that stem from inherent limitations of the underlying learning algorithm.
To exploit these vulnerabilities, attackers utilize an attack technique referred to as adversarial machine learning (AML), which has already been demonstrated in security and ML research~\cite{carlini2017adversarial,szegedy2013intriguing,huang2011adversarial}.
For example, in the \textit{evasion threat}~\cite{carlini2017adversarial,goodfellow2014explaining,kurakin2016adversarial}, the adversary may exploit the ML model by generating an input sample that is very similar to some other correctly classified input but is incorrectly classified by the  model~\cite{carlini2017adversarial}.
In this case, the adversary's main objective is to compromise the integrity of the ML model by causing the model to provide incorrect outputs for a specific input generated by the attacker.
The evasion threat is just one example of AML.
Other studies have demonstrated additional threats to ML models, including (a) privacy threats (such as membership inference~\cite{shokri2017membership,nasr2019comprehensive}), data property inference~\cite{ateniese2015hacking,ganju2018property,song2020information}, data reconstruction~\cite{fredrikson2014privacy,hidano2017model}, and model extraction~\cite{chandrasekaran2020exploring,juuti2019prada,oh2019towards,papernot2017practical}; and (b) availability threats (such as model corruption~\cite{kravchik2021poisoning,mei2015using,jagielski2018manipulating}) and resource exhaustion~\cite{shumailov2020sponge,shapira2023phantom}.

%\vspace{-15pt}
\subsection{\label{subsec:scope}Scope and Purpose}
%\vspace{-5pt}
Previous works have presented general taxonomies on AML attacks~\cite{tabassi2019taxonomy,huang2011adversarial,bitton2021evaluating}, and others discussed the application of such attacks in 5G~\cite{sagduyu2021adversarial,liu2020adversarial,usama2019black}.
However, these works focused on specific usecases (applications) and were not aimed at providing a thorough and systematic risk analysis (that includes the complete pipeline of a ML application) of AML attacks on 5G.
In addition, the unique properties of the \oran, namely open environment, cloudification, and virtualization, introduce new threat models (i.e., threat actors and attack vectors) that were not relevant to 5G and therefore were not considered before.

%\newText{RON: In addition, the technical merit is weak. The paper lacks a clear delineation to existing taxonomies on adversarial machine learning (for example [17] ,[B])
%[B] A Taxonomy and Terminology of Adversarial Machine Learning. NIST IR, 2019
%}
%\newText{Reviewer4: What are new attacks which are specific to O-RAN and not presented in 5G (ML have been used in 5G network).-- 
%Add introductory paragraph here that says that previous works introduced AML in 5G like [a], [d],[e] however,....they are focused on specific attack, no systemathic...
%Although previous works deal with AML in 5G, for example [A], (1) no systematic analysis that refers to the complete pipeline exists (for example, attack on the FE components) and (2) due to the openess, cloudification introduces new threat models that were not considered
%[A] Adversarial machine learning for 5G communications security. Game Theory and  Machine Learning for Cyber Security, 270-288. Adversarial Machine Learning in 5G [D, E], 
%[D] Adversarial attack on DL-based massive MIMO CSI feedback,” Journal of  Communications and Networks, vol. 22, no. 3, pp. 230–235, Jun. 2020,
%[E] Black-box Adversarial Machine Learning Attack on Network Traffic Classification,” in IWCMC 2019.
%}

In this research, we present a systematic AML threat analysis of the \oran.
Our analysis considers the complete pipeline of a ML application in \oran as well as the unique properties of the \oran environment.
We begin by describing the \oran architecture , including its components and interfaces (Section~\ref{sec:background}).
Then, we review the various ML use cases applicable to the \oran and analyze the different deployment scenarios of ML workflows in the \oran (Section~\ref{sec:ml_dep_oran}).
Next, we describe the ontology used in this threat analysis (Section \ref{sec:ontology}).
In Section~\ref{sec:threat_model}, we describe the proposed threat model, which outlines the capabilities needed by an attacker to perform AML attacks.
In that section, we also identify potential threat actors in the \oran ecosystem and map them to the abovementioned capabilities.
In Section~\ref{sec:threat_category}, we describe the main AML threat categories according to the adversary's goal.
In Section~\ref{sec:attack_tech}, we evaluate the various AML threats applicable to the \oran; for each threat, also we review various attack techniques.
Section~\ref{sec:demonstration}, we demonstrate the applicability of an AML attack on the traffic steering use case implemented in the \oran reference implementation.
Moreover, in Section~\ref{sec:countermeasures}, we review the various AML countermeasures and analyze them according to a methodology we defined, which focuses on information security aspects (e.g., required access privileges for implementing a countermeasure) and the degree of influence on the basic assumptions of the \oran environment (e.g., runtime and memory overhead within different \oran hosts). 
This approach is intended to link a defensive need and a relevant countermeasure while considering the defender's degree of privileges and the model's limitations within the \oran environment.
Finally, in Section~\ref{sec:secexamplegeneric}, we present risk assessment process
that can be applied to any ML use case in O-RAN and we present a tool through which the evaluation process can be systematically carried out.

%\vspace{-10pt}
\subsection{\label{subsec:contributions}Contributions}
%\vspace{-3pt}
The main contributions of this paper can be summarized as follows:
\begin{enumerate}%[leftmargin=*,noitemsep,topsep=0pt]
    \item a comprehensive threat assessment of ML usecases within \oran according to a common cybersecurity risk assessment ontology;
    \item systematic mapping of possible ML deployments in \oran, threat actors and threat actors capabilities specific to the \oran, AML families and attack techniques in \oran, and finally, enumeration of AML threats in \oran;
    \item a novel taxonomy of AML countermeasures and their applicability to \oran;
    \item a novel procedure and a practical tool for the concrete identification and prioritization AML threats for specific ML usecase in \oran as well as enabling effective countermeasure planning for the high risk AML threats; the procedure and tool are based on the general threat assessment of AML threats in \oran presented in this paper;
    \item demonstration of practical AML attack and the application of the risk assessment procedure and tool for a selected ML usecase in \oran -- \emph{traffic steering}.
\end{enumerate}

% \vspace{-10pt}
\section{\label{sec:background}The O-RAN Architecture}
% \vspace{-8pt}
The RAN provides wireless connectivity to mobile devices and acts as the final link between the cellular network and user equipment. 
A typical RAN is composed of a radio unit (RU) and a base station (containing baseband units/BBUs).
In older mobile network generations (prior to 4G), the electronic equipment of the RU and BBU were coupled together at the bottom of the mobile antenna towers, and RF cables were used to connect the RU to the antennas at the top of the towers.
However, this approach was inefficient in terms of the signal performance, and eventually, the cellular industry relocated the RU equipment to the top of the tower.

A traditional RAN is vendor-specific.
That is, the interfaces between the RU and the BBU are defined by the vendor, and the applications running on them are tailored and optimized to the specific vendor's equipment. 
Vendor-specific solutions allowed vendors to provide optimized, integrated solutions; however, it may be sub-optimal.
In addition, a proprietary RAN requires the vendor to develop all of the components, which increases the cost of the RAN for operators and increases the operators' dependence on the vendor (vendor lock).

During the 4G and 5G, the RAN architecture has evolved, becoming less centralized and more disaggregated while transitioning from the use of dedicated equipment and software to the use of general-purpose hardware, virtualization, and the adoption of cloud-native technologies.
While decentralization mainly reduces costs, disaggregation allows deployment flexibility and accommodates the diversity of 5G use cases.

\begin{figure}[h]
    \centering
    \includegraphics[width=0.5\textwidth]{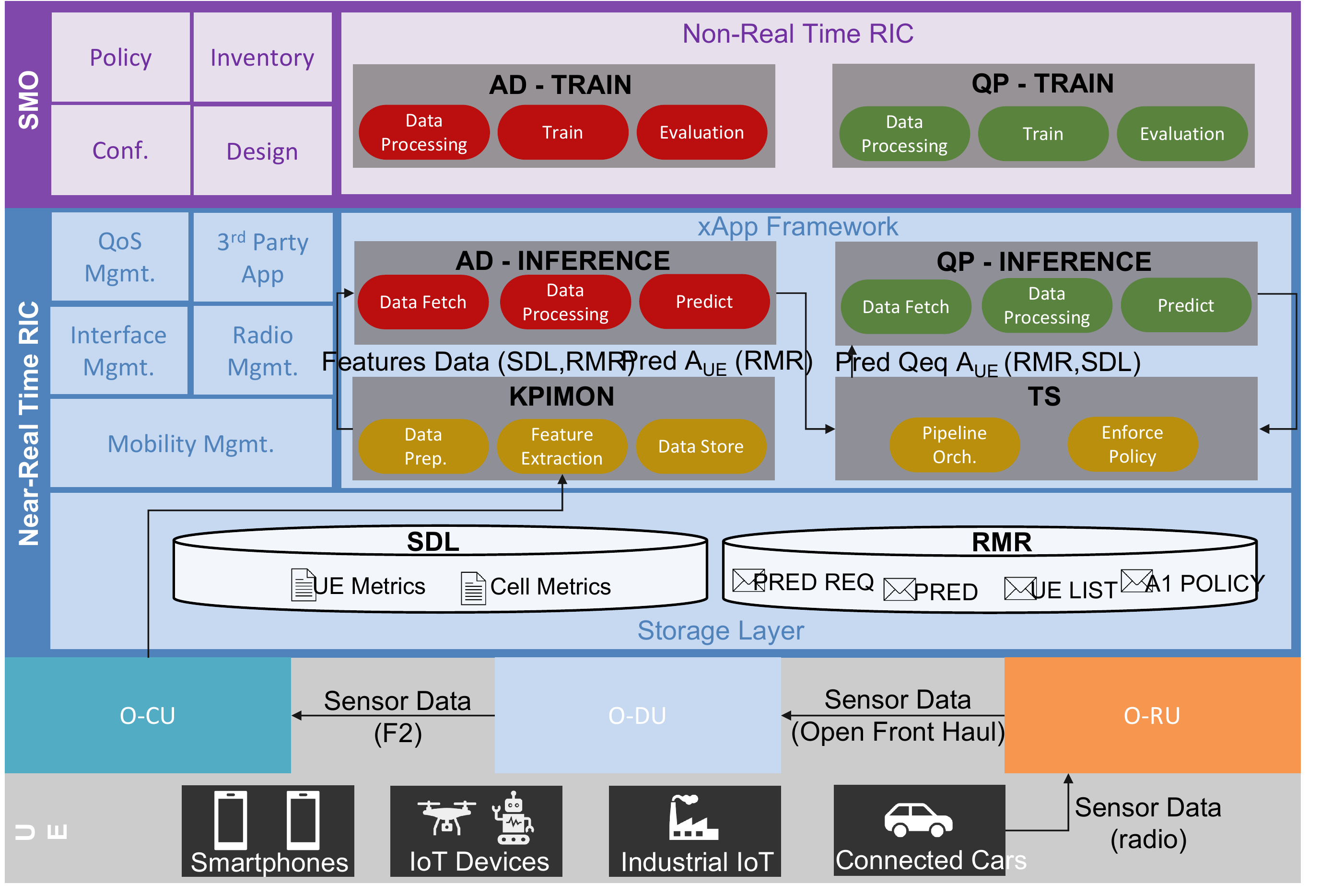}
    \caption{An illustration of the O-RAN reference architecture and the traffic steering process.}
    \label{fig:oran_arch}
\end{figure}

In Figure~\ref{fig:oran_arch}, we present the disaggregated and open architecture introduced by the O-RAN alliance.
A brief description of the various components and protocols introduced in O-RAN is presented in Table~\ref{tab:oran_comp}.

%\vspace{-8pt}
    
\begin{table}[h]
    \centering
%    \scriptsize
%     \resizebox{\textwidth}{!}{%
    \begin{tabular}{|p{0.10\textwidth}|p{0.35\textwidth}|}
    \Xhline{3\arrayrulewidth}
        \textbf{Component} & \textbf{Description}  \\
    \Xhline{3\arrayrulewidth}
        \rowcolor{Grey}
    \multicolumn{2}{|c|}{\textbf{O-RAN Components}}\\ 

        \textbf{User equipment (UE)} & The UE is the end user's device (e.g., smartphone, connected car, smart IoT sensor), which consumes network services from the cellular network over a radio channel.\\\hline
        
        \textbf{\oran~radio unit (O-RU):} & The RU is responsible for the broadcast and transmission of radio frequency signals, and it is usually part of the antenna. 
        It provides the physical layer functionality.\\\hline
        
        \textbf{\oran distributed unit (O-DU)} & The DU is responsible for real-time scheduling functions.\\\hline 
        
        \textbf{\oran~centralized unit (O-CU)} & The CU is composed of two logical nodes: central unit-control plane (CU-CP) and central unit-user plane (CU-UP). 
        The CU-CP hosts the radio resource control and the control plane part of the packet data convergence protocol (PDCP). 
        The CU-UP hosts the service data adaptation protocol and the UP part of PDCP. 
        The rationale behind this separation is to improve the placement of different RAN functions, thereby accommodating different situations and performance needs. 
        \\\hline
        
        \textbf{Near-RT RIC} & An intelligent controller that enables near real-time control and optimization of RAN elements and resources. 
        It is implemented as a (private/public) cloud platform, which support low latency.\\\hline
        
        \textbf{Service management \& orchestration (SMO)} & This component holds management services for fault, configuration, accounting, performance, and security. \\\hline % (FCAPS).\\\hline
        
        \textbf{Non-RT RIC} & This module supports intelligent RAN optimization. It provides policy-based guidance, ML pipeline management (e.g., hosting, training, updating), and enrichment information for the \nrt.\\\hline
        
        \rowcolor{Grey}
\multicolumn{2}{|c|}{\textbf{O-RAN Interfaces}}\\ 

        \textbf{A1} & The interface between the \nonrt~and the \nrt. This interface supports: policy management, enrichment information, and ML-model management.\\\hline
        
        \textbf{E2} &  The interface between the \nrt~and an E2 node (e.g., O-CU, O-DU). This interface supports \nrt~services (report, insert, control, and policy), \nrt~support functions (e.g., setup, reset, reporting of general errors) and \nrt~service update.\\\hline
        
        \textbf{Open Fronthaul M-Plane} & The interface between the O-DU and the O-RU; it includes the control user synchronization plane and management plane (M-Plane).\\\hline
       
      \textbf{O1} & The interface between the SMO and  the management entity (e.g., O-DU, O-RU, O-CU). 
    This interface supports FCAPS management, physical network function (PNF) software management, and file management.\\\hline
              
     \textbf{O2} & The interface between the SMO and the O-Cloud. 
    This interface supports infrastructure management services and deployment management services.\\\hline
    \Xhline{3\arrayrulewidth}
    \multicolumn{2}{c}{}\\
    \end{tabular}
    %\caption{The various components and interfaces in O-RAN.}
%    }
%    \vspace{-3pt}
    \caption{\oran components and interfaces.}
    \label{tab:oran_comp}
\end{table}
%%\vspace{-25pt}
\section{Machine Learning in the O-RAN \label{sec:ml_dep_oran}}
\begin{figure}[h]
    \centering
    \includegraphics[width=0.50\textwidth]{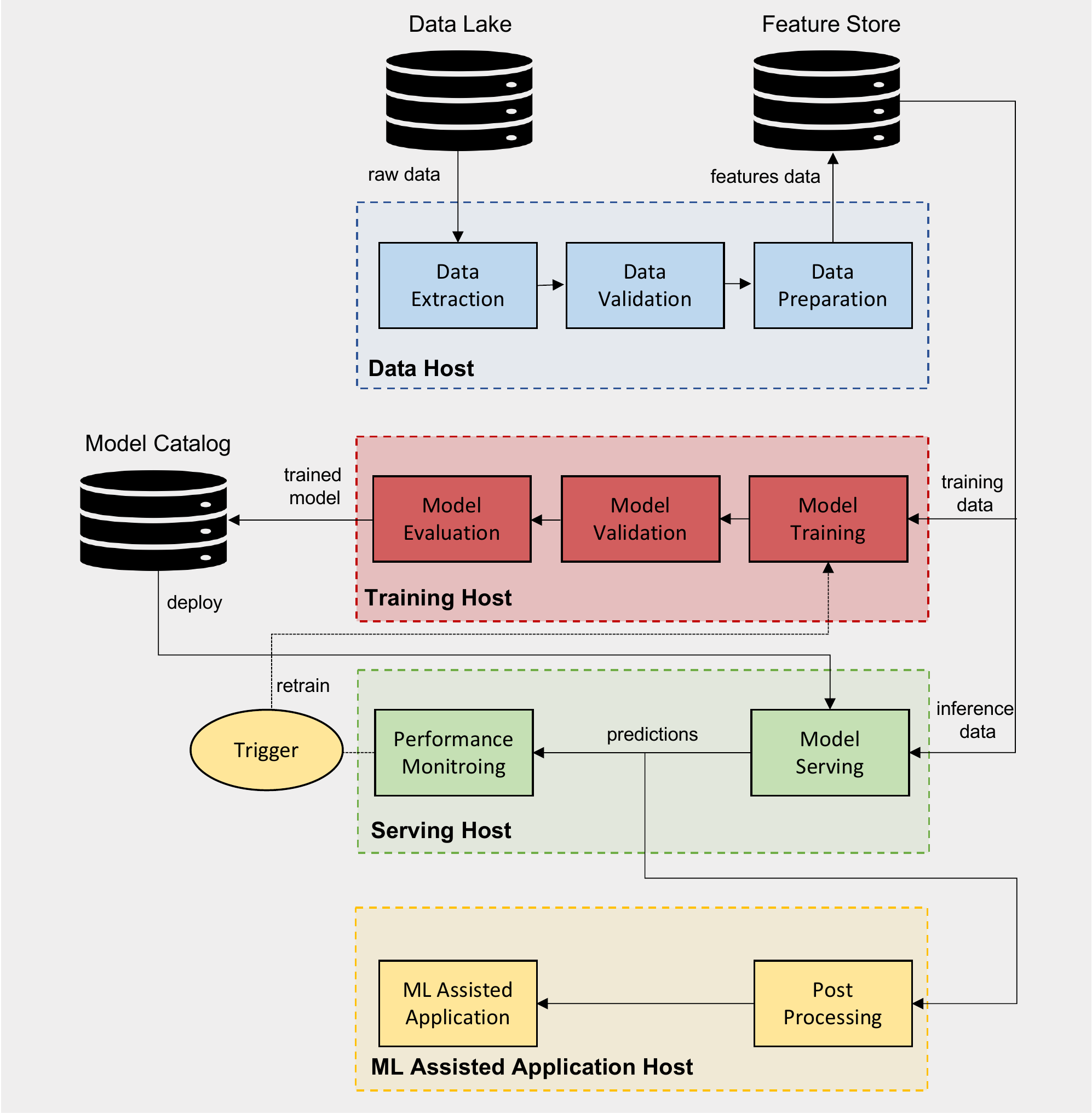}
    \caption{General ML pipeline in \oran.}
    \label{fig:ml_dep}
\end{figure}

%%\vspace{-8pt}
%Artificial intelligence (AI) and 
Machine learning (ML) plays a crucial role in the 5G network in general and in the vision of O-RAN in particular~\cite{masur2021artificial}.
For example, in 5G networks, ML is used for network planning, automation of network operations (such as provisioning, optimization, and fault prediction), network slicing, and quality of service prediction. 
In this section, we review the deployment of ML workflows in the O-RAN. 
In Table~\ref{tab:ml_comp}, we review the various components of a typical ML workflow. 
These components are also illustrated in Figure~\ref{fig:ml_dep}, grouped by the virtual machine hosting them: Data Host (for data preprocessing and preparation), Training Host (for training, validating and evaluating ML models), Serving Host (for enabling the use and monitoring of the developed ML models) and the ML Assisted Application Host (an application which utilizes a specific ML model).

\begin{table}[h]
    \centering
%    \scriptsize
    \begin{tabular}{|p{0.07\textwidth}|p{0.36\textwidth}|}
    \Xhline{3\arrayrulewidth}
        \textbf{Component} & \textbf{Description}  \\
    \Xhline{3\arrayrulewidth}
        
        \textbf{Data Lake} & A big data architecture responsible for storing all of the raw data required for the ML application.\\\hline
        
        \textbf{Data Host} & %A component 
        Responsible for extracting and integrating data from all of the data sources in the data lake, validating the data, and preparing a dataset for the model.\\\hline
        
        \textbf{Feature Store} & A big data architecture responsible for storing the extracted features used by the different ML apps.\\\hline 
        
        \textbf{Training Host} &  %A component 
        Responsible for training ML models with datasets retrieved from the feature store, performing model validation, and evaluating a model's performance before its deployment. \\\hline
        
        \textbf{Model Catalog} & A repository that stores all of the ML app resources including the model files, selected features, and models' metadata.\\\hline
        
        \textbf{Model} &  The trained model file, which is produced by feeding training data to the learning algorithm. The model itself is saved in the model catalog and deployed to the serving host. \\\hline
        
        \textbf{Serving Host} & Implements both ML compiling host and ML inference host functionalities. 
        Responsible for deploying a model from the model catalog and providing predictions. 
        The serving host is also responsible for online learning. \\\hline
        
        \makecell{\shortstack[l]{\textbf{ML-assisted}\\\textbf{App Host}}} &  A component responsible for executing all of the application's workflow, i.e., receiving the new input and preparing it for the inference stage and activating the appropriate action, according to the model's output, in the system.\\
    \Xhline{3\arrayrulewidth}
    \end{tabular}
    \caption{The various components of a typical ML workflow.}
    \label{tab:ml_comp}
\end{table}

%\subsection{AI/ML Tasks supported by the O-RAN \label{subsec:ai_tasks}}
%ML applications can broadly be classified into three main tasks: (i) supervised learning, (ii) unsupervised learning, and (iii) reinforcement learning. 

%\oran architecture supports all these three types of ML tasks.
%\begin{itemize}
   
%\item\textbf{Supervised Machine Learning Task.}
%In a supervised learning task, the algorithm must be provided with the desired output for each data sample (i.e., data labels).
%The algorithm learns a function mapping the input to the desired output (learning phase).
%The learned function (denoted as the model) can later be used to predict the label of novel data (serving phase).

%\item\textbf{Unsupervised Machine Learning Task.}
%In unsupervised learning task, the algorithm do not provided with the desired output for each data sample (i.e., data labels).
%The algorithm learns a function mapping the input to the desired output (learning phase).
%The learned function (denoted as the model) can later be used to predict the label of novel data (serving phase).

%\item\textbf{Reinforcement Learning Task.}
%The purpose of this ML task is to optimize a long-term objective based on a trial and error process with a defined environment.

%\end{itemize}

%%\vspace{-15pt}
\subsection{The Deployment of ML workflows in the O-RAN \label{subsec:ai_mapping}}
\begin{figure*}[h]
    \centering
    \includegraphics[width=0.75\linewidth] {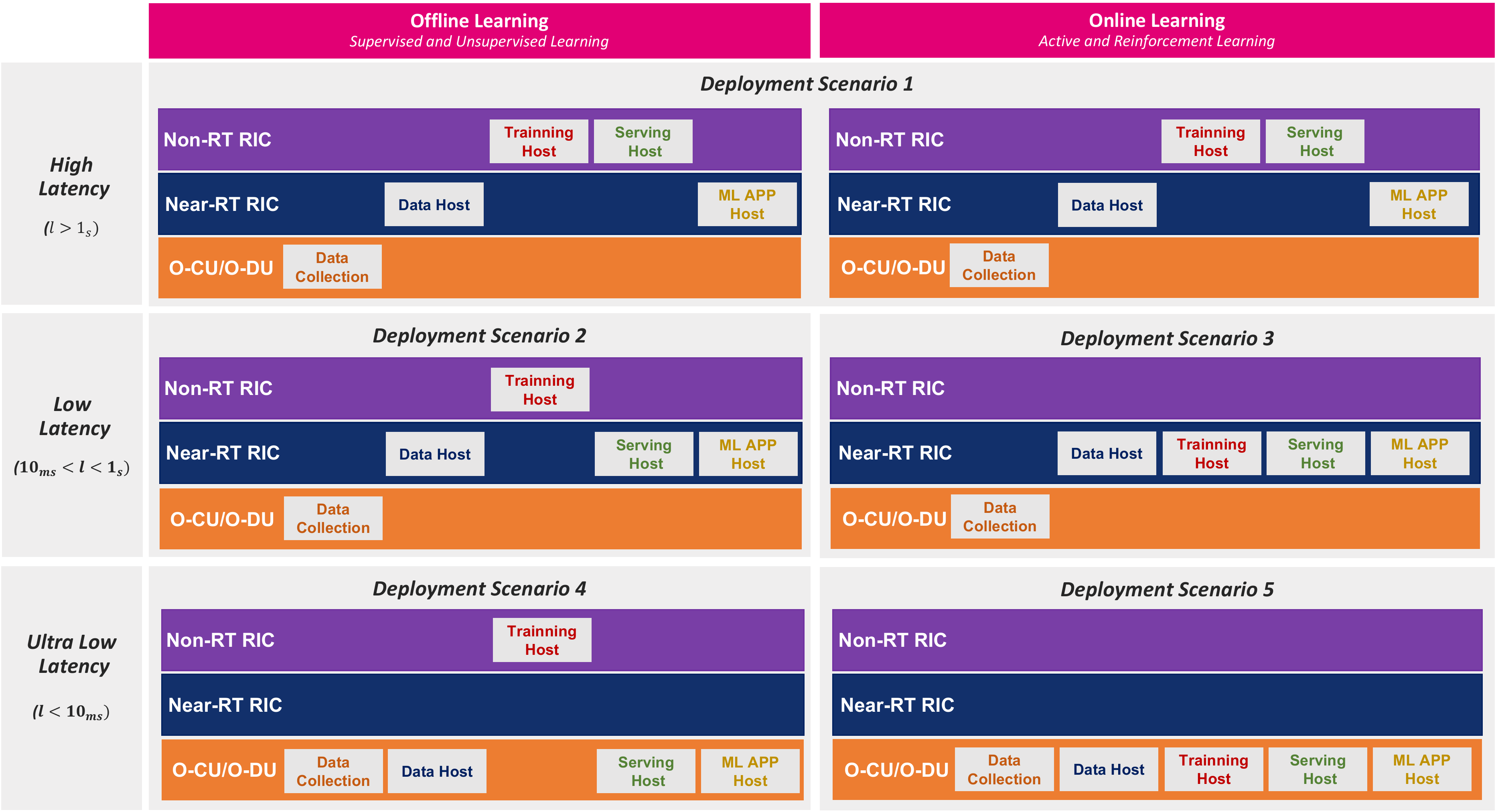}
    \caption{\oran~ML Deployment Scenarios.}
    \label{fig:ml_dep_scenario}
\end{figure*}

%%\vspace{-5pt}
As presented in Fig.~\ref{fig:ml_dep_scenario}, there are five different ML-based application deployment scenarios (DS) in the \oran which are defined by the type of ML task and the application's latency requirements. 
Specifically, the \oran architecture supports three main types of ML tasks: supervised learning, unsupervised learning, and reinforcement learning, and three levels of latency requirements: high latency ($time > 1s$), low latency ($10ms < time <= 1s$), and ultra-low latency ($time <= 10ms$). 

According to the current design document of the \oran, the various ML components can be implemented on the following four \oran units: (1) the \nonrt, (2) the \nrt, (3) the O-CU (inference only), and (4) the O-DU (inference only); currently, implementing ML components on O-RUs is not supported (as stated in the \oran alliance technical report~\cite{WG2AIML2020}).

In Figure~\ref{fig:ml_dep_scenario}, we present the deployment of the different ML components (data collection, data host, training host, serving host, and ML APP host) within the different layers of the \oran (O-CU/O-DU, \nrt, and \nonrt) for each deployment scenario according to the ML task and required latency.
Since \oran is currently at an early development stage and not all the technological aspects are clearly defined, it is important to note that the presented layouts of ML workflow within O-RAN and deployment of the ML components are based on best practices and the design documents of the \oran alliance.
Additional DS may assume that parts of the workflow (e.g., the data lake, host, and training) can reside outside the \oran architecture.

%\vspace{-12pt}
\subsection{\label{subsec:ai_usecases}\oran ML Use Cases}
%\vspace{-5pt}
The \oran ML use cases can be divided into two types: network management applications and third-party applications. 
The network management applications are deployed as part of the \oran's core functionality and are used to automatically and efficiently manage a system's resources. 
The third-party applications can be deployed by every provider that has a 5G-based service they want to provide to connected UEs. 
Some examples of network management ML-based applications provided in the \oran are: 
traffic steering (identify the optimal cell for each UE in order to ensure acceptable quality of experience), V2X handover management (optimize handover sequences on the UE level), and resource allocation optimization (predict traffic demands at different times and locations).

\section{Cybersecurity Risk Assessment Ontology \label{sec:ontology}}
\begin{figure}[h]
\centering 
\includegraphics[width=0.49\textwidth]{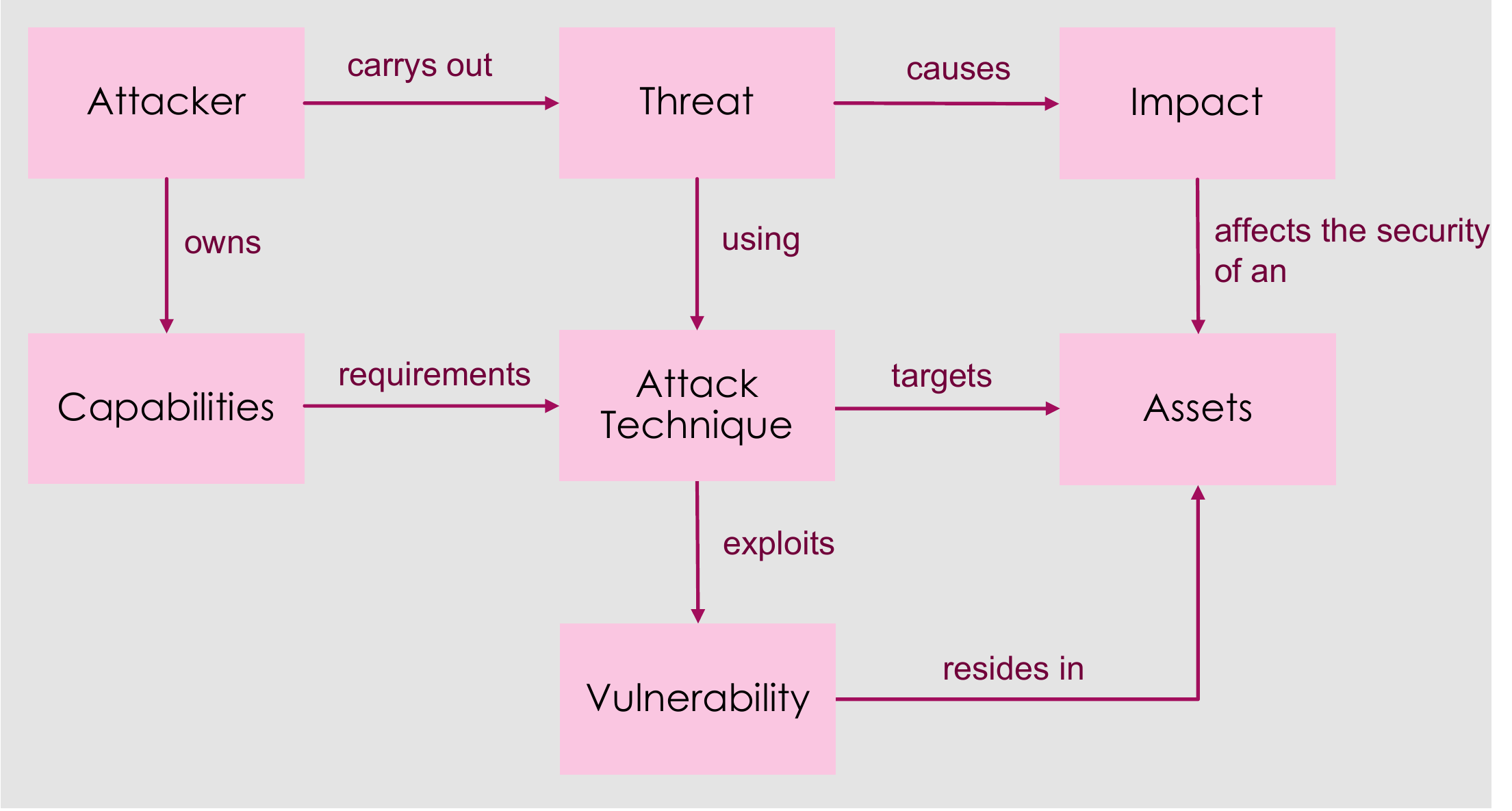}  
\caption{Threat assessment ontology.}
\label{fig:threat_ontology}
\end{figure}

%\vspace{-20pt}
In this paper, in order to provide a systematic AML threat analysis of the \oran, we follow the threat analysis ontology presented in~\cite{bitton2021mlrisk}, which is based on the NIST ontology for evaluating enterprise security risk.\footnote{\url{https://nvlpubs.nist.gov/nistpubs/legacy/sp/nistspecialpublication800-30r1.pdf}}
The ontology, which includes the entities summarized in Table~\ref{tab:ontology}, is used in the sections that follows in order to conduct a systematic risk assessment for ML usecases implemented within \oran.
We begin by enumerating the assets of a typical ML production system; identifying these assets is a crucial step in threat modeling.
Then, we describe the threat model, which outlines the potential adversaries, their capabilities, and their main goals.
Next, we describe the various threats to ML production systems.
Finally, we enumerate the attacks that can materialize these threats.
The ontology (presented in Figure~\ref{fig:threat_ontology}) includes the entities summarized in Table~\ref{tab:ontology}.

%\begin{comment}
%\end{comment}

\begin{table}[h]
    \centering
%    \scriptsize
    \begin{tabular}{|p{0.07\textwidth}|p{0.38\textwidth}|}
    \Xhline{3\arrayrulewidth}
        \textbf{Entity} & \textbf{Description}  \\
    \Xhline{3\arrayrulewidth}
        
        \textbf{Assets} & The main data, components, processes, and services that comprise an ML production pipeline and should be protected.\\\hline
        
        \textbf{Vulnerability} & In the context of AML attacks, a vulnerability Refers to the inherent ability to systematically manipulate the input to the ML model.\\\hline
        
        \textbf{Attacker (threat actor)} & An individual, group, or state responsible for an incident that impacts, or has the potential to impact, the security or safety of the ML.
        Threat actors may have different capabilities and resources and may perform different types of attacks.\\\hline
        
         \textbf{Capabilities} & Refers to the capabilities that are available to the attacker. 
        In the context of AML attackers, we distinguish between access capabilities and knowledge capabilities.\\\hline
        
        \textbf{Impact (attacker goal)} & By exploiting vulnerabilities, threats are able to have security impact on the vulnerable assets, e.g., violate the confidentiality of the data used to train a model or tamper with the model's integrity. 
        Security impact may lead to operational impact.\\\hline
        
         \textbf{Technique} & An act or method that violates the security policy of a system.\\\hline
         
         \textbf{Threat} & %A threat represents a 
         A potential violation of a security property, such as integrity, confidentiality, or availability.
        In a threat, an attacker that possesses the required capabilities exploits a vulnerability by utilizing a relevant attack technique in order to perform an attack.\\
    \Xhline{3\arrayrulewidth}
    \end{tabular}
    \caption{The entities included in the threat analysis ontology.}
    \label{tab:ontology}
\end{table}
This ontology is used in the sections that follows in order to conduct a systematic risk assessment for ML usecases implemented within \oran.
We begin by enumerating the assets of a typical ML production system; identifying these assets is a crucial step in threat modeling.
Then, we describe the threat model, which outlines the potential adversaries, their capabilities, and their main goals.
Next, we describe the various threats to ML production systems.
Finally, we enumerate the attacks that can materialize these threats.

%\vspace{-28pt}
\section{Threat Model \label{sec:threat_model}}
%\vspace{-8pt}
In this section, we define the threat model within the context of \oran: (1) identifying potential threat actors in the \oran ecosystem, (2) defining the main goals of the attacker, and (3) specifying the attacker capabilities (i.e., attacker model).

%\vspace{-13pt}
\subsection{Attacker Model\label{subsec:attacermodel}}
\begin{figure}[h]
    \centering
    \includegraphics[width=0.48\textwidth,trim={0 0 8.7cm 0},clip]{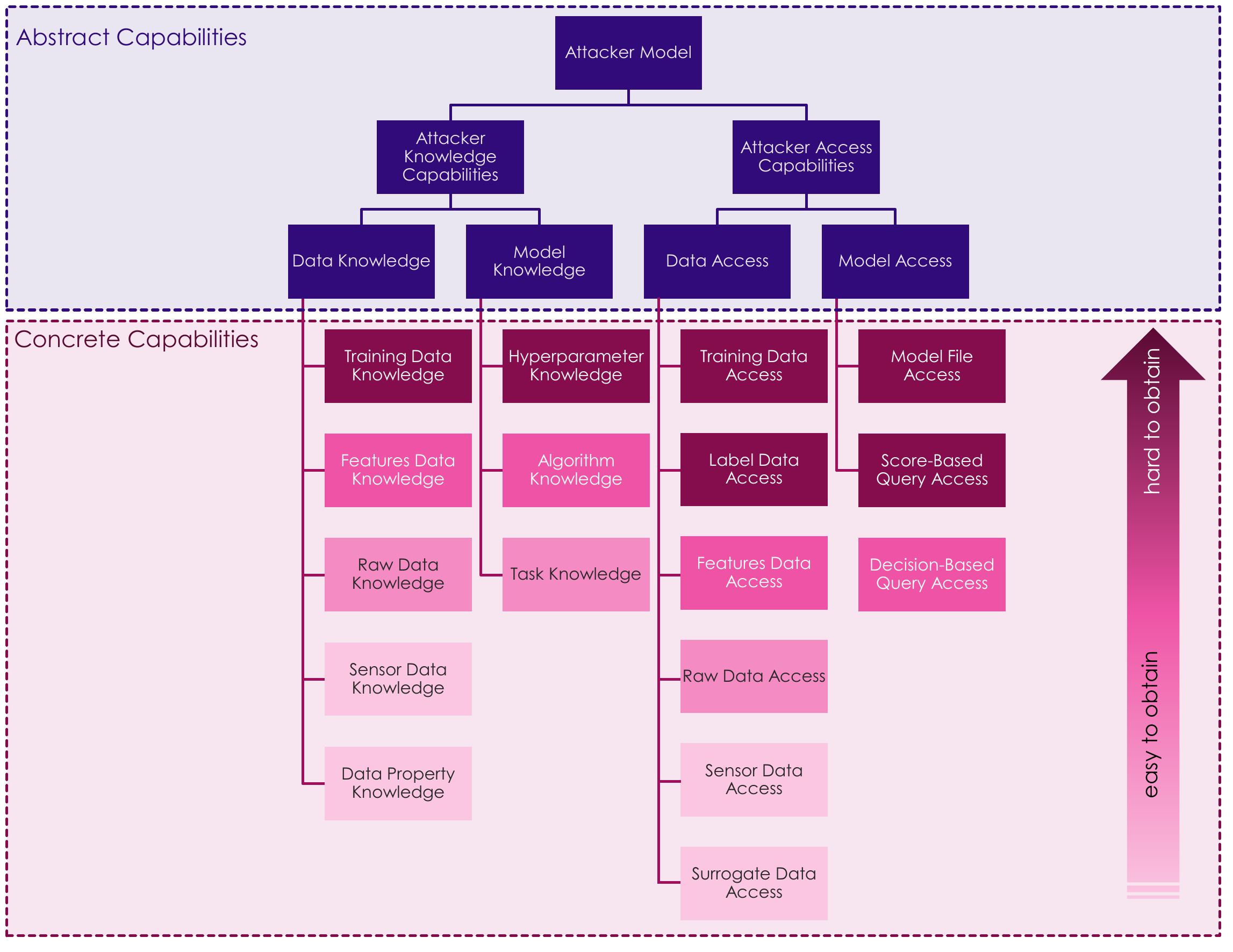}
    \caption{Adversarial capabilities.}
    \label{fig:adversarial_capabilities}
\end{figure}

%\vspace{-5pt}
Each AML attack may require different attacker capabilities in order to be executed.
As can be seen in Figure~\ref{fig:adversarial_capabilities}, we distinguish between two types of adversarial capabilities: access capabilities (AC) and knowledge capabilities (AK).
%In addition, we define a partially-ordered set of those capabilities.
These set of capabilities are partially-ordered. 

%\vspace{-13pt}
\subsubsection{\label{subsubsec:attacker_capabilities}Access Capabilities (AC)} 
%\vspace{-6pt}
The set of assets/capabilities possessed by the attacker within the \oran ML deployment are (see Figure~\ref{fig:ml_dep}):

\noindent \textbf{Model Access (ACM).} An adversary that has access to the ML model used in the specific use case (in the \nonrt /\nrt).

\noindent \textbf{[ACM1] Score-Based Query Access.} An adversary with the ability to query the trained model in the \mlaa's model inference host and obtain the model's probability (confidence) vector (the model's confidence for each class).

\noindent \textbf{[ACM2] Decision-Based Query Access.} An adversary with the ability to query the trained model in the \nrt~and obtain the model's decision (the final classification); i.e., no access to the model's probability vector.

%\noindent \textbf{[ACM3] Model Access.} In this threat model, we assume an adversary that has access to the exact model used in the target pipeline.\\

\noindent \textbf{Data Access (ACD).} An adversary with access to the data used in the pipeline of the specific use case (in the \nonrt /\nrt).

\noindent \textbf{[ACD1] Training Data Access.} An adversary with access to the processed dataset in the \mlaa's model's training host which is used to train the \mlaa's model. 
In this case, it is assumed that the adversary knows the exact features and samples used to train the model.

\noindent \textbf{[ACD2] Feature Data Access.} An adversary with access to the raw data and feature transformation functions.

\noindent \textbf{[ACD3] Raw Data Access.} An adversary with access to the raw data used by the training host to train the \mlaa's model.

\noindent \textbf{[ACD4] Sensor Data Access.} An adversary with the ability to manipulate data sent from a UE (or multiple UEs) that is accessible to the adversary.
This can be conducted either by direct access to the UE itself or the data sent from the UE.

\noindent \textbf{[ACD5] Labeled Data Access.} An adversary with access to the labels of the dataset used for training the target model (i.e., the adversary is able to manipulate the labels of training instances).

\noindent \textbf{[ACD6] Surrogate Data Access.} An adversary that has access to a reference dataset with similar characteristics and data distribution to the dataset used to train the model.

%\vspace{-15pt}
\subsubsection{\label{subsubsec:attacker_knowledge}Attacker's Knowledge (AK)} 
%\vspace{-5pt}
Information regarding the \oran architecture and deployment that is available to the attacker and can be used to more effectively generate AML attacks.

\noindent \textbf{Model Knowledge (AKM).} An adversary that knows the exact model used by the \mlaa. %'s pipeline.

\noindent \textbf{[AKM1] Hyperparameter Knowledge.} An adversary that knows the exact algorithm and hyperparameters used to train the \mlaa's models, e.g., for an artificial neural network, the hyperparameters include the network architecture, number of epochs used to train the model, learning rate, etc. 

\noindent \textbf{[AKM2] Algorithm Knowledge.} An adversary that knows the algorithm used to train the \mlaa's model but does not know the specific hyperparameters of the model.

\noindent \textbf{[AKM3] Task Knowledge.} An adversary that has general knowledge of the ML task, including the type of inputs and outputs, e.g., an attacker that knows that the ML pipeline is used to predict the quality of experience (QoE) for a given UE considering the current state of the cells.

\noindent \textbf{Data Knowledge (AKD).} An adversary that knows the exact data used to train the model in the \mlaa's pipeline.

\noindent \textbf{[AKD1] Training Data Knowledge.} An adversary that knows all/part of the training data used for training the \mlaa's model (including the specific feature transformations applied on the raw data).

\noindent \textbf{[AKD2] Features Data Knowledge.} An adversary that knows what raw data and feature transformations are applied to the data used to train the model.

\noindent \textbf{[AKD3] Raw Data Knowledge.} An adversary that knows what raw data is used to train the \mlaa's model (e.g., when a target model is being trained on a public dataset) but is not aware of the specific feature transformations applied to the raw data.

\noindent \textbf{[AKD4] Data Property Knowledge.} An adversary that knows the statistical properties (distribution) of the data %(data distribution) 
used to train the \mlaa's model.

%\vspace{-5pt}
%\vspace*{-\baselineskip}

%\vspace{-25pt}
\subsection{\label{subsec:threatactor}O-RAN Threat Actors}
%\vspace{-3pt}
In this section, we describe O-RAN's threat actors and map them to the relevant adversarial capabilities.
The threat actors that we present are aligned with the threat actors mapped by the \oran alliance's security analysis report~\cite{SecOpenThreat2022}.

%\newText{RON: 
%Reviewer3: I miss a clear discussion in how realistic are these threats in a real deployment. For instance for threat A2 how realistic it is to assume that a core developer of the infrastructure compromises, on purpose, one of the components of the architecture? Do not the internal best practices of these organizations mitigate this already? For threat A5 which is arguably the broader one since it covers the end user, how realistic is it to assume that the actor has direct query access to the ML model? Providing a few concrete examples, and if available pointing to concrete instances where these attacks did happen, would greatly benefit the paper.
%Response: 1. Taking the threat actors in the ORAN alliance, 2. Explain that the current threat actors means also unintentional (e.g., external attack). 3. examples of real/reported/research attacks that are relevant to the threat actors (e.g., log4j, ???)}

The proposed threat model includes two type of threat actors: internal (those that reside within the RAN) and external (those that reside outside the RAN).
External threat actors include: script kiddies, hacktivists, cyber-criminals, cyber-terrorists, and nation-state.
Internal threat actors include any stake-holder in the system.
In our analysis, we assume that internal threat actors (e.g., core developer or hardware infrastructure provider) do not necessarily compromise the \oran infrastructure on purpose, and they may serve as an attack vector to other malicious (external) entities, for example via a supply chain attack, as was recently demonstrated in the Solarwind and log4j attacks.
Table~\ref{tab:threat_actors_capabilities} summarizes the adversarial capabilities of the various threat actors.

% In Figure~\ref{fig:internal_threat_actors}, we provide a high-level overview of the \nrt~cluster as well as a mapping to the relevant internal threat actors, namely, the O-RAN's software application developer, software infrastructure developer, containerization software infrastructure provider, and hardware provider. 

% \begin{figure}[h]
%             \centering
%             \includegraphics[width=0.3\textwidth]{sections/figures/internal_threat_actors.pdf}
%              \caption{The O-RAN's internal threat actors.}
%             \label{fig:internal_threat_actors}
% \end{figure}

%%\vspace{-5pt}
\begin{table}[h]
    \centering
    \includegraphics[trim={0.7cm 3cm 17cm 3cm},clip,width=0.5\textwidth]{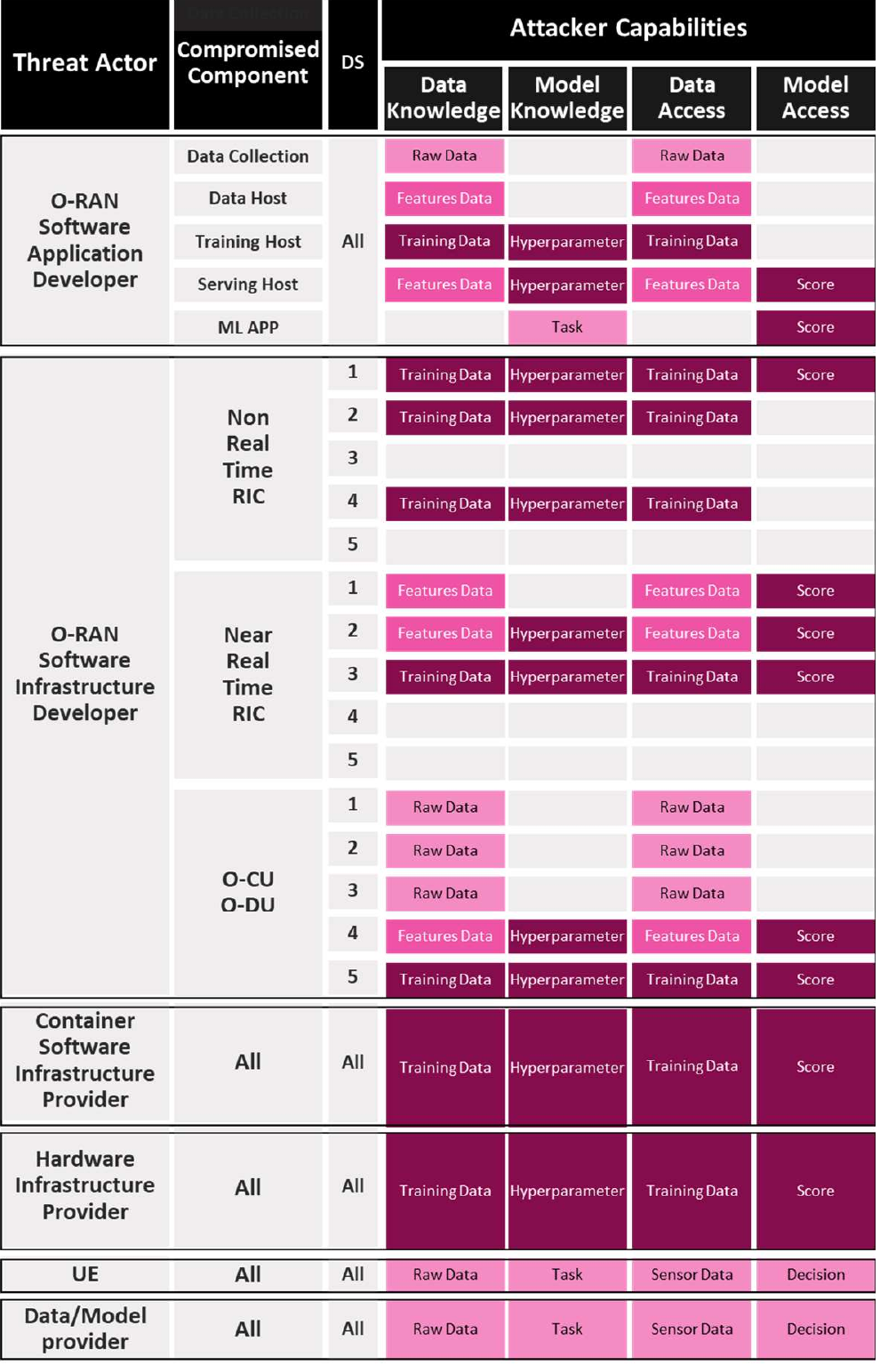}
    \caption{The O-RAN's threat actors and their adversarial capabilities (for each threat actor, we mention the adversarial capability that it is the most difficult to obtain).}
    \label{tab:threat_actors_capabilities}
\end{table}

\begin{comment}
\noindent\textbf{[A1] O-RAN Software Application Developer.}  Develops O-RAN applications, with access to RAN resources and interfaces. 
Adversarial capabilities depend on the specific application developed.

\noindent \textbf{[A2] O-RAN Software Infrastructure Developer.} Develops O-RAN infrastructure, with ability to compromise any application deployed within the infrastructure. Adversarial capabilities depend on location of compromised infrastructure and deployment scenario.

\noindent \textbf{[A3] Containerization Software Infrastructure Provider.} Provides virtualized/containerization software infrastructure, with ability to compromise any application deployed within the infrastructure.

\noindent \textbf{[A4] Hardware Infrastructure Provider.} Provides hardware infrastructure used in O-RAN, with ability to compromise any software running on the infrastructure.

\noindent\textbf{[A5] User Equipment (UE). } User equipment connected to O-RAN, able to change behavior and attack machine learning models, and has decision-based query access to deployed models.
\end{comment}
\noindent\textbf{[A1] O-RAN Software Application Developer.} 
This threat actor is responsible for developing O-RAN applications, i.e., software code deployed within the O-RAN components, such as the \nonrt~(rApp platform), \nrt~(xApp platform), and O-CU/O-DU (yet to be developed).
%As presented in Figure~\ref{fig:internal_threat_actors}, 
O-RAN applications are implemented as a Kubernetes Pod %(currently restricted to have one container) 
and have access to RAN resources and interfaces via infrastructure utility functions, which provide APIs for reading telemetry data from RAN interfaces, reading from (or writing to) RAN shared storage (SDL), sending and receiving messages, etc.
Since this threat actor operates at the application level, we assume that he/she can only compromise the \textit{specific} application developed by the adversary.
Specifically, we assume that this adversary has full control of the host (container) running the compromised component.
Thus, the adversarial capabilities of this threat actor depend on the specific application developed (and therefore controlled) by the threat actor.

\noindent \textbf{[A2] O-RAN Software Infrastructure Developer.}
Responsible for developing the O-RAN infrastructure which includes the following core components: % (see Figure~\ref{fig:internal_threat_actors}):
(1) scheduling functions (alarm and subscription managers); (2) utility functions (database communication, message routing, rest-full API, and logging); (3) application frameworks (for rapid development of Non-RT RIC, Near-RT RIC, O-CU, and O-DU applications); and (4) interface support, which includes the A1, O1, O2, and E2.
Since this threat actor develops the core infrastructure, we assume that he/she can compromise \textit{any} application deployed within the infrastructure.
Specifically, we assume that this adversary has full control over the compromised infrastructure and as a result, full control of the applications that use that software infrastructure.
The adversarial capabilities of this threat actor depend on the location of the compromised infrastructure within the O-RAN (i.e., \nonrt, \nrt, and O-CU/O-DU) and the deployment scenario (which is based on the learning task and latency constraint).

\noindent \textbf{[A3] Containerization Software Infrastructure Provider.}
Provides virtualized/containerization software infrastructure such as the Kubernetes and Docker infrastructure. % (see Figure~\ref{fig:internal_threat_actors}).
We assume that this threat actor can compromise \textit{any} application (container) deployed within the compromised infrastructure.
Specifically, we assume that this adversary has full control of the compromised infrastructure and as a result, full control of the containers running on the compromised infrastructure.
To simplify the assessment, we assume that all O-RAN containerization infrastructures are provided by the same  provider; thus, the adversarial capabilities of this threat actor are not dependent on the deployment scenario.

\noindent \textbf{[A4] Hardware Infrastructure Provider.}
This threat actor provides the hardware infrastructure used in the O-RAN.
We assume that this threat actor can compromise \textit{any} software running on the compromised hardware infrastructure.
Specifically, we assume that this adversary has full control of the compromised hardware and as a result, full control of the software running on the compromised infrastructure.
To simplify the assessment, we assume that all O-RAN components are provided by the same provider; thus, the adversarial capabilities of this threat actor are not dependent on the deployment scenario.

\noindent\textbf{[A5] User Equipment (UE). } User equipment that is connected to the O-RAN.
We assume that this threat actor can change its own behavior (i.e., is able to manipulate sensor data) in order to attack ML models deployed within the O-RAN.
In addition we assume that this threat actor has decisions-based query access to the deployed model, i.e., the adversary is affected by model's decisions.
%\end{comment}

\noindent\textbf{[A6] 3rd Party Data and/or Model Provider. } A third party entity that provides data that is used to train a model %that will be deployed in \oran 
and/or the model itself.
It may also include other service providers in the world of ML such data verifiers, labelers, and model verifiers.
We assume that this threat actor has full knowledge and control over the training set, the training process and model. 
In addition, this threat actor can change the behavior of UEs (controlled/owned by this threat actor) in order to attack ML models deployed within the O-RAN.

%\vspace{-15pt}
\subsection{\label{subsec:attacker_goal}Attacker's Goal}
%\vspace{-5pt}
\noindent \textbf{[AG1] Tampering.}
This type of threat is associated with malicious activities that compromise the \textit{integrity} of the \oran system.

\noindent \textbf{[AG2] Denial of Service.}
This type of threat is associated with malicious activities that compromise the \textit{availability} of the \oran system.
This goal can be achieved by causing, for example, the QoE model to make an incorrect prediction and infer that a specific cell is better than all of the others for all UEs; therefore, the model will assign all of the users to the same cell.

\noindent \textbf{[AG3] Information Disclosure.} This type of threat is associated with malicious activities that compromise the \textit{privacy} of the data/models used in the O-RAN.

%%\vspace{-12pt}
\section{Threat Categories\label{sec:threat_category}}
%\vspace{-10pt}
%%\vspace*{-\baselineskip}
Following is a brief description of the main AML threat categories.

\begin{comment}
\noindent\textbf{[T1] Evasion.} Adversary causes incorrect outputs for specific input, compromising integrity of ML model.\\
\noindent \textbf{[T2] ML Model Corruption.} Adversary causes incorrect decisions for targeted samples, compromising integrity and availability of ML model.\\
\noindent \textbf{[T3] Membership Inference.} Adversary attempts to extract information about input sample's existence in training set, compromising privacy of ML model.\\
\noindent \textbf{[T4] Data Property Inference.} Adversary extracts information to learn about general properties of training dataset, compromising privacy of ML model.\\
\noindent\textbf{[T5] Data Reconstruction (theft).} Adversary causes information leak of data samples used to train model, compromising privacy of ML model.\\
\noindent \textbf{[T6] Model Extraction.} Adversary attempts to extract information to train replica of model, compromising privacy of ML model.\\
\noindent \textbf{[T7] Resource Exhaustion.} Adversary causes ML model to use more resources, compromising availability of ML model.
\end{comment}

\noindent\textbf{[T1] Evasion.}
An adversary that causes the ML model to provide incorrect outputs for a specific input, e.g., causing the QoE model to classify an \textit{excellent} signal as a \textit{poor} signal.
This threat mainly compromises the \textit{integrity} of the machine learning model.

\noindent \textbf{[T2] ML Model Corruption.} An adversary that causes the ML model to make incorrect decisions for targeted samples, e.g., causing the QoE model to allocate all UE to a certain cell.
This threat can compromise both the \textit{integrity} and \textit{availability} of the ML model.

\noindent \textbf{[T3] Membership Inference.} An adversary that attempts to extract information regarding the existence of a given input sample in the model's training set, e.g., exposing the fact that certain UE was connected to a certain cellular cell.
This threat compromises the \textit{privacy} of the ML model.

\noindent \textbf{[T4] Data Property Inference.} An adversary that attempts to extract information from the ML model in order to learn about general properties of the model's training dataset (such as the feature distribution or input types), e.g., exposing the average quality of service of various UE or cellular cells.
This threat compromises the \textit{privacy} of the ML model.

\noindent\textbf{[T5] Data Reconstruction (theft).} An adversary that causes the ML model to leak information in such a way that some of the data samples used to train the model are exposed, e.g., exposing timeframes in which UE results in a poor QoE.
This threat compromises the \textit{privacy} of the ML model.

\noindent \textbf{[T6] Model Extraction.} An adversary that attempts to extract information about the model (usually by querying the model) in order to train a replica of the model, e.g., replicating the model used for QoE classification.
This threat compromises the \textit{privacy} of the ML model.

\noindent \textbf{[T7] Resource Exhaustion.} An adversary that causes the ML model to use more resources at inference time, e.g., increasing the latency of the QoE model when classifying targeted examples.
This threat compromises the \textit{availability} of the machine learning model.

%\vspace*{-\baselineskip}
\section{Attack Families \label{sec:attack_tech}}
\begin{figure}[h]
    \centering
    \includegraphics[width=0.48\textwidth]{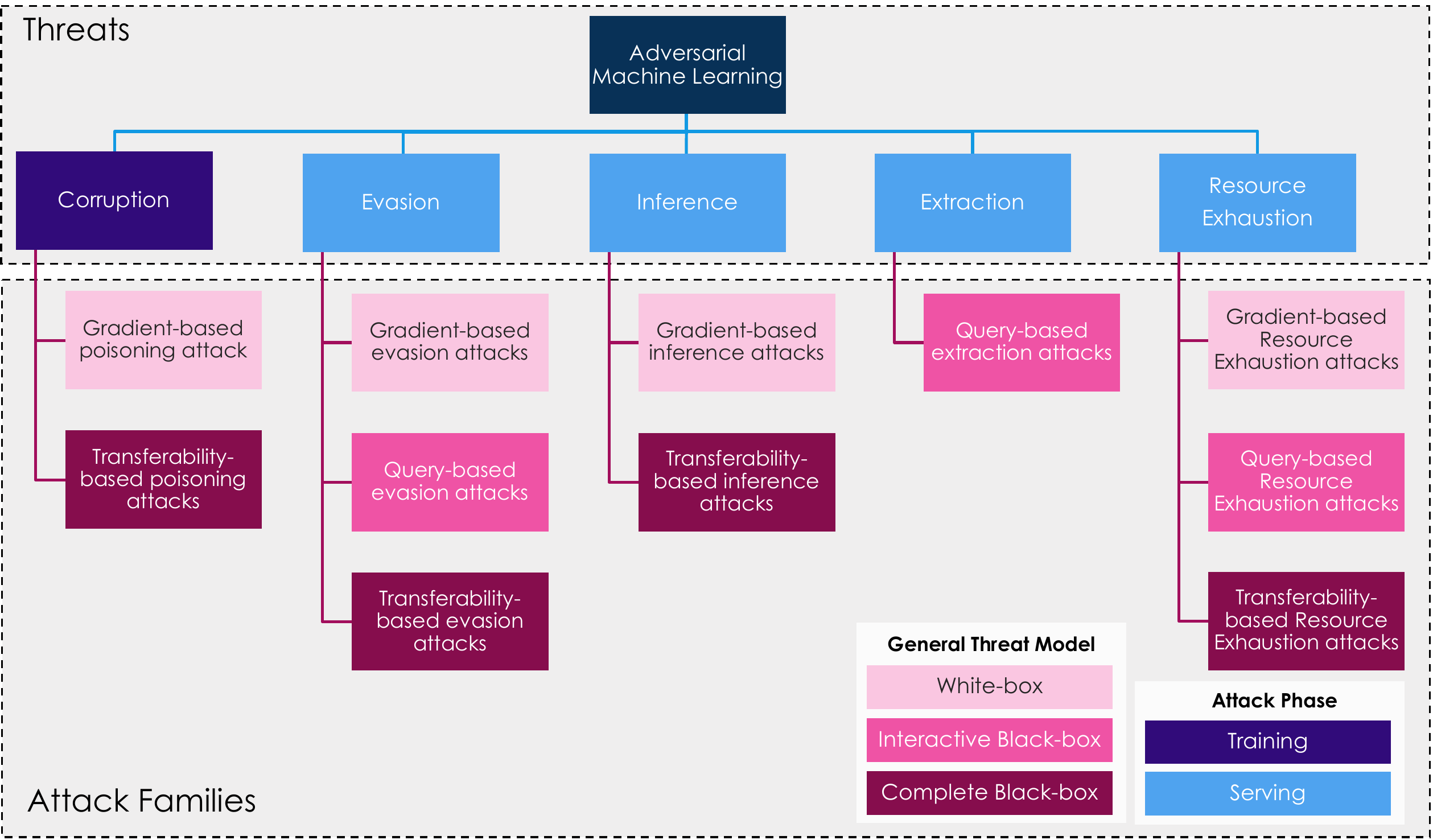}
    \caption{The main AML attack families.}
    \label{fig:aml_at}
\end{figure}

%\vspace*{-\baselineskip}
%%\vspace{-5pt}
We review various attack families against ML systems. 
For each attack family, we provide a brief description of the attack method, identify the adversarial capabilities required to successfully execute the attack, and map these capabilities to the relevant threat actors and impact.
In addition, we present a taxonomy of AML attack (see Figure~\ref{fig:aml_at}), which classifies each attack family according to the following three criteria:

\noindent \textbf{1) Threat category:} The threat materialized by the attack (i.e., corruption, evasion, inference, extraction, and resource exhaustion).
    
\noindent \textbf{2) General threat model:} The general threat model classification, which is based on the privileges required by the attacker to successfully execute the attack (white-box refers to attacks that require access to the model's internal parameters; interactive black-box refers to attacks that do not require access to the model's internal parameters but require the ability to query the model; and complete black-box refers to attacks that do not require any access to the model);
    
\noindent \textbf{3) Attack phase:} Whether the attack compromises the model's training or the serving (inference) phase.

%%\vspace*{-\baselineskip}
%\begin{comment}
   
%%\vspace*{-\baselineskip}
%\end{comment}

%%\vspace{-10pt}
%\vspace*{-\baselineskip}
%%\vspace{-13pt}
\subsection{Evasion attacks \label{subsubsec:evasion_attacks}} 
In evasion attacks the adversary exploits the ML model by generating a crafted input sample (an adversarial example) which is very similar to some other correctly classified input but is incorrectly classified by the ML model.
The adversary's main objective is to compromise the integrity \textit{[AG1]} of the ML model \textit{[F6]} by causing the ML system to provide incorrect outputs for a specific input \textit{[T1]}. 
Evasion attacks can be broadly classified based on the following criteria: attack technique, which characterizes the technique used by the attacker to craft the adversarial example (e.g., gradient-based, boundary-based, and transferability-based); the threat model, which characterizes the attacker’s access capabilities and attacker's knowledge of the ML-based system  (e.g., white-box, black-box); and the attack's specificity, which characterizes the goal of the attacker (e.g., targeted vs untargeted).  

\noindent \textbf{[AF1] Gradient-based evasion attacks (white-box).}
In these types of attacks, the adversary manipulates an input sample in such a way that the classification loss is maximized.
The specific manipulation is determined by calculating the gradients of the classification loss with respect to the input sample~\cite{goodfellow2014explaining,kurakin2016adversarial,carlini2017adversarial}.
For example, in the FGSM attack~\cite{goodfellow2014explaining}, the adversarial example is calculated by adding noise to the input sample, in the direction of the gradient of the classification loss with respect to the input sample.
Since these types of attacks are based on gradient computation, 
they consider a white-box adversary with perfect knowledge of the target pipeline \textit{[AKM1] / [AKD1]}.
In addition, they further assume that the adversary has the ability to query the trained model \textit{[ACM1] / [ACM2]}; this requirement is necessary for executing the attack (i.e., sending the crafted adversarial example for classification).

\noindent \textbf{[AF2] Query-based evasion attacks (interactive black-box).}
Similar to \textit{[AF1]}, these types of attacks generate the adversarial example based on the gradients of the classification loss with respect to the input sample, however in contrast to \textit{[AF1]} that \textit{calculate} the gradients, these attacks \textit{estimate} the gradients.
For example, in the ZOO attack~\cite{chen2017zoo}, the adversary utilizes zeroth-order optimization methods to directly estimate the gradients of the target model.
Since these attacks are not based on gradient computation, they consider an adversary with general knowledge of the task \textit{[AKM3]} and black-box query access \textit{[ACM1]}.
In Section~\ref{sec:demonstration} we present demonstration of a query-based evasion attack within the traffic steering use case.

\noindent \textbf{[AF3] Transferability-based evasion attacks (complete black-box)}
These types of attacks include the following three main phases:
First, the adversary creates a surrogate model. The surrogate model can be constructed by training a learning algorithm on a reference dataset with similar characteristics and distribution as the training set \textit{[AKD4]}.
Second, the adversary generates adversarial examples by executing white-box gradient-based attacks on the surrogate model.
Third, the adversary uses the adversarial examples (generated against the surrogate model) to attack the target model.
These attacks exploit the \textit{transferability} property of ML models~\cite{papernot2016transferability,szegedy2013intriguing,goodfellow2014explaining}, i.e., adversarial examples that affect one model can often affect other models, even if they trained using different learning algorithms, hyperparameters, or training sets, as long as all models were trained to perform the same task (e.g., image classification).
For example in~\cite{papernot2016transferability}, the adversary utilizes Jacobian data augmentation to generate a surrogate dataset given an initial substitute training set of limited size.  
Since the gradient-based attacks are executed on a surrogate model, they can be executed by a black-box attacker with very limited query access \textit{[ACM2]} as long as the adversary knows the general task \textit{[AKM3]} and data properties [AKD4] and has access to a reference dataset \textit{[AKD4]}.

%\vspace*{-\baselineskip}%%\vspace{-15pt}
\subsection{Poisoning attacks}
%\vspace{-5pt}
\noindent \textbf{[AF4] Gradient-based poisoning attacks (white-box).}
In these types of attacks, the adversary manipulates a small number of training samples in such a way that the classification loss of some other set of samples targeted by the attacker is maximized \textit{[T1],[AG1]}.
The specific manipulation is calculated by solving, using gradient-optimization techniques, a bilevel optimization problem where the outer optimization aims at manipulating the malicious input to maximize the loss function on a dataset that includes the samples targeted by the attacker, while the inner optimization corresponds to retraining the learning algorithm on a dataset that includes the malicious examples~\cite{kravchik2021poisoning,mei2015using,jagielski2018manipulating}.
Since these types of attacks are based on the adversary's ability to manipulate a small number of training samples, the adversary must have the ability to write to the dataset used to train the model \textit{[ACD1]}.   
In addition, since these attacks are based on gradient computation, the adversary must possess perfect knowledge of the target pipeline (including the model architecture and parameters) \textit{[AKM1], [AKD1]}. 
Furthermore, to execute the attack, the adversary must have the ability to query the trained model \textit{[ACM2]}. 

\noindent \textbf{[AF5] Transferability-based poisoning attacks (complete black-box).}
Similar to \textit{[AF4]}, the adversary manipulates a small number of training samples in a such a way that the classification loss of some other set of samples targeted by the attacker is maximized \textit{[T1],[AG1]}.
Therefore, these types of attacks also require the adversary to have the ability to write into the dataset used to train the model \textit{[ACD1]}.   
However, in contrast to \textit{[AF6]} which calculate the gradients using the target model (which requires perfect knowledge of the system), these attacks calculate the gradients using a surrogate model~\cite{biggio2012poisoning,munoz2017towards,jagielski2018manipulating} (created by training a learning algorithm on a reference dataset whose  characteristics %and distribution 
are similar to the training set) \textit{[AKD4]}.
As a result, they can be executed by a black-box attacker with very limited query access \textit{[ACM2]}, as long as the adversary knows the general task \textit{[AKM3]} and data properties [AKD4].

\subsection{Membership Inference attacks \label{subsubsec:inference_attacks}}
%\vspace{-5pt}
\noindent \textbf{[AF6] Gradient-based membership inference attack (white-box).}
In these types of attacks, an adversary has partial knowledge on the training data \textit{[AKD1]}, \textit{[ACD1]}; and complete access \textit{[ACM]} and knowledge \textit{[AKM]} on the target system's model. Membership inference \textit{[T3]} is performed in these types of attacks based on gradient behavior observed during the training phase~\cite{nasr2019comprehensive}.

\noindent \textbf{[AF7] Query-based membership inference attacks (interactive black-box).}
In these types of attacks, given an ML model \textit{[T6]} and a record, the adversary's goal is to determine whether the record was used as part of the model's training dataset \textit{[T3]} and thereby compromise the privacy of that record \textit{[AG3]}. These attacks exploit the fact that ML models often behave differently on the data that they were trained on than they behave on the test data~\cite{shokri2017membership}. 
Specifically, as introduced in~\cite{shokri2017membership}, these attacks utilize shadow models that only require access to the prediction vector \textit{[ACM1]} of the target ML system. However, these attacks assume the adversary has partial information about the target system's training data by exploiting either partial access to raw \textit{[ACD3]} and training data \textit{[ACD1]} or by having prior partial knowledge on the training data \textit{[AKD1]}.

% \noindent \textbf{[AT16] Shadow training-based property inference attacks.}
% In these types of attacks, an adversary extracts information about the features that are not correlated with the learning task \textit{[T4]}.  These attacks can be performed using shadow models, leveraging the adversary's access to the target model's predictions (\textit{[ACM1]},\textit{[ACM2]}), and partial knowledge of the training data, leveraging \textit{[ACD1]}, and the target system's task \textit{[AKM3]}~\cite{ateniese2015hacking,ganju2018property,song2020information}.

%\vspace{-15pt}
\subsection{Data Reconstruction}
% \noindent \textbf{[AT17] Maximum a posteriori-based gray-box data reconstruction attacks.}
% In these types of attacks~\cite{fredrikson2014privacy,hidano2017model} falling under \textit{[T5]}, an adversary reconstructs training samples, and their respective labels \textit{[F1],[F3]}, of the target system's ML model \textit{[F6]}. 
% The adversary uses a maximum a posteriori (MAP) estimate of the attribute that maximizes the probability of observing the known parameters while assuming partial access \textit{[ACD1], [ACM1]} and knowledge \textit{[AKD1], [AKM3]} of the target model \textit{[F6]} and feature information from the training data.
%\vspace{-5pt}
\noindent \textbf{[AF8] Gradient-based data reconstruction attacks (white-box).}
In these attacks \cite{fredrikson2015model}, the adversary utilizes gradient optimization methods to  solve an optimization problem, which tries to recover the input data point for a given class \textit{[T5]}.
Since these attacks are based on gradient computation, they require a complete model knowledge \textit{[AKM]}, task knowledge \textit{[AKM3]}, and partial training data knowledge \textit{[AKD1]}. 
It should be mentioned that the adversary can obtain the required knowledge by exploiting \textit{[ACM1]} and \textit{[ACD1]}.

\noindent \textbf{[AF9] Query-based data reconstruction attacks (interactive black-box).}
In these types of attacks~\cite{yang2019neural}, an adversary reconstructs training samples \textit{[T5]} while having limited training task information [AKM3] and only query access \textit{[ACM1]} to the target model. These attacks use an autoencoder setting where the target model plays the role of an encoder, and the trainable decoder network tries to reconstruct the training sample on the prediction vector \textit{[ACM1]} of the target model.
%%\vspace{-10pt}
%\vspace*{-\baselineskip}
\subsection{Model Extraction}
%\vspace{-5pt}
\noindent \textbf{[AF10] Query-based model extraction attacks (black-box).}
In these types of attacks~\cite{chandrasekaran2020exploring,juuti2019prada,oh2019towards,papernot2017practical}, an adversary tries to extract complete information of a target ML model \textit{[T6]} by fully reconstructing it or creating a substitute model that behaves very similarly \textit{[T6]}.
These attacks are preformed by querying the model with specific inputs crafted by the adversary.
Therefore, they require query access \textit{[ACM1]} to the target model and limited knowledge of the training task \textit{[AKM2],[AKD1],[AKM3]}.

%%\vspace*{-\baselineskip}
%\vspace{-10pt}
\subsection{Resource Exhaustion}
%\vspace{-5pt}
\noindent \textbf{[AF11] Gradient-based resource exhaustion attacks (white-box)} 
In these types of attacks~\cite{shumailov2020sponge,shapira2023phantom},
the adversary searches (using gradient optimization method) for inputs that increase activation values of the model across all of the layers simultaneously.
High activation values prevent hardware optimization and therefore increase latency.
Since these types of attacks are based on gradient computation, they consider a white-box adversary with perfect knowledge of the target pipeline \textit{[AKM1] / [AKD1]}.
In addition, they %further 
assume that the adversary has the ability to query the trained model \textit{[ACM1] / [ACM2]}; this requirement is necessary for executing the attack (i.e., sending the crafted %adversarial 
example for classification).

\noindent \textbf{[AF12] Query-based resource exhaustion attacks (interactive black-box)} 
In these types of attacks \cite{shumailov2020sponge},
the adversary searches (using a genetic algorithm) for inputs that increase model's latency.
In contrast to the gradient-based method, this method does not require access to the model's parameters.
Therefore, they require query access \textit{[ACM1]} to the target model.

\noindent \textbf{[AF13] Transferability-based resource exhaustion attacks (complete black-box)} 
In this type of attack~\cite{shumailov2020sponge}, the adversary creates a surrogate model and searches (using a genetic algorithm or gradient-based optimization) for inputs that increase models' latency.
Since the gradient-based optimization are executed on a surrogate model, they can be executed by a black-box attacker with %very 
limited query access \textit{[ACM2]} as long as the adversary knows the general task \textit{[AKM3]} and data properties [AKD4] and has access to a reference dataset \textit{[AKD4]}.

\subsection{Enumeration of Attack Techniques}
In Table~\ref{tab:attack_techniques_with_actors} we enumerate the AML threats to a general ML pipeline in \oran by mapping all attack techniques to the relevant threat model and threat actors.
In each row we define a specific threat by mapping an attack technique (which belongs to one of the attack families [AF1] - [AF13] described in Section~\ref{sec:attack_tech}), and evaluating the required capabilities that attackers must possess to successfully carry out an attack (described in Section~\ref{subsec:attacermodel}), and the impact of the attack (listed in Section~\ref{subsec:attacker_goal}); these are indicated by the `threat model` columns. 
Furthermore, we list the potential threat actors (described in Section~\ref{subsec:threatactor}) who possess the required capabilities to use the attack technique.

For example, in the \textit{Gradient-based evasion} attack family, there are two possible attack techniques: the first, is ``training data and hyperparameter knowledge" and the second is ``model knowledge".
In the ``training data and hyperparameter knowledge" attack technique, the attacker generates a surrogate model using a surrogate dataset which is similar to the training set used to train the target model, and based on a prior knowledge on the target model's hyperparameters.
Then, the attacker generates the adversarial example on the surrogate model, and send the adversarial example to the target model.
Therefore, for the ``training data and hyperparameter knowledge" attack technique, `Training Data Knowledge' and Hyperparameter Knowledge' are required for generating the adversarial example and `Sensor Data Access,' `Label Data Access,' Surrogate Data Access,' and `Decision-Based Query Access' are the capabilities required for executing the attack.
The impact of such attack technique is `Tempering' with the model's decision.
Such capabilities are available, for example, to the \textit{[A1]} threat actor with access to the training host at every in every deployment scenario (DS).

\setlength\tabcolsep{4.05pt}
\begin{table*}[tb]
\tiny
\centering
\hspace*{-1.5cm}
% \rotatebox{90}
%\resizebox{1.15\textwidth}{0.42\textheight}
{\begin{tabular}{@{}
|m{0.10\textwidth}"  % Threat Category
m{0.24\textwidth}|  % Attack Technique
m{0.001\textwidth}m{0.001\textwidth}m{0.001\textwidth}m{0.001\textwidth}|  % AKD
m{0.001\textwidth}m{0.001\textwidth}m{0.001\textwidth}m{0.001\textwidth}|  % AKM
m{0.001\textwidth}m{0.001\textwidth}m{0.001\textwidth}m{0.001\textwidth}m{0.001\textwidth}m{0.001\textwidth}|  % ACD
m{0.001\textwidth}m{0.001\textwidth}|  % ACM
m{0.001\textwidth}m{0.001\textwidth}m{0.001\textwidth}"  % Impact
% ||m{0.001\textwidth}|  % Help Titles
m{0.001\textwidth}m{0.001\textwidth}m{0.001\textwidth}m{0.001\textwidth}m{0.001\textwidth}|  % SW App Dev
m{0.001\textwidth}m{0.001\textwidth}m{0.001\textwidth}m{0.001\textwidth}m{0.001\textwidth}|  % SW ISX Dev-nonRT
m{0.001\textwidth}m{0.001\textwidth}m{0.001\textwidth}m{0.001\textwidth}m{0.001\textwidth}|  % SW ISX Dev-nRT
m{0.001\textwidth}m{0.001\textwidth}m{0.001\textwidth}m{0.001\textwidth}m{0.001\textwidth}|  % SW ISX Dev-CUDU
c|  % Contatiner SW ISX Prov
c|  % HW ISX Prov
c|  % UE
c| % 3rd party data provider
% @{}  % Last col space=0
}
\Xhline{3\arrayrulewidth}

\multirow{5}{*}{\makecell{\textbf{Attack}\\\textbf{Family}}} &
\multirow{5}{*}{\textbf{Attack Technique}} &
\multicolumn{19}{c"}{\textbf{Threat Model}} & 
\multicolumn{24}{c|}{\textbf{Threat Actor}} \\
\cline{3-45} 

& & 
\multicolumn{8}{c|}{Knowledge} &
\multicolumn{8}{c|}{Access} & 
\multicolumn{3}{c"}{Impact} & 
\multicolumn{5}{c|}{\textbf{\textit{A1}} } &  % O-RAN SW-App Dev
\multicolumn{15}{c|}{\textbf{\textit{A2}} } &  % O-RAN SW-ISX Dev
\textbf{\textit{A3}} &  % Containerization SW-ISX Prov
\textbf{\textit{A4}} &  % HW-IDX Prov
\textbf{\textit{A5}} &  % UE
\textbf{\textit{A6}} \\  % 3rd Party

\cline{3-45}

& & 
\multicolumn{4}{c|}{AKD} &  % Knowledge
\multicolumn{4}{c|}{AKM} &  % Knowledge
\multicolumn{6}{c|}{ACD} &  % Access
\multicolumn{2}{c|}{ACM} &  % Access
\multicolumn{3}{c"}{} &  % Impact
\rotatebox{90}{Data Collection} &  % SW-App Dev
\rotatebox{90}{Data Host} &  % SW-App Dev
\rotatebox{90}{Training Host} &  % SW-App Dev
\rotatebox{90}{Serving Host} &  % SW-App Dev
\rotatebox{90}{ML App} &  % SW-App Dev
\multicolumn{5}{c|}{\makecell{Non-RT\\RIC}} &  % SW-ISX Dev
\multicolumn{5}{c|}{\makecell{Near-RT\\RIC}} &  % SW-ISX Dev
\multicolumn{5}{c|}{\makecell{O-CU / \\O-DU}} &  % SW-ISX Dev
\rotatebox{90}{\makecell{All}} &  % C14n SW-ISX Prov
\rotatebox{90}{\makecell{All}} &  % HW-ISX Prov
\rotatebox{90}{\makecell{All}} & % UE 
\rotatebox{90}{\makecell{All}} \\ % 3rd Party data provider
\cline{3-45} 

& & 
% Data Knowledge
\rotatebox{90}{Training Data Knowledge} & 
\rotatebox{90}{Features Data Knowledge} & 
\rotatebox{90}{Raw Data Knowledge} & 
\rotatebox{90}{Data Property Knowledge} &

% Model Knowledge
\rotatebox{90}{Model Knowledge} &
\rotatebox{90}{Hyperparameter Knowledge} &
\rotatebox{90}{Algorithm Knowledge} &
\rotatebox{90}{Task Knowledge} &

% Data Access
\rotatebox{90}{Training Data Access} &
\rotatebox{90}{Features Data Access} &
\rotatebox{90}{Raw Data Access} &
\rotatebox{90}{Sensor Data Access} &
\rotatebox{90}{Label Data Access} &
\rotatebox{90}{Surrogate Data Access} &

% Model Access
\rotatebox{90}{Score-Based Query Access} &
\rotatebox{90}{Decision-Based Query Access} &

% Impact
\rotatebox{90}{Tampering} &
\rotatebox{90}{Denial of Service} &
\rotatebox{90}{Information Disclosure} & 

% O-RAN SW-App Dev (DS)
\multicolumn{5}{c|}{\rotatebox{90}{\makecell{All}}} & 

% O-RAN SW-ISX Dev (DS)
\rotatebox{90}{\textit{DS 1}} & \rotatebox{90}{\textit{DS 2}} & \rotatebox{90}{\textit{DS 3}} & \rotatebox{90}{\textit{DS 4}} & \rotatebox{90}{\textit{DS 5}} &  %  Non-RT RIC
\rotatebox{90}{\textit{DS 1}} & \rotatebox{90}{\textit{DS 2}} & \rotatebox{90}{\textit{DS 3}} & \rotatebox{90}{\textit{DS 4}} & \rotatebox{90}{\textit{DS 5}} &  %  Near-RT RIC
\rotatebox{90}{\textit{DS 1}} & \rotatebox{90}{\textit{DS 2}} & \rotatebox{90}{\textit{DS 3}} & \rotatebox{90}{\textit{DS 4}} & \rotatebox{90}{\textit{DS 5}} &  %  O-CU/O-DU
\rotatebox{90}{All} &  % C14n SW-IDX Prov
\rotatebox{90}{All} &  % HW-IDX Prov
\rotatebox{90}{All} &  % UE
\rotatebox{90}{All} \\  % 3rd party data provider

\Xhline{3\arrayrulewidth}

% Evasion
	
\rowcolor{Grey}
\multicolumn{45}{|c|}{\textbf{Threat Category:} Evasion}\\ 
\multirow{2}{*}{\textbf{\shortstack[l]{Gradient-based \\evasion attacks}}} 
&
training data and hyperparameter knowledge &
$\bullet$ & $\circ$ & $\circ$  & $\circ$ &  % AKD
$\circ$ & $\bullet$ & $\circ$ & $\circ$ &  % AKM
$\circ$ & $\circ$ & $\circ$ & $\odot$ & $\odot$ & $\odot$ &% ACD
$\circ$ & $\odot$ &  % ACM
$\bullet$ & $\circ$ & $\circ$ &  % AG
$\circ$ & $\circ$ & $\bullet$ & $\circ$ & $\circ$ &  % O-RAN SW-App Dev
$\bullet$ & $\bullet$ & $\circ$ & $\bullet$ & $\circ$ &  % O-RAN SW-ISX Dev - Non
$\circ$ & $\bullet$ & $\bullet$ & $\circ$ & $\circ$ &  % O-RAN SW-ISX Dev - Near
$\circ$ & $\circ$ & $\circ$ & $\circ$ & $\bullet$ &  % O-RAN SW-ISX Dev - CUDU
$\bullet$ &  % C14n SW-ISX Prov
$\bullet$ &  % HW-ISX Prov
$\circ$ &  % UE
$\bullet$ \\  % 3rd party data provider

&
model knowledge &
$\circ$ & $\circ$ & $\circ$  & $\circ$ &  % AKD
$\bullet$ & $\circ$ & $\circ$ & $\circ$ &  % AKM
$\circ$ & $\circ$ & $\circ$ & $\odot$ & $\odot$ & $\odot$ & % ACD
$\circ$ & $\odot$ &  % ACM
$\bullet$ & $\circ$ & $\circ$ &  % AG
$\circ$ & $\circ$ & $\bullet$ & $\bullet$ & $\circ$ &  % O-RAN SW-App Dev
$\bullet$ & $\bullet$ & $\circ$ & $\bullet$ & $\circ$ &  % O-RAN SW-ISX Dev - Non
$\circ$ & $\bullet$ & $\bullet$ & $\circ$ & $\circ$ &  % O-RAN SW-ISX Dev - Near
$\circ$ & $\circ$ & $\circ$ & $\bullet$ & $\bullet$ &  % O-RAN SW-ISX Dev - CUDU
$\bullet$ &  % C14n SW-ISX Prov
$\bullet$ &  % HW-ISX Prov
$\circ$ &  % UE
$\bullet$ \\  % 3rd party data provider

\hline

\multirow{2}{*}{\textbf{\shortstack[l]{Query-based \\evasion attacks}}} 
&
model's score (i.e., vector of probabilities) &
$\circ$ & $\circ$ & $\circ$  & $\circ$  &  % AKD
$\circ$ & $\circ$ & $\circ$ & $\bullet$ &  % AKM
$\circ$ & $\circ$ & $\circ$ & $\bullet$ & $\odot$ & $\odot$ & % ACD
$\bullet$ & $\circ$ &  % ACM
$\bullet$ & $\circ$ & $\circ$ &  % AG
$\circ$ & $\circ$ & $\bullet$ & $\bullet$ & $\bullet$ &  % O-RAN SW-App Dev
$\bullet$ & $\bullet$ & $\circ$ & $\bullet$ & $\circ$ &  % O-RAN SW-ISX Dev - Non
$\bullet$ & $\bullet$ & $\bullet$ & $\circ$ & $\circ$ &  % O-RAN SW-ISX Dev - Near
$\circ$ & $\circ$ & $\circ$ & $\bullet$ & $\bullet$ &  % O-RAN SW-ISX Dev - CUDU
$\bullet$ &  % C14n SW-ISX Prov
$\bullet$ &  % HW-ISX Prov
$\circ$ &  % UE
$\bullet$ \\  % 3rd party data provider

&
model's decision &
$\circ$ & $\circ$ & $\circ$  & $\circ$  &  % AKD
$\circ$ & $\circ$ & $\circ$ & $\bullet$ &  % AKM
$\circ$ & $\circ$ & $\circ$ & $\bullet$ & $\odot$ & $\odot$ & % ACD
$\circ$ & $\bullet$ &  % ACM
$\bullet$ & $\circ$ & $\circ$ &  % AG
$\circ$ & $\circ$ & $\bullet$ & $\bullet$ & $\bullet$ &  % O-RAN SW-App Dev
$\bullet$ & $\bullet$ & $\circ$ & $\bullet$ & $\circ$ &  % O-RAN SW-ISX Dev - Non
$\bullet$ & $\bullet$ & $\bullet$ & $\circ$ & $\circ$ &  % O-RAN SW-ISX Dev - Near
$\circ$ & $\circ$ & $\circ$ & $\bullet$ & $\bullet$ &  % O-RAN SW-ISX Dev - CUDU
$\bullet$ &  % C14n SW-ISX Prov
$\bullet$ &  % HW-ISX Prov
$\bullet$ &  % UE
$\bullet$ \\  % 3rd party data provider

\hline

\multirow{12}{*}{\textbf{\shortstack[l]{Transferability-based\\evasion attacks}}} 
&
training data and hyperparameter knowledge &
$\bullet$ & $\circ$ & $\circ$  & $\circ$  &  % AKD
$\circ$ & $\bullet$ & $\circ$ & $\circ$ &  % AKM
$\circ$ & $\circ$ & $\circ$ & $\odot$ & $\odot$ & $\odot$ & % ACD
$\circ$ & $\odot$ &  % ACM
$\bullet$ & $\circ$ & $\circ$ &  % AG
$\circ$ & $\circ$ & $\bullet$ & $\circ$ & $\circ$ &  % O-RAN SW-App Dev
$\bullet$ & $\bullet$ & $\circ$ & $\bullet$ & $\circ$ &  % O-RAN SW-ISX Dev - Non
$\circ$ & $\bullet$ & $\bullet$ & $\circ$ & $\circ$ &  % O-RAN SW-ISX Dev - Near
$\circ$ & $\circ$ & $\circ$ & $\bullet$ & $\bullet$ &  % O-RAN SW-ISX Dev - CUDU
$\bullet$ &  % C14n SW-ISX Prov
$\bullet$ &  % HW-ISX Prov
$\circ$ &  % UE
$\bullet$ \\  % 3rd party data provider

&
features data and hyperparameter knowledge &
$\circ$ & $\bullet$ & $\circ$  & $\circ$  &  % AKD
$\circ$ & $\bullet$ & $\circ$ & $\circ$ &  % AKM
$\circ$ & $\circ$ & $\circ$ & $\odot$ & $\odot$ & $\odot$ & % ACD
$\circ$ & $\odot$ &  % ACM
$\bullet$ & $\circ$ & $\circ$ &  % AG
$\circ$ & $\circ$ & $\bullet$ & $\bullet$ & $\circ$ &  % O-RAN SW-App Dev
$\bullet$ & $\bullet$ & $\circ$ & $\bullet$ & $\circ$ &  % O-RAN SW-ISX Dev - Non
$\circ$ & $\bullet$ & $\bullet$ & $\circ$ & $\circ$ &  % O-RAN SW-ISX Dev - Near
$\circ$ & $\circ$ & $\circ$ & $\bullet$ & $\bullet$ &  % O-RAN SW-ISX Dev - CUDU
$\bullet$ &  % C14n SW-ISX Prov
$\bullet$ &  % HW-ISX Prov
$\circ$ &  % UE
$\bullet$ \\  % 3rd party data provider

&
raw data and hyperparameter knowledge &
$\circ$ & $\circ$ & $\bullet$  & $\circ$  &  % AKD
$\circ$ & $\bullet$ & $\circ$ & $\circ$ &  % AKM
$\circ$ & $\circ$ & $\circ$ & $\odot$ & $\odot$ & $\odot$ & % ACD
$\circ$ & $\odot$ &  % ACM
$\bullet$ & $\circ$ & $\circ$ &  % AG
$\circ$ & $\circ$ & $\bullet$ & $\bullet$ & $\circ$ &  % O-RAN SW-App Dev
$\bullet$ & $\bullet$ & $\circ$ & $\bullet$ & $\circ$ &  % O-RAN SW-ISX Dev - Non
$\circ$ & $\bullet$ & $\bullet$ & $\circ$ & $\circ$ &  % O-RAN SW-ISX Dev - Near
$\circ$ & $\circ$ & $\circ$ & $\bullet$ & $\bullet$ &  % O-RAN SW-ISX Dev - CUDU
$\bullet$ &  % C14n SW-ISX Prov
$\bullet$ &  % HW-ISX Prov
$\circ$ &  % UE
$\bullet$ \\  % 3rd party data provider

& 
data property and hyperparameter knowledge &
$\circ$ & $\circ$ & $\circ$  & $\bullet$  &  % AKD
$\circ$ & $\bullet$ & $\circ$ & $\circ$ &  % AKM
$\circ$ & $\circ$ & $\circ$ & $\odot$ & $\odot$ & $\odot$ & % ACD
$\circ$ & $\odot$ &  % ACM
$\bullet$ & $\circ$ & $\circ$ &  % AG
$\circ$ & $\circ$ & $\bullet$ & $\bullet$ & $\circ$ &  % O-RAN SW-App Dev
$\bullet$ & $\bullet$ & $\circ$ & $\bullet$ & $\circ$ &  % O-RAN SW-ISX Dev - Non
$\circ$ & $\bullet$ & $\bullet$ & $\circ$ & $\circ$ &  % O-RAN SW-ISX Dev - Near
$\circ$ & $\circ$ & $\circ$ & $\bullet$ & $\bullet$ &  % O-RAN SW-ISX Dev - CUDU
$\bullet$ &  % C14n SW-ISX Prov
$\bullet$ &  % HW-ISX Prov
$\circ$ &  % UE
$\bullet$ \\  % 3rd party data provider

&
training data and algorithm knowledge &
$\bullet$ & $\circ$ & $\circ$  & $\circ$  &  % AKD
$\circ$ & $\circ$ & $\bullet$ & $\circ$ &  % AKM
$\circ$ & $\circ$ & $\circ$ & $\odot$ & $\odot$ & $\odot$ & % ACD
$\circ$ & $\odot$ &  % ACM
$\bullet$ & $\circ$ & $\circ$ &  % AG
$\circ$ & $\circ$ & $\bullet$ & $\circ$ & $\circ$ &  % O-RAN SW-App Dev
$\bullet$ & $\bullet$ & $\circ$ & $\bullet$ & $\circ$ &  % O-RAN SW-ISX Dev - Non
$\circ$ & $\bullet$ & $\bullet$ & $\circ$ & $\circ$ &  % O-RAN SW-ISX Dev - Near
$\circ$ & $\circ$ & $\circ$ & $\bullet$ & $\bullet$ &  % O-RAN SW-ISX Dev - CUDU
$\bullet$ &  % C14n SW-ISX Prov
$\bullet$ &  % HW-ISX Prov
$\circ$ &  % UE
$\bullet$ \\  % 3rd party data provider

&
features data and algorithm knowledge &
$\circ$ & $\bullet$ & $\circ$  & $\circ$  &  % AKD
$\circ$ & $\circ$ & $\bullet$ & $\circ$ &  % AKM
$\circ$ & $\circ$ & $\circ$ & $\odot$ & $\odot$ & $\odot$ & % ACD
$\circ$ & $\odot$ &  % ACM
$\bullet$ & $\circ$ & $\circ$ &  % AG
$\circ$ & $\circ$ & $\bullet$ & $\circ$ & $\circ$ &  % O-RAN SW-App Dev
$\bullet$ & $\bullet$ & $\circ$ & $\bullet$ & $\circ$ &  % O-RAN SW-ISX Dev - Non
$\circ$ & $\bullet$ & $\bullet$ & $\circ$ & $\circ$ &  % O-RAN SW-ISX Dev - Near
$\circ$ & $\circ$ & $\circ$ & $\bullet$ & $\bullet$ &  % O-RAN SW-ISX Dev - CUDU
$\bullet$ &  % C14n SW-ISX Prov
$\bullet$ &  % HW-ISX Prov
$\circ$ &  % UE
$\bullet$ \\  % 3rd party data provider

&
raw data and algorithm knowledge &
$\circ$ & $\circ$ & $\bullet$  & $\circ$  &  % AKD
$\circ$ & $\circ$ & $\bullet$ & $\circ$ &  % AKM
$\circ$ & $\circ$ & $\circ$ & $\odot$ & $\odot$ & $\odot$ & % ACD
$\circ$ & $\odot$ &  % ACM
$\bullet$ & $\circ$ & $\circ$ &  % AG
$\circ$ & $\circ$ & $\bullet$ & $\circ$ & $\circ$ &  % O-RAN SW-App Dev
$\bullet$ & $\bullet$ & $\circ$ & $\bullet$ & $\circ$ &  % O-RAN SW-ISX Dev - Non
$\circ$ & $\bullet$ & $\bullet$ & $\circ$ & $\circ$ &  % O-RAN SW-ISX Dev - Near
$\circ$ & $\circ$ & $\circ$ & $\bullet$ & $\bullet$ &  % O-RAN SW-ISX Dev - CUDU
$\bullet$ &  % C14n SW-ISX Prov
$\bullet$ &  % HW-ISX Prov
$\circ$ &  % UE
$\bullet$ \\  % 3rd party data provider

&
data property and algorithm knowledge &
$\circ$ & $\circ$ & $\circ$  & $\bullet$  &  % AKD
$\circ$ & $\circ$ & $\bullet$ & $\circ$ &  % AKM
$\circ$ & $\circ$ & $\circ$ & $\odot$ & $\odot$ & $\odot$ & % ACD
$\circ$ & $\odot$ &  % ACM
$\bullet$ & $\circ$ & $\circ$ &  % AG
$\circ$ & $\circ$ & $\bullet$ & $\bullet$ & $\circ$ &  % O-RAN SW-App Dev
$\bullet$ & $\bullet$ & $\circ$ & $\bullet$ & $\circ$ &  % O-RAN SW-ISX Dev - Non
$\circ$ & $\bullet$ & $\bullet$ & $\circ$ & $\circ$ &  % O-RAN SW-ISX Dev - Near
$\circ$ & $\circ$ & $\circ$ & $\bullet$ & $\bullet$ &  % O-RAN SW-ISX Dev - CUDU
$\bullet$ &  % C14n SW-ISX Prov
$\bullet$ &  % HW-ISX Prov
$\circ$ &  % UE
$\bullet$ \\  % 3rd party data provider

&
training data and task knowledge &
$\bullet$ & $\circ$ & $\circ$  & $\circ$  &  % AKD
$\circ$ & $\circ$ & $\circ$ & $\bullet$ &  % AKM
$\circ$ & $\circ$ & $\circ$ & $\odot$ & $\odot$ & $\odot$ & % ACD
$\circ$ & $\odot$ &  % ACM
$\bullet$ & $\circ$ & $\circ$ &  % AG
$\circ$ & $\circ$ & $\bullet$ & $\circ$ & $\circ$ &  % O-RAN SW-App Dev
$\bullet$ & $\bullet$ & $\circ$ & $\bullet$ & $\circ$ &  % O-RAN SW-ISX Dev - Non
$\bullet$ & $\bullet$ & $\bullet$ & $\circ$ & $\circ$ &  % O-RAN SW-ISX Dev - Near
$\circ$ & $\circ$ & $\circ$ & $\bullet$ & $\bullet$ &  % O-RAN SW-ISX Dev - CUDU
$\bullet$ &  % C14n SW-ISX Prov
$\bullet$ &  % HW-ISX Prov
$\circ$ &  % UE
$\bullet$ \\  % 3rd party data provider

&
features data and task knowledge &
$\circ$ & $\bullet$ & $\circ$  & $\circ$  &  % AKD
$\circ$ & $\circ$ & $\circ$ & $\bullet$ &  % AKM
$\circ$ & $\circ$ & $\circ$ & $\odot$ &  $\odot$ & $\odot$ & % ACD
$\circ$ & $\odot$ &  % ACM
$\bullet$ & $\circ$ & $\circ$ &  % AG
$\circ$ & $\bullet$ & $\bullet$ & $\bullet$ & $\circ$ &  % O-RAN SW-App Dev
$\bullet$ & $\bullet$ & $\circ$ & $\bullet$ & $\circ$ &  % O-RAN SW-ISX Dev - Non
$\bullet$ & $\bullet$ & $\bullet$ & $\circ$ & $\circ$ &  % O-RAN SW-ISX Dev - Near
$\circ$ & $\circ$ & $\circ$ & $\bullet$ & $\bullet$ &  % O-RAN SW-ISX Dev - CUDU
$\bullet$ &  % C14n SW-ISX Prov
$\bullet$ &  % HW-ISX Prov
$\circ$ &  % UE
$\bullet$ \\  % 3rd party data provider

&
raw data and task knowledge &
$\circ$ & $\circ$ & $\bullet$  & $\circ$  &  % AKD
$\circ$ & $\circ$ & $\circ$ & $\bullet$ &  % AKM
$\circ$ & $\circ$ & $\circ$ & $\odot$ &  $\odot$ & $\odot$ & % ACD
$\circ$ & $\odot$ &  % ACM
$\bullet$ & $\circ$ & $\circ$ &  % AG
$\bullet$ & $\bullet$ & $\bullet$ & $\bullet$ & $\bullet$ &  % O-RAN SW-App Dev
$\bullet$ & $\bullet$ & $\circ$ & $\bullet$ & $\circ$ &  % O-RAN SW-ISX Dev - Non
$\bullet$ & $\bullet$ & $\bullet$ & $\circ$ & $\circ$ &  % O-RAN SW-ISX Dev - Near
$\bullet$ & $\bullet$ & $\bullet$ & $\bullet$ & $\bullet$ &  % O-RAN SW-ISX Dev - CUDU
$\bullet$ &  % C14n SW-ISX Prov
$\bullet$ &  % HW-ISX Prov
$\circ$ &  % UE
$\bullet$ \\  % 3rd party data provider

&
data property and task knowledge &
$\circ$ & $\circ$ & $\circ$  & $\bullet$  &  % AKD
$\circ$ & $\circ$ & $\circ$ & $\bullet$ &  % AKM
$\circ$ & $\circ$ & $\circ$ & $\odot$ &  $\odot$ & $\odot$ & % ACD
$\circ$ & $\odot$ &  % ACM
$\bullet$ & $\circ$ & $\circ$ &  % AG
$\bullet$ & $\bullet$ & $\bullet$ & $\bullet$ & $\bullet$ &  % O-RAN SW-App Dev
$\bullet$ & $\bullet$ & $\circ$ & $\bullet$ & $\circ$ &  % O-RAN SW-ISX Dev - Non
$\bullet$ & $\bullet$ & $\bullet$ & $\circ$ & $\circ$ &  % O-RAN SW-ISX Dev - Near
$\bullet$ & $\bullet$ & $\bullet$ & $\bullet$ & $\bullet$ &  % O-RAN SW-ISX Dev - CUDU
$\bullet$ &  % C14n SW-ISX Prov
$\bullet$ &  % HW-ISX Prov
$\circ$ &  % UE
$\bullet$ \\  % 3rd party data provider

\Xhline{3\arrayrulewidth}

\rowcolor{Grey}
\multicolumn{45}{|c|}{\textbf{Threat Category:} Model Corruption}\\ 

\textbf{\shortstack[l]{Gradient-based\\poisoning attacks}}
&
model knowledge &
$\circ$ & $\circ$ & $\circ$  & $\circ$  &  % AKD
$\bullet$ & $\circ$ & $\circ$ & $\circ$ &  % AKM
$\bullet$ & $\circ$ & $\circ$ & $\circ$ &  $\odot$ & $\odot$ & % ACD
$\circ$ & $\circ$ &  % ACM
$\bullet$ & $\bullet$ & $\circ$ &  % AG
$\circ$ & $\circ$ & $\bullet$ & $\circ$ & $\circ$ &  % O-RAN SW-App Dev
$\bullet$ & $\bullet$ & $\circ$ & $\bullet$ & $\circ$ &  % O-RAN SW-ISX Dev - Non
$\circ$ & $\bullet$ & $\bullet$ & $\circ$ & $\circ$ &  % O-RAN SW-ISX Dev - Near
$\circ$ & $\circ$ & $\circ$ & $\bullet$ & $\bullet$ &  % O-RAN SW-ISX Dev - CUDU
$\bullet$ &  % C14n SW-ISX Prov
$\bullet$ &  % HW-ISX Prov
$\circ$ &  %UE 
$\bullet$ \\  % 3rd party data provider
\hline

\multirow{12}{*}{\textbf{\shortstack[l]{Transferability-based \\poisoning attacks}}} 
&
training data and hyperparameter knowledge &
$\bullet$ & $\circ$ & $\circ$  & $\circ$  &  % AKD
$\circ$ & $\bullet$ & $\circ$ & $\circ$ &  % AKM
$\bullet$ & $\circ$ & $\circ$ & $\circ$ & $\odot$ & $\odot$ & % ACD
$\circ$ & $\circ$ &  % ACM
$\bullet$ & $\bullet$ & $\circ$ &  % AG
$\circ$ & $\circ$ & $\bullet$ & $\circ$ & $\circ$ &  % O-RAN SW-App Dev
$\bullet$ & $\bullet$ & $\circ$ & $\bullet$ & $\circ$ &  % O-RAN SW-ISX Dev - Non
$\circ$ & $\bullet$ & $\bullet$ & $\circ$ & $\circ$ &  % O-RAN SW-ISX Dev - Near
$\circ$ & $\circ$ & $\circ$ & $\bullet$ & $\bullet$ &  % O-RAN SW-ISX Dev - CUDU
$\bullet$ &  % C14n SW-ISX Prov
$\bullet$ &  % HW-ISX Prov
$\circ$ &  % UE
$\bullet$ \\  % 3rd party data provider

&
features data and hyperparameter knowledge &
$\circ$ & $\bullet$ & $\circ$  & $\circ$  &  % AKD
$\circ$ & $\bullet$ & $\circ$ & $\circ$ &  % AKM
$\bullet$ & $\circ$ & $\circ$ & $\circ$ & $\odot$ & $\odot$ & % ACD
$\circ$ & $\circ$ &  % ACM
$\bullet$ & $\bullet$ & $\circ$ &  % AG
$\circ$ & $\circ$ & $\bullet$ & $\circ$ & $\circ$ &  % O-RAN SW-App Dev
$\bullet$ & $\bullet$ & $\circ$ & $\bullet$ & $\circ$ &  % O-RAN SW-ISX Dev - Non
$\circ$ & $\bullet$ & $\bullet$ & $\circ$ & $\circ$ &  % O-RAN SW-ISX Dev - Near
$\circ$ & $\circ$ & $\circ$ & $\bullet$ & $\bullet$ &  % O-RAN SW-ISX Dev - CUDU
$\bullet$ &  % C14n SW-ISX Prov
$\bullet$ &  % HW-ISX Prov
$\circ$ &  % UE
$\bullet$ \\  % 3rd party data provider

&
raw data and hyperparameter knowledge &
$\circ$ & $\circ$ & $\bullet$  & $\circ$  &  % AKD
$\circ$ & $\bullet$ & $\circ$ & $\circ$ &  % AKM
$\bullet$ & $\circ$ & $\circ$ & $\circ$ & $\odot$ & $\odot$ & % ACD
$\circ$ & $\circ$ &  % ACM
$\bullet$ & $\bullet$ & $\circ$ &  % AG
$\circ$ & $\circ$ & $\bullet$ & $\circ$ & $\circ$ &  % O-RAN SW-App Dev
$\bullet$ & $\bullet$ & $\circ$ & $\bullet$ & $\circ$ &  % O-RAN SW-ISX Dev - Non
$\circ$ & $\bullet$ & $\bullet$ & $\circ$ & $\circ$ &  % O-RAN SW-ISX Dev - Near
$\circ$ & $\circ$ & $\circ$ & $\bullet$ & $\bullet$ &  % O-RAN SW-ISX Dev - CUDU
$\bullet$ &  % C14n SW-ISX Prov
$\bullet$ &  % HW-ISX Prov
$\circ$ &  % UE
$\bullet$ \\  % 3rd party data provider

&
data property and hyperparameter knowledge &
$\circ$ & $\circ$ & $\circ$  & $\bullet$  &  % AKD
$\circ$ & $\bullet$ & $\circ$ & $\circ$ &  % AKM
$\bullet$ & $\circ$ & $\circ$ & $\circ$ & $\odot$ & $\odot$ & % ACD
$\circ$ & $\circ$ &  % ACM
$\bullet$ & $\bullet$ & $\circ$ &  % AG
$\circ$ & $\circ$ & $\bullet$ & $\circ$ & $\circ$ &  % O-RAN SW-App Dev
$\bullet$ & $\bullet$ & $\circ$ & $\bullet$ & $\circ$ &  % O-RAN SW-ISX Dev - Non
$\circ$ & $\bullet$ & $\bullet$ & $\circ$ & $\circ$ &  % O-RAN SW-ISX Dev - Near
$\circ$ & $\circ$ & $\circ$ & $\bullet$ & $\bullet$ &  % O-RAN SW-ISX Dev - CUDU
$\bullet$ &  % C14n SW-ISX Prov
$\bullet$ &  % HW-ISX Prov
$\circ$ &  % UE
$\bullet$ \\  % 3rd party data provider

&
training data and algorithm knowledge &
$\bullet$ & $\circ$ & $\circ$  & $\circ$  &  % AKD
$\circ$ & $\circ$ & $\bullet$ & $\circ$ &  % AKM
$\bullet$ & $\circ$ & $\circ$ & $\circ$ & $\odot$ & $\odot$ & % ACD
$\circ$ & $\circ$ &  % ACM
$\bullet$ & $\bullet$ & $\circ$ &  % AG
$\circ$ & $\circ$ & $\bullet$ & $\circ$ & $\circ$ &  % O-RAN SW-App Dev
$\bullet$ & $\bullet$ & $\circ$ & $\bullet$ & $\circ$ &  % O-RAN SW-ISX Dev - Non
$\circ$ & $\bullet$ & $\bullet$ & $\circ$ & $\circ$ &  % O-RAN SW-ISX Dev - Near
$\circ$ & $\circ$ & $\circ$ & $\bullet$ & $\bullet$ &  % O-RAN SW-ISX Dev - CUDU
$\bullet$ &  % C14n SW-ISX Prov
$\bullet$ &  % HW-ISX Prov
$\circ$ &  % UE
$\bullet$ \\  % 3rd party data provider

&
features data and algorithm knowledge &
$\circ$ & $\bullet$ & $\circ$  & $\circ$  &  % AKD
$\circ$ & $\circ$ & $\bullet$ & $\circ$ &  % AKM
$\bullet$ & $\circ$ & $\circ$ & $\circ$ & $\odot$ & $\odot$ & % ACD
$\circ$ & $\circ$ &  % ACM
$\bullet$ & $\bullet$ & $\circ$ &  % AG
$\circ$ & $\circ$ & $\bullet$ & $\circ$ & $\circ$ &  % O-RAN SW-App Dev
$\bullet$ & $\bullet$ & $\circ$ & $\bullet$ & $\circ$ &  % O-RAN SW-ISX Dev - Non
$\circ$ & $\bullet$ & $\bullet$ & $\circ$ & $\circ$ &  % O-RAN SW-ISX Dev - Near
$\circ$ & $\circ$ & $\circ$ & $\bullet$ & $\bullet$ &  % O-RAN SW-ISX Dev - CUDU
$\bullet$ &  % C14n SW-ISX Prov
$\bullet$ &  % HW-ISX Prov
$\circ$ &  % UE
$\bullet$ \\  % 3rd party data provider

&
raw data and algorithm knowledge &
$\circ$ & $\circ$ & $\bullet$  & $\circ$  &  % AKD
$\circ$ & $\circ$ & $\bullet$ & $\circ$ &  % AKM
$\bullet$ & $\circ$ & $\circ$ & $\circ$ & $\odot$ & $\odot$ & % ACD
$\circ$ & $\circ$ &  % ACM
$\bullet$ & $\bullet$ & $\circ$ &  % AG
$\circ$ & $\circ$ & $\bullet$ & $\circ$ & $\circ$ &  % O-RAN SW-App Dev
$\bullet$ & $\bullet$ & $\circ$ & $\bullet$ & $\circ$ &  % O-RAN SW-ISX Dev - Non
$\circ$ & $\bullet$ & $\bullet$ & $\circ$ & $\circ$ &  % O-RAN SW-ISX Dev - Near
$\circ$ & $\circ$ & $\circ$ & $\bullet$ & $\bullet$ &  % O-RAN SW-ISX Dev - CUDU
$\bullet$ &  % C14n SW-ISX Prov
$\bullet$ &  % HW-ISX Prov
$\circ$ &  % UE
$\bullet$ \\  % 3rd party data provider

&
data property and algorithm knowledge &
$\circ$ & $\circ$ & $\circ$  & $\bullet$  &  % AKD
$\circ$ & $\circ$ & $\bullet$ & $\circ$ &  % AKM
$\bullet$ & $\circ$ & $\circ$ & $\circ$ & $\odot$ & $\odot$ & % ACD
$\circ$ & $\circ$ &  % ACM
$\bullet$ & $\bullet$ & $\circ$ &  % AG
$\circ$ & $\circ$ & $\bullet$ & $\bullet$ & $\circ$ &  % O-RAN SW-App Dev
$\bullet$ & $\bullet$ & $\circ$ & $\bullet$ & $\circ$ &  % O-RAN SW-ISX Dev - Non
$\circ$ & $\bullet$ & $\bullet$ & $\circ$ & $\circ$ &  % O-RAN SW-ISX Dev - Near
$\circ$ & $\circ$ & $\circ$ & $\bullet$ & $\bullet$ &  % O-RAN SW-ISX Dev - CUDU
$\bullet$ &  % C14n SW-ISX Prov
$\bullet$ &  % HW-ISX Prov
$\circ$ &  % UE
$\bullet$ \\  % 3rd party data provider

&
training data and task knowledge &
$\bullet$ & $\circ$ & $\circ$  & $\circ$  &  % AKD
$\circ$ & $\circ$ & $\circ$ & $\bullet$ &  % AKM
$\bullet$ & $\circ$ & $\circ$ & $\circ$ & $\odot$ & $\odot$ & % ACD
$\circ$ & $\circ$ &  % ACM
$\bullet$ & $\bullet$ & $\circ$ &  % AG
$\circ$ & $\bullet$ & $\bullet$ & $\bullet$ & $\circ$ &  % O-RAN SW-App Dev
$\bullet$ & $\bullet$ & $\circ$ & $\bullet$ & $\circ$ &  % O-RAN SW-ISX Dev - Non
$\bullet$ & $\bullet$ & $\bullet$ & $\circ$ & $\circ$ &  % O-RAN SW-ISX Dev - Near
$\circ$ & $\circ$ & $\circ$ & $\bullet$ & $\bullet$ &  % O-RAN SW-ISX Dev - CUDU
$\bullet$ &  % C14n SW-ISX Prov
$\bullet$ &  % HW-ISX Prov
$\circ$ &  % UE
$\bullet$ \\  % 3rd party data provider

&
features data and task knowledge &
$\circ$ & $\bullet$ & $\circ$  & $\circ$  &  % AKD
$\circ$ & $\circ$ & $\circ$ & $\bullet$ &  % AKM
$\bullet$ & $\circ$ & $\circ$ & $\circ$ & $\odot$ & $\odot$ & % ACD
$\circ$ & $\circ$ &  % ACM
$\bullet$ & $\bullet$ & $\circ$ &  % AG
$\circ$ & $\bullet$ & $\bullet$ & $\bullet$ & $\circ$ &  % O-RAN SW-App Dev
$\bullet$ & $\bullet$ & $\circ$ & $\bullet$ & $\circ$ &  % O-RAN SW-ISX Dev - Non
$\bullet$ & $\bullet$ & $\bullet$ & $\circ$ & $\circ$ &  % O-RAN SW-ISX Dev - Near
$\circ$ & $\circ$ & $\circ$ & $\bullet$ & $\bullet$ &  % O-RAN SW-ISX Dev - CUDU
$\bullet$ &  % C14n SW-ISX Prov
$\bullet$ &  % HW-ISX Prov
$\circ$ &  % UE
$\bullet$ \\  % 3rd party data provider

&
raw data and task knowledge &
$\circ$ & $\circ$ & $\bullet$  & $\circ$  &  % AKD
$\circ$ & $\circ$ & $\circ$ & $\bullet$ &  % AKM
$\bullet$ & $\circ$ & $\circ$ & $\circ$ & $\odot$ & $\odot$ & % ACD
$\circ$ & $\circ$ &  % ACM
$\bullet$ & $\bullet$ & $\circ$ &  % AG
$\bullet$ & $\bullet$ & $\bullet$ & $\bullet$ & $\circ$ &  % O-RAN SW-App Dev
$\bullet$ & $\bullet$ & $\circ$ & $\bullet$ & $\circ$ &  % O-RAN SW-ISX Dev - Non
$\bullet$ & $\bullet$ & $\bullet$ & $\circ$ & $\circ$ &  % O-RAN SW-ISX Dev - Near
$\odot$ & $\bullet$ & $\bullet$ & $\bullet$ & $\bullet$ &  % O-RAN SW-ISX Dev - CUDU
$\bullet$ &  % C14n SW-ISX Prov
$\bullet$ &  % HW-ISX Prov
$\circ$ &  % UE
$\bullet$ \\  % 3rd party data provider

&
data property and task knowledge &
$\circ$ & $\circ$ & $\circ$  & $\bullet$  &  % AKD
$\circ$ & $\circ$ & $\circ$ & $\bullet$ &  % AKM
$\bullet$ & $\circ$ & $\circ$ & $\circ$ & $\odot$ & $\odot$ & % ACD
$\circ$ & $\circ$ &  % ACM
$\bullet$ & $\bullet$ & $\circ$ &  % AG
$\bullet$ & $\bullet$ & $\bullet$ & $\bullet$ & $\circ$ &  % O-RAN SW-App Dev
$\bullet$ & $\bullet$ & $\circ$ & $\bullet$ & $\circ$ &  % O-RAN SW-ISX Dev - Non
$\bullet$ & $\bullet$ & $\bullet$ & $\circ$ & $\circ$ &  % O-RAN SW-ISX Dev - Near
$\bullet$ & $\bullet$ & $\bullet$ & $\bullet$ & $\bullet$ &  % O-RAN SW-ISX Dev - CUDU
$\bullet$ &  % C14n SW-ISX Prov
$\bullet$ &  % HW-ISX Prov
$\circ$ &  % UE
$\bullet$ \\  % 3rd party data provider

\Xhline{3\arrayrulewidth}

% Inference and Reconstruction
\rowcolor{Grey}
\multicolumn{45}{|c|}{\textbf{Threat Category:} Membership Inference}\\ 
\textbf{\shortstack[l]{Gradient-based\\inference attacks}}
&
model knowledge (e.g., when using public models) &
$\circ$ & $\circ$ & $\circ$  & $\circ$  &  % AKD
$\bullet$ & $\circ$ & $\circ$ & $\circ$ &  % AKM
$\circ$ & $\circ$ & $\circ$ & $\circ$ & $\odot$ & $\odot$ & % ACD
$\circ$ & $\circ$ &  % ACM
$\circ$ & $\circ$ & $\bullet$ &  % AG
$\circ$ & $\circ$ & $\bullet$ & $\bullet$ & $\circ$ &  % O-RAN SW-App Dev
$\bullet$ & $\bullet$ & $\circ$ & $\bullet$ & $\circ$ &  % O-RAN SW-ISX Dev - Non
$\circ$ & $\bullet$ & $\bullet$ & $\circ$ & $\circ$ &  % O-RAN SW-ISX Dev - Near
$\circ$ & $\circ$ & $\circ$ & $\bullet$ & $\bullet$ &  % O-RAN SW-ISX Dev - CUDU
$\bullet$ &  % C14n SW-ISX Prov
$\bullet$ &  % HW-ISX Prov
$\circ$ &  % UE
$\bullet$ \\  % 3rd party data provider

\hline

\textbf{\shortstack[l]{Query-based\\inference attacks}} 
&
query access and data property knowledge&
$\circ$ & $\circ$ & $\circ$  & $\bullet$  &  % AKD
$\circ$ &  $\circ$ & $\circ$ & $\bullet$ &  % AKM
$\circ$ & $\circ$ & $\circ$ & $\bullet$ & $\odot$ & $\odot$ & % ACD
$\bullet$ & $\circ$ &  % ACM
$\circ$ & $\circ$ & $\bullet$ &  % AG
$\circ$ & $\circ$ & $\bullet$ & $\bullet$ & $\bullet$ &  % O-RAN SW-App Dev
$\bullet$ & $\bullet$ & $\circ$ & $\bullet$ & $\circ$ &  % O-RAN SW-ISX Dev - Non
$\bullet$ & $\bullet$ & $\bullet$ & $\circ$ & $\circ$ &  % O-RAN SW-ISX Dev - Near
$\circ$ & $\circ$ & $\circ$ & $\bullet$ & $\bullet$ &  % O-RAN SW-ISX Dev - CUDU
$\bullet$ &  % C14n SW-ISX Prov
$\bullet$ &  % HW-ISX Prov
$\circ$ &  % UE
$\bullet$ \\  % 3rd party data provider

\Xhline{3\arrayrulewidth}

%  Data Reconstruction
\rowcolor{Grey}
\multicolumn{45}{|c|}{\textbf{Threat Category:} Data Reconstruction}\\ 
\textbf{\shortstack[l]{Gradient-based Data \\Reconstruction attacks}}
& 
model knowledge  &
$\circ$ & $\circ$ & $\circ$  & $\circ$  &  % AKD
$\bullet$ & $\circ$ & $\circ$ & $\circ$ &  % AKM
$\circ$ & $\circ$ & $\circ$ & $\circ$ & $\odot$ & $\odot$ & % ACD
$\circ$ & $\circ$ &  % ACM
$\circ$ & $\circ$ & $\bullet$ &  % AG
$\circ$ & $\circ$ & $\bullet$ & $\bullet$ & $\circ$ &  % O-RAN SW-App Dev
$\bullet$ & $\bullet$ & $\circ$ & $\bullet$ & $\circ$ &  % O-RAN SW-ISX Dev - Non
$\circ$ & $\bullet$ & $\bullet$ & $\circ$ & $\circ$ &  % O-RAN SW-ISX Dev - Near
$\circ$ & $\circ$ & $\circ$ & $\bullet$ & $\bullet$ &  % O-RAN SW-ISX Dev - CUDU
$\bullet$ &  % C14n SW-ISX Prov
$\bullet$ &  % HW-ISX Prov
$\circ$ &  % UE
$\bullet$ \\  % 3rd party data provider

\hline
\textbf{\shortstack[l]{Query-based Data\\Reconstruction attacks}} 

&
query access and data property knowledge&
$\circ$ & $\circ$ & $\circ$  & $\bullet$  &  % AKD
$\circ$ &  $\circ$ & $\circ$ & $\bullet$ &  % AKM
$\circ$ & $\circ$ & $\circ$ & $\bullet$ & $\odot$ & $\odot$ & % ACD
$\bullet$ & $\circ$ &  % ACM
$\circ$ & $\circ$ & $\bullet$ &  % AG
$\circ$ & $\circ$ & $\bullet$ & $\bullet$ & $\bullet$ &  % O-RAN SW-App Dev
$\bullet$ & $\bullet$ & $\circ$ & $\bullet$ & $\circ$ &  % O-RAN SW-ISX Dev - Non
$\bullet$ & $\bullet$ & $\bullet$ & $\circ$ & $\circ$ &  % O-RAN SW-ISX Dev - Near
$\circ$ & $\circ$ & $\circ$ & $\bullet$ & $\bullet$ &  % O-RAN SW-ISX Dev - CUDU
$\bullet$ &  % C14n SW-ISX Prov
$\bullet$ &  % HW-ISX Prov
$\circ$ &  % UE
$\bullet$ \\  % 3rd party data provider

\Xhline{3\arrayrulewidth}

\rowcolor{Grey}
\multicolumn{45}{|c|}{\textbf{Threat Category:} Model Extraction}\\ 

\multirow{2}{*}{\textbf{\shortstack[l]{Query-based model \\extraction attacks}}} 
&
score-based query access &
$\circ$ & $\circ$ & $\circ$  & $\circ$  &  % AKD
$\circ$ & $\circ$ & $\circ$ & $\circ$ &  % AKM
$\circ$ & $\circ$ & $\circ$ & $\bullet$& $\odot$ & $\odot$ & % ACD
$\bullet$ & $\circ$ &  % ACM
$\circ$ & $\circ$ & $\bullet$ &  % AG
$\circ$ & $\circ$ & $\bullet$ & $\bullet$ & $\bullet$ &  % O-RAN SW-App Dev
$\bullet$ & $\bullet$ & $\circ$ & $\bullet$ & $\circ$ &  % O-RAN SW-ISX Dev - Non
$\bullet$ & $\bullet$ & $\bullet$ & $\circ$ & $\circ$ &  % O-RAN SW-ISX Dev - Near
$\circ$ & $\circ$ & $\circ$ & $\bullet$ & $\bullet$ &  % O-RAN SW-ISX Dev - CUDU
$\bullet$ &  % C14n SW-ISX Prov
$\bullet$ &  % HW-ISX Prov
$\circ$ &  % UE
$\bullet$ \\  % 3rd party data provider

&
decision-based query access &
$\circ$ & $\circ$ & $\circ$  & $\circ$  &  % AKD
$\circ$ & $\circ$ & $\circ$ & $\circ$ &  % AKM
$\circ$ & $\circ$ & $\circ$ & $\bullet$& $\odot$ & $\odot$ & % ACD
$\circ$ & $\bullet$ &  % ACM
$\circ$ & $\circ$ & $\bullet$ &  % AG
$\circ$ & $\circ$ & $\bullet$ & $\bullet$ & $\bullet$ &  % O-RAN SW-App Dev
$\bullet$ & $\bullet$ & $\circ$ & $\bullet$ & $\circ$ &  % O-RAN SW-ISX Dev - Non
$\bullet$ & $\bullet$ & $\bullet$ & $\circ$ & $\circ$ &  % O-RAN SW-ISX Dev - Near
$\circ$ & $\circ$ & $\circ$ & $\bullet$ & $\bullet$ &  % O-RAN SW-ISX Dev - CUDU
$\bullet$ &  % C14n SW-ISX Prov
$\bullet$ &  % HW-ISX Prov
$\bullet$ &  % UE
$\bullet$ \\  % 3rd party data provider

\Xhline{3\arrayrulewidth}

\rowcolor{Grey}
\multicolumn{45}{|c|}{\textbf{Threat Category:} Resource Exhaustion}\\ 
\multirow{2}{*}{\textbf{\shortstack[l]{Gradient-based resource \\exhaustion attacks}}} 
&
training data and hyperparameter knowledge &
$\bullet$ & $\circ$ & $\circ$  & $\circ$  &  % AKD
$\circ$ & $\bullet$ & $\circ$ & $\circ$ &  % AKM
$\circ$ & $\circ$ & $\circ$ & $\circ$ & $\odot$ & $\odot$ & % ACD
$\circ$ & $\odot$ &  % ACM
$\circ$ & $\bullet$ & $\circ$ &  % AG
$\circ$ & $\circ$ & $\bullet$ & $\circ$ & $\circ$ &  % O-RAN SW-App Dev
$\bullet$ & $\bullet$ & $\circ$ & $\bullet$ & $\circ$ &  % O-RAN SW-ISX Dev - Non
$\circ$ & $\bullet$ & $\bullet$ & $\circ$ & $\circ$ &  % O-RAN SW-ISX Dev - Near
$\circ$ & $\circ$ & $\circ$ & $\bullet$ & $\bullet$ &  % O-RAN SW-ISX Dev - CUDU
$\bullet$ &  % C14n SW-ISX Prov
$\bullet$ &  % HW-ISX Prov
$\circ$ &  % UE
$\bullet$ \\  % 3rd party data provider

&
model's knowledge &
$\circ$ & $\circ$ & $\circ$  & $\circ$  &  % AKD
$\bullet$ & $\circ$ & $\circ$ & $\circ$ &  % AKM
$\circ$ & $\circ$ & $\circ$ & $\circ$ & $\odot$ & $\odot$ & % ACD
$\circ$ & $\odot$ &  % ACM
$\circ$ & $\bullet$ & $\circ$ &  % AG
$\circ$ & $\circ$ & $\bullet$ & $\bullet$ & $\circ$ &  % O-RAN SW-App Dev
$\bullet$ & $\bullet$ & $\circ$ & $\bullet$ & $\circ$ &  % O-RAN SW-ISX Dev - Non
$\circ$ & $\bullet$ & $\bullet$ & $\circ$ & $\circ$ &  % O-RAN SW-ISX Dev - Near
$\circ$ & $\circ$ & $\circ$ & $\bullet$ & $\bullet$ &  % O-RAN SW-ISX Dev - CUDU
$\bullet$ &  % C14n SW-ISX Prov
$\bullet$ &  % HW-ISX Prov
$\circ$ &  % UE
$\bullet$ \\  % 3rd party data provider

\hline

\shortstack[l]{\textbf{Query-based resource}\\\textbf{exhaustion attacks}}
&
query based access and model's latency measurements &
$\circ$ & $\circ$ & $\circ$  & $\circ$  &  % AKD
$\circ$ & $\circ$ & $\circ$ & $\bullet$ &  % AKM
$\circ$ & $\circ$ & $\circ$ & $\bullet$ & $\odot$ & $\odot$ & % ACD
$\circ$ & $\bullet$ &  % ACM
$\circ$ & $\bullet$ & $\circ$ &  % AG
$\circ$ & $\bullet$ & $\bullet$ & $\bullet$ & $\bullet$ &  % O-RAN SW-App Dev
$\bullet$ & $\bullet$ & $\circ$ & $\bullet$ & $\circ$ &  % O-RAN SW-ISX Dev - Non
$\bullet$ & $\bullet$ & $\bullet$ & $\circ$ & $\circ$ &  % O-RAN SW-ISX Dev - Near
$\circ$ & $\circ$ & $\circ$ & $\bullet$ & $\bullet$ &  % O-RAN SW-ISX Dev - CUDU
$\bullet$ &  % C14n SW-ISX Prov
$\bullet$ &  % HW-ISX Prov
$\bullet$ &  % UE
$\bullet$ \\  % 3rd party data provider

\hline

\multirow{12}{*}{\textbf{\shortstack[l]{Transferability-based\\resource exhaustion\\attacks}}} 

&
training data and hyperparameter knowledge &
$\bullet$ & $\circ$ & $\circ$  & $\circ$  &  % AKD
$\circ$ & $\bullet$ & $\circ$ & $\circ$ &  % AKM
$\circ$ & $\circ$ & $\circ$ & $\odot$ & $\odot$ & $\odot$ & % ACD
$\circ$ & $\odot$ &  % ACM
$\circ$ & $\bullet$ & $\circ$ &  % AG
$\circ$ & $\circ$ & $\bullet$ & $\circ$ & $\circ$ &  % O-RAN SW-App Dev
$\bullet$ & $\bullet$ & $\circ$ & $\bullet$ & $\circ$ &  % O-RAN SW-ISX Dev - Non
$\circ$ & $\bullet$ & $\bullet$ & $\circ$ & $\circ$ &  % O-RAN SW-ISX Dev - Near
$\circ$ & $\circ$ & $\circ$ & $\bullet$ & $\bullet$ &  % O-RAN SW-ISX Dev - CUDU
$\bullet$ &  % C14n SW-ISX Prov
$\bullet$ &  % HW-ISX Prov
$\circ$ &  % UE
$\bullet$ \\  % 3rd party data provider

&
features data and hyperparameter knowledge &
$\circ$ & $\bullet$ & $\circ$  & $\circ$  &  % AKD
$\circ$ & $\bullet$ & $\circ$ & $\circ$ &  % AKM
$\circ$ & $\circ$ & $\circ$ & $\odot$ & $\odot$ & $\odot$ & % ACD
$\circ$ & $\odot$ &  % ACM
$\circ$ & $\bullet$ & $\circ$ &  % AG
$\circ$ & $\circ$ & $\bullet$ & $\bullet$ & $\circ$ &  % O-RAN SW-App Dev
$\bullet$ & $\bullet$ & $\circ$ & $\bullet$ & $\circ$ &  % O-RAN SW-ISX Dev - Non
$\circ$ & $\bullet$ & $\bullet$ & $\circ$ & $\circ$ &  % O-RAN SW-ISX Dev - Near
$\circ$ & $\circ$ & $\circ$ & $\bullet$ & $\bullet$ &  % O-RAN SW-ISX Dev - CUDU
$\bullet$ &  % C14n SW-ISX Prov
$\bullet$ &  % HW-ISX Prov
$\circ$ &  % UE
$\bullet$ \\  % 3rd party data provider

&
raw data and hyperparameter knowledge &
$\circ$ & $\circ$ & $\bullet$  & $\circ$  &  % AKD
$\circ$ & $\bullet$ & $\circ$ & $\circ$ &  % AKM
$\circ$ & $\circ$ & $\circ$ & $\odot$ & $\odot$ & $\odot$ & % ACD
$\circ$ & $\odot$ &  % ACM
$\circ$ & $\bullet$ & $\circ$ &  % AG
$\circ$ & $\circ$ & $\bullet$ & $\bullet$ & $\circ$ &  % O-RAN SW-App Dev
$\bullet$ & $\bullet$ & $\circ$ & $\bullet$ & $\circ$ &  % O-RAN SW-ISX Dev - Non
$\circ$ & $\bullet$ & $\bullet$ & $\circ$ & $\circ$ &  % O-RAN SW-ISX Dev - Near
$\circ$ & $\circ$ & $\circ$ & $\bullet$ & $\bullet$ &  % O-RAN SW-ISX Dev - CUDU
$\bullet$ &  % C14n SW-ISX Prov
$\bullet$ &  % HW-ISX Prov
$\circ$ &  % UE
$\bullet$ \\  % 3rd party data provider

&
data property and hyperparameter knowledge &
$\circ$ & $\circ$ & $\circ$  & $\bullet$  &  % AKD
$\circ$ & $\bullet$ & $\circ$ & $\circ$ &  % AKM
$\circ$ & $\circ$ & $\circ$ & $\odot$ & $\odot$ & $\odot$ & % ACD
$\circ$ & $\odot$ &  % ACM
$\circ$ & $\bullet$ & $\circ$ &  % AG
$\circ$ & $\circ$ & $\bullet$ & $\bullet$ & $\circ$ &  % O-RAN SW-App Dev
$\bullet$ & $\bullet$ & $\circ$ & $\bullet$ & $\circ$ &  % O-RAN SW-ISX Dev - Non
$\circ$ & $\bullet$ & $\bullet$ & $\circ$ & $\circ$ &  % O-RAN SW-ISX Dev - Near
$\circ$ & $\circ$ & $\circ$ & $\bullet$ & $\bullet$ &  % O-RAN SW-ISX Dev - CUDU
$\bullet$ &  % C14n SW-ISX Prov
$\bullet$ &  % HW-ISX Prov
$\circ$ &  % UE
$\bullet$ \\  % 3rd party data provider

&
training data and algorithm knowledge &
$\bullet$ & $\circ$ & $\circ$  & $\circ$  &  % AKD
$\circ$ & $\circ$ & $\bullet$ & $\circ$ &  % AKM
$\circ$ & $\circ$ & $\circ$ & $\odot$ & $\odot$ & $\odot$ & % ACD
$\circ$ & $\odot$ &  % ACM
$\circ$ & $\bullet$ & $\circ$ &  % AG
$\circ$ & $\circ$ & $\bullet$ & $\circ$ & $\circ$ &  % O-RAN SW-App Dev
$\bullet$ & $\bullet$ & $\circ$ & $\bullet$ & $\circ$ &  % O-RAN SW-ISX Dev - Non
$\circ$ & $\bullet$ & $\bullet$ & $\circ$ & $\circ$ &  % O-RAN SW-ISX Dev - Near
$\circ$ & $\circ$ & $\circ$ & $\bullet$ & $\bullet$ &  % O-RAN SW-ISX Dev - CUDU
$\bullet$ &  % C14n SW-ISX Prov
$\bullet$ &  % HW-ISX Prov
$\circ$ &  % UE
$\bullet$ \\  % 3rd party data provider

&
features data and algorithm knowledge &
$\circ$ & $\bullet$ & $\circ$  & $\circ$  &  % AKD
$\circ$ & $\circ$ & $\bullet$ & $\circ$ &  % AKM
$\circ$ & $\circ$ & $\circ$ & $\odot$ & $\odot$ & $\odot$ & % ACD
$\circ$ & $\odot$ &  % ACM
$\circ$ & $\bullet$ & $\circ$ &  % AG
$\circ$ & $\circ$ & $\bullet$ & $\circ$ & $\circ$ &  % O-RAN SW-App Dev
$\bullet$ & $\bullet$ & $\circ$ & $\bullet$ & $\circ$ &  % O-RAN SW-ISX Dev - Non
$\circ$ & $\bullet$ & $\bullet$ & $\circ$ & $\circ$ &  % O-RAN SW-ISX Dev - Near
$\circ$ & $\circ$ & $\circ$ & $\bullet$ & $\bullet$ &  % O-RAN SW-ISX Dev - CUDU
$\bullet$ &  % C14n SW-ISX Prov
$\bullet$ &  % HW-ISX Prov
$\circ$ &  % UE
$\bullet$ \\  % 3rd party data provider

&
raw data and algorithm knowledge &
$\circ$ & $\circ$ & $\bullet$  & $\circ$  &  % AKD
$\circ$ & $\circ$ & $\bullet$ & $\circ$ &  % AKM
$\circ$ & $\circ$ & $\circ$ & $\odot$ & $\odot$ & $\odot$ & % ACD
$\circ$ & $\odot$ &  % ACM
$\circ$ & $\bullet$ & $\circ$ &  % AG
$\circ$ & $\circ$ & $\bullet$ & $\circ$ & $\circ$ &  % O-RAN SW-App Dev
$\bullet$ & $\bullet$ & $\circ$ & $\bullet$ & $\circ$ &  % O-RAN SW-ISX Dev - Non
$\circ$ & $\bullet$ & $\bullet$ & $\circ$ & $\circ$ &  % O-RAN SW-ISX Dev - Near
$\circ$ & $\circ$ & $\circ$ & $\bullet$ & $\bullet$ &  % O-RAN SW-ISX Dev - CUDU
$\bullet$ &  % C14n SW-ISX Prov
$\bullet$ &  % HW-ISX Prov
$\circ$ &  % UE
$\bullet$ \\  % 3rd party data provider

&
data property and algorithm knowledge &
$\circ$ & $\circ$ & $\circ$  & $\bullet$  &  % AKD
$\circ$ & $\circ$ & $\bullet$ & $\circ$ &  % AKM
$\circ$ & $\circ$ & $\circ$ & $\odot$ & $\odot$ & $\odot$ & % ACD
$\circ$ & $\odot$ &  % ACM
$\circ$ & $\bullet$ & $\circ$ &  % AG
$\circ$ & $\circ$ & $\bullet$ & $\bullet$ & $\circ$ &  % O-RAN SW-App Dev
$\bullet$ & $\bullet$ & $\circ$ & $\bullet$ & $\circ$ &  % O-RAN SW-ISX Dev - Non
$\circ$ & $\bullet$ & $\bullet$ & $\circ$ & $\circ$ &  % O-RAN SW-ISX Dev - Near
$\circ$ & $\circ$ & $\circ$ & $\bullet$ & $\bullet$ &  % O-RAN SW-ISX Dev - CUDU
$\bullet$ &  % C14n SW-ISX Prov
$\bullet$ &  % HW-ISX Prov
$\circ$ &  % UE
$\bullet$ \\  % 3rd party data provider

&
training data and task knowledge &
$\bullet$ & $\circ$ & $\circ$  & $\circ$  &  % AKD
$\circ$ & $\circ$ & $\circ$ & $\bullet$ &  % AKM
$\circ$ & $\circ$ & $\circ$ & $\odot$ & $\odot$ & $\odot$ & % ACD
$\circ$ & $\odot$ &  % ACM
$\circ$ & $\bullet$ & $\circ$ &  % AG
$\circ$ & $\circ$ & $\bullet$ & $\circ$ & $\circ$ &  % O-RAN SW-App Dev
$\bullet$ & $\bullet$ & $\circ$ & $\bullet$ & $\circ$ &  % O-RAN SW-ISX Dev - Non
$\bullet$ & $\bullet$ & $\bullet$ & $\circ$ & $\circ$ &  % O-RAN SW-ISX Dev - Near
$\circ$ & $\circ$ & $\circ$ & $\bullet$ & $\bullet$ &  % O-RAN SW-ISX Dev - CUDU
$\bullet$ &  % C14n SW-ISX Prov
$\bullet$ &  % HW-ISX Prov
$\circ$ &  % UE
$\bullet$ \\  % 3rd party data provider

&
features data and task knowledge &
$\circ$ & $\bullet$ & $\circ$  & $\circ$  &  % AKD
$\circ$ & $\circ$ & $\circ$ & $\bullet$ &  % AKM
$\circ$ & $\circ$ & $\circ$ & $\odot$ & $\odot$ & $\odot$ & % ACD
$\circ$ & $\odot$ &  % ACM
$\circ$ & $\bullet$ & $\circ$ &  % AG
$\circ$ & $\bullet$ & $\bullet$ & $\bullet$ & $\circ$ &  % O-RAN SW-App Dev
$\bullet$ & $\bullet$ & $\circ$ & $\bullet$ & $\circ$ &  % O-RAN SW-ISX Dev - Non
$\bullet$ & $\bullet$ & $\bullet$ & $\circ$ & $\circ$ &  % O-RAN SW-ISX Dev - Near
$\circ$ & $\circ$ & $\circ$ & $\bullet$ & $\bullet$ &  % O-RAN SW-ISX Dev - CUDU
$\bullet$ &  % C14n SW-ISX Prov
$\bullet$ &  % HW-ISX Prov
$\circ$ &  % UE
$\bullet$ \\  % 3rd party data provider

&
raw data and task knowledge &
$\circ$ & $\circ$ & $\bullet$  & $\circ$  &  % AKD
$\circ$ & $\circ$ & $\circ$ & $\bullet$ &  % AKM
$\circ$ & $\circ$ & $\circ$ & $\odot$ & $\odot$ & $\odot$ & % ACD
$\circ$ & $\odot$ &  % ACM
$\circ$ & $\bullet$ & $\circ$ &  % AG
$\bullet$ & $\bullet$ & $\bullet$ & $\bullet$ & $\bullet$ &  % O-RAN SW-App Dev
$\bullet$ & $\bullet$ & $\circ$ & $\bullet$ & $\circ$ &  % O-RAN SW-ISX Dev - Non
$\bullet$ & $\bullet$ & $\bullet$ & $\circ$ & $\circ$ &  % O-RAN SW-ISX Dev - Near
$\bullet$ & $\bullet$ & $\bullet$ & $\bullet$ & $\bullet$ &  % O-RAN SW-ISX Dev - CUDU
$\bullet$ &  % C14n SW-ISX Prov
$\bullet$ &  % HW-ISX Prov
$\circ$ &  % UE
$\bullet$ \\  % 3rd party data provider

&
data property and task knowledge &
$\circ$ & $\circ$ & $\circ$  & $\bullet$  &  % AKD
$\circ$ & $\circ$ & $\circ$ & $\bullet$ &  % AKM
$\circ$ & $\circ$ & $\circ$ & $\odot$ & $\odot$ & $\odot$ & % ACD
$\circ$ & $\odot$ &  % ACM
$\circ$ & $\bullet$ & $\circ$ &  % AG
$\bullet$ & $\bullet$ & $\bullet$ & $\bullet$ & $\bullet$ &  % O-RAN SW-App Dev
$\bullet$ & $\bullet$ & $\circ$ & $\bullet$ & $\circ$ &  % O-RAN SW-ISX Dev - Non
$\bullet$ & $\bullet$ & $\bullet$ & $\circ$ & $\circ$ &  % O-RAN SW-ISX Dev - Near
$\bullet$ & $\bullet$ & $\bullet$ & $\bullet$ & $\bullet$ &  % O-RAN SW-ISX Dev - CUDU
$\bullet$ &  % C14n SW-ISX Prov
$\bullet$ &  % HW-ISX Prov
$\circ$ &  % UE
$\bullet$ \\  % 3rd party data provider

\Xhline{3\arrayrulewidth}

\multicolumn{45}{l}{Threat model -  \textbf{$\circ$}: does not require such type of access/knowledge, \textbf{$\bullet$}: such type of access/knowledge is required for generating the attack, \textbf{$\odot$}: such type of access/knowledge is required for executing the attack.} \\

\multicolumn{45}{l}{Impact -  \textbf{$\circ$}: The attacker can achieve such type of goal, by executing this attack, \textbf{$\bullet$}: The attacker can't achieve such type of goal, by executing this attack.} \\
\multicolumn{45}{l}{Threat Actor -  \textbf{$\bullet$}: The threat actor can obtain the capabilities required to execute the attack ,\textbf{$\circ$}:The threat actor cannot obtain the capabilities required to execute the attack.} \\

\end{tabular}}
\caption{Adversarial machine learning attack techniques and threat actors.}
\label{tab:attack_techniques_with_actors}
\end{table*}

\begin{comment}
\begin{figure}[h]
    \centering
    \includegraphics[width=0.48\textwidth]{sections/figures/attack_tech.pdf}
    \caption{The main AML attack families.}
    \label{fig:aml_at}
\end{figure}
\end{comment}
\section{Attack Demonstration \label{sec:demonstration}}

\subsection{Traffic Steering Overview}
In this section, we demonstrate the applicability of an AML attack on the traffic steering use case.
The demonstration is based on the \oran SC's third open software release, "Cherry," which implements the \textit{\nrt}---a software-based near-real-time platform for hosting and running microservice applications (xApps) on Linux/Kubernetes. 
The \nrt~platform provides the application development platform (xApp), the design and implementation of the open A1/E2 communication interfaces, and the development of cellular network management infrastructure.

\begin{figure}[h]
    \centering
    \includegraphics[width=0.48 \textwidth]{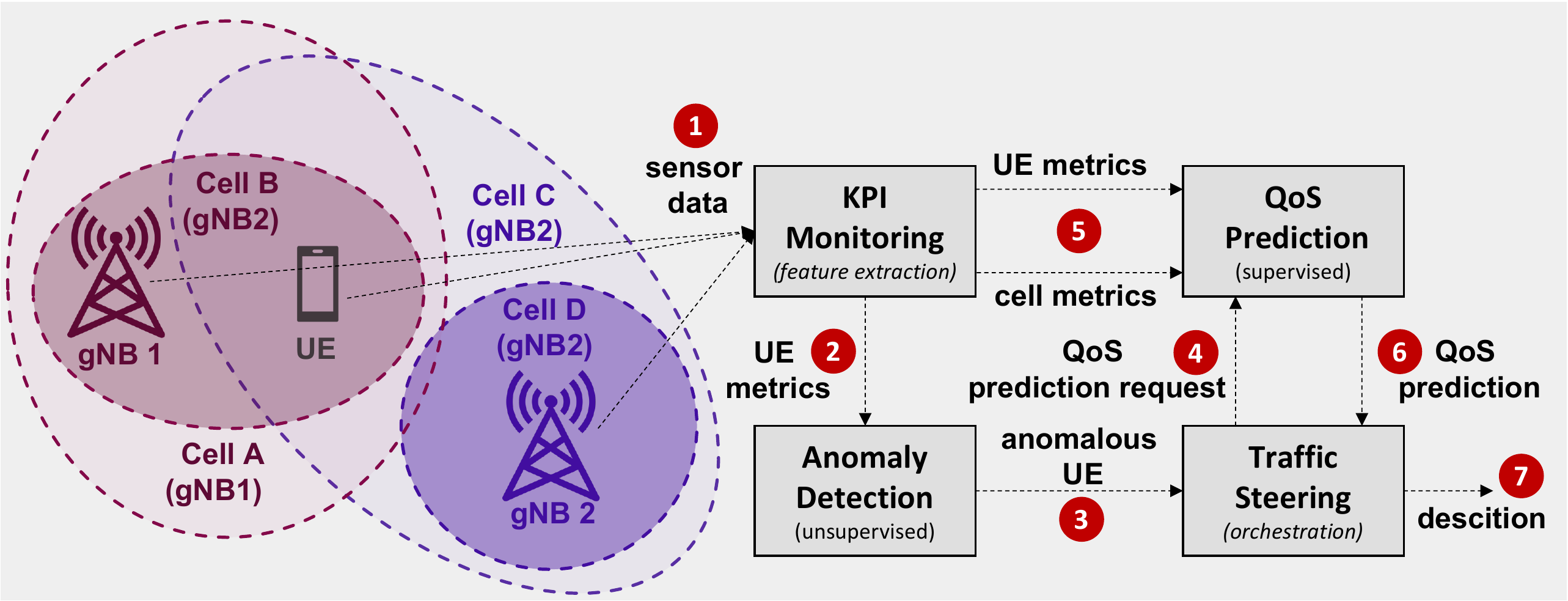}
    \caption{Illustration of the traffic steering process.}% in O-RAN.}
    \label{fig:ts_process}
\end{figure}

The traffic steering task is illustrated in Figure~\ref{fig:ts_process}.
The UE is located within the reception range of Cells A, B, and C and can be connected by each of these cells.
The network provider needs to make a decision regarding which cell the UE should be connected to so that the UE has an acceptable quality of experience (QoE) without compromising the QoE of other UEs.
This challenge is addressed by the traffic steering ML-based application.

%Furthermore, as presented in Figure~\ref{fig:ts_process}, the traffic steering (TS) use case includes four main components: \textit{(i)} the anomaly detection (AD) xApp---detects UE with an anomalous QoE level;  \textit{(ii)} the KPI monitor (KPIMON) xApp---transforms the communication data into UE and cell metrics by extracting relevant features; \textit{(iii)} the QoE prediction (QP) xApp---predicts the QoE level of a given UE on a given cell in the current state; and \textit{(iv)} the traffic steering (TS) xApp---executes the process of allocating UE to new (better) cells based on the QP xApp's results.
        
%\subsubsection{Anomaly Detection}
% The TS process is illustrated in Figure~\ref{fig:ts_process}: 
The \textbf{KPI  monitor (KPIMON) xApp} continuously collects data from the O-CUs and O-DUs, computes UE and cell metrics (i.e., features), and stores them in the \nrt~for other xApps' usage (step 1).
%The UE metrics include: the UE ID and serving cell ID (as identifiers), packet data convergence protocol (PDCP) report timestamp, PDCP aggregation period, UE PDCP downlink and uplink, resource block (PRB) report timestamp, PRB aggregation period, UE PRB downlink and uplink ratios, serving cell report timestamp, Reference Signal Received Power (RSRP), Reference Signal Received Quality (RSRQ), and signal-to-noise Ratio (SINR).
%The cell metrics include: the cell ID, PCDP report timestamp, PCDP aggregation period, PCDP downlink and uplink as the PCDP, PRB report timestamp, PRB aggregation period, and PRB downlink and uplink ratios.

The \textbf{Anomaly detection (AD) xApp}, which is scheduled to run every $10 ms$, detects UE with an anomalous QoE (step 2).
Anomaly detection is performed based on the UE metrics extracted by the KPIMON xApp and stored in the \nrt.
This list of anomalies is sent over the remote message router (RMR) API to the \textbf{traffic steering (TS) xApp} for reallocation of the anomalous UE (step 3).
Then, the \textbf{QoS prediction (QP) xApp} is called by the TS xApp, and for each anomalous UE, it receives the UE and cell metrics and predicts the QoS of the UE in a given cell (steps 4 and 5). 
Finally, based on the predictions provided by the QP xApp and the given A1 policy (which includes configuration information for the ML-assisted application, e.g., the threshold for cell reallocation, in traffic steering), the TS xApp decides whether to allocate anomalous UE to a new cell (steps 6 and 7).
The deployment of the various components of the TS use case within the \oran framework is presented in Figure~\ref{fig:oran_arch}.
%%\vspace{-5pt}

%\vspace{-25pt}
\subsection{Attack Scenario}
While there are various AML attacks that can be performed, in the TS use case, 
we demonstrate an attack in which the threat actor is a malicious UE.
Based on the attacker model described in Section~\ref{sec:threat_model}, we assume that the attacker has the following adversarial capabilities:
(a) Sensor data access--the adversary can manipulate the behavior (data) of it's own UE;
(b) Decision-based query access--the adversary knows its current serving cell;
and (c) Task knowledge--the adversary knows the traffic steering task's input and goal.
The goal of the adversary is to receive a high QoE while spoofing KPIs such that the signal is classified as poor by the ML application used for QoE prediction.
Such an attack may result in alert fatigue (many alerts about UEs with anomalous QoE) and in the exhaustion of resources due to the fact that the system will have to analyze more UE and reallocate it (UE handover from one cell to another is a resource-consuming operation).
In this attack scenario, the  adversary  generates a crafted signal by manipulating its own behavior; this manipulation is initiated by an AML attack.  
The signal parameters are propagated through the O-RU, O-DU, and O-CU, and the KPIMON xApp extracts KPIs from the manipulated signal.
The AD xApp and QP xApp produce the estimated QoE for each neighbor cell and send that information to the traffic steering xApp which updates the serving cell accordingly.
The adversary receives the cell update status and updates the crafted signal.
This process is repeated until the cell serving the UE is changed.

%\vspace{-10pt}
\subsection{Evasion Attack Implementation}
The TS use case in the Cherry version has the following limitations: \textit{(1)} the SMO logic is not implemented, and therefore the AD xApp and QP xApp in the \nrt~implement the whole ML workflow (i.e., data processing, training, prediction); \textit{(2)} the QP xApp is not implemented; \textit{(3)} the anomaly detection process (the AD xApp) is implemented in two phases: first, a signal classification model is used to predict for a given UE the QoE category of a given signal (which can be excellent, good, average, or poor) and then, a naive rule-based approach is used for selecting the anomalous UEs; \textit{(4)} only an offline implementation of the use case is available, i.e., the data is not dynamic, and only a static dataset for training and evaluation is available.
Given the above limitations, the implemented attack focused on the signal classification model (implemented as part of the AD xApp):
% \begin{tcolorbox}[colback=gray!5!white,colframe=gray!75!black,fonttitle=\small,left=0pt,right=0pt,title=The Signal Classification Model]
% \footnotesize

    \noindent\textbf{Task:} Predict for a given UE the QoE category of a signal.

    \noindent\textbf{Training set:} %The dataset used to train the signal classification model was 
    Collected using a simulator which was created specifically to simulate the activity of multiple UEs in a predefined geographical area. 
    %As mentioned above, the use case is implemented in an offline mode (i.e., the data is static).

    \noindent\textbf{Features:} The KPIMON xApp extracts two feature groups:
    \textit{UE metrics} that include: the UE ID and serving cell ID, %packet data convergence protocol (PDCP) 
    PDCP report timestamp, PDCP aggregation period, UE PDCP downlink/uplink, resource block (PRB) report timestamp, PRB aggregation period, UE PRB downlink/uplink ratios, serving cell report timestamp, reference signal received power, reference signal received quality, and signal-to-noise ratio; and, \textit{Cell metrics} that include: the cell ID, PCDP report timestamp, PCDP aggregation period, PCDP downlink/uplink as the PCDP, PRB report timestamp, PRB aggregation period, and PRB downlink/uplink ratios.
    
    \noindent\textbf{Labels:} The true labels (i.e., QoE categories) of the signals were assigned using an expert-based procedure which defines KPI thresholds for each category.
    
    \noindent\textbf{Algorithm:} Random Forest model.
   
    %\noindent\textbf{Goal of the attack:} To fool the signal classification model.
% \end{tcolorbox}
For the attack implementation we select HopSkipJump~\cite{chen2020hopskipjumpattack}, a query-based evasion attack technique.
The attack begins by sampling two signals from the training set: a poor signal and an excellent signal.
Those signals are the \textit{initial sample} and \textit{target sample}, which are required for the execution of the HopSkipJump attack.
In each step of the attack, the adversary (1) performs a binary search between the initial signal and the target signal; (2) adjusts the resulting sample based on the gradient with respect to the boundary.

\begin{figure}[h]
    \centering
     \includegraphics[width=0.5 \textwidth]{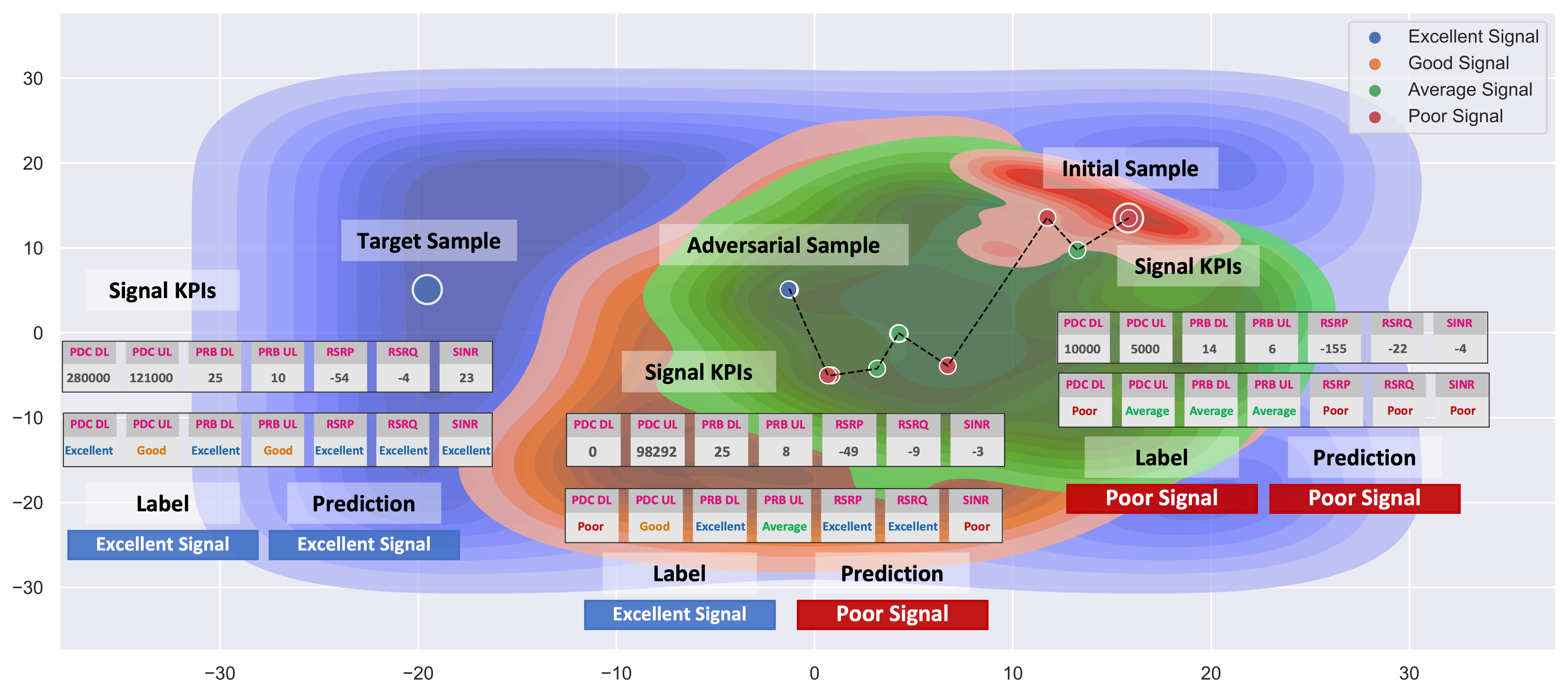}
    \caption{Demonstrating an attack strategy using the HopSkipJump attack technique.}
    \label{fig:attack_strategy}
\end{figure}

Figure~\ref{fig:attack_strategy} illustrates the classifier's decision boundary, and presents a single case where we applied the HopSkipJump attack.
The illustration was created by monotonically sampling the input space of the classifier (UMAP was used for dimensionality reduction). 
As can be seen, the decision boundaries of the different categories overlap. 
This observation demonstrates the feasibility of performing an evasion attack on the RF model. 
We also present the path from the initial sample (a poor signal) to the adversarial sample.
As can be seen, the adversarial example is classified as a poor signal but labeled (based on the expert thresholds) as an excellent signal.
This process can be repeated to generate many adversarial examples.

%\vspace{-15pt}
\subsection{Poisoning Attack Implementation}

Poisoning attacks exploits the ML training procedure by injecting crafted data samples into the dataset used for training the learning algorithm.
The adversary must have the ability to write to the dataset used to train the model.
Within the context of the traffic steering usecase, a poisoning attack can be executed by exploiting the following inherent vulnerabilities: 
(1) Both Cell metrics and UE metrics are not validated against AML attacks.
(2) To maintain accuracy, the traffic steering models are retrained on collected data.
Thus, by utilizing a gradient-based poisoning attack (white box attack), compromised UE(s) can generate manipulated signals that will eventually result in wrong detection of anomalies and wrong classifications of the QoE.
All data (manipulated signals and wrong classifications) are stored in the data repository and used later for training the models.
The model trained on the poisoned data will have an impact on the classifications of legitimate UEs.

%\vspace{-15pt}
\section{\label{sec:countermeasures} Mitigating AML Threats in \oran}
%\vspace{-3pt}
In this section, we present the methodology we have defined to evaluate the criteria for AML security countermeasures in \oran. 
The proposed methodology includes various properties (e.g., protection phase and security type).
Through our analysis, we reviewed the various countermeasures and grouped them into four categories, where the unique property of each category is the stage in which the countermeasure should be integrated throughout the model's life pipeline.
This analysis practical ability is to match a defensive need with applied methods while distinguishing the characteristics required for its implementation.

%\vspace{-10pt}
\subsection{\label{subsec:categories}Method %ology 
for Countermeasure Evaluation}
The methodology we have defined aims to evaluate the impact of various countermeasures on constraints and features that are important for the model in the \oran environment, e.g., model's ability to operate in real-time or duration of the training phase.
In addition, the methodology aims to examine the various countermeasures in light of the threat analysis perspective(i.e., the defender's perspective)
, as the models embedded in \oran have different access and can sometimes interface as a white box and sometimes as a black box to which access is minimal.
This approach is intended to link a defensive need and a relevant countermeasure while considering the defender's degree of privileges and the model's limitations in a specific deployment scenario.

We consider four categories of countermeasures implemented at different stages of the model's pipeline and six properties to measure their effectiveness.
The four countermeasure categories are as follows:

\noindent \textbf{[CC1] Data Pre-processing} consists of techniques that manipulate, process, and filter data to ensure its purity and diversity during the data collection phase and before the training phase begins. 

\noindent \textbf{[CC2] Model Robustness Enhancement} consists of techniques designed to strengthen the model's robustness by intervening in the training phase, architecture changes, hyperparameters, and learning techniques to improve the model's resistance to various attacks.

\noindent \textbf{[CC3] Model and Data Privacy Preserving} consists of techniques designed to prevent the disclosure of valuable information about the model's parameters, structure, or the data used for training. 

\noindent \textbf{[CC4] Auxiliary Models}  consists of techniques designed to aid the original model with preliminary identification of adversarial examples, restriction of adversarial examples, and cleaning of adversarial perturbations. 

%\subsection{\label{sec:properties}Properties of the Countermeasures}
%In order to 
% To analyze the countermeasures and defense techniques, we defined an ontology (presented in Figure~\ref{fig:countermeasures_ontology}) to examine the properties that distinguish each method.

%The metrics that extends the properties are presented in Figure~\ref{fig:countermeasures_taxonomy}.

\subsection{\label{sec:properties}Properties of the Countermeasures}
\begin{figure}[h]
    \centering
    \includegraphics[trim={0.05cm 0 0.1cm 0},clip,width=0.5\textwidth]{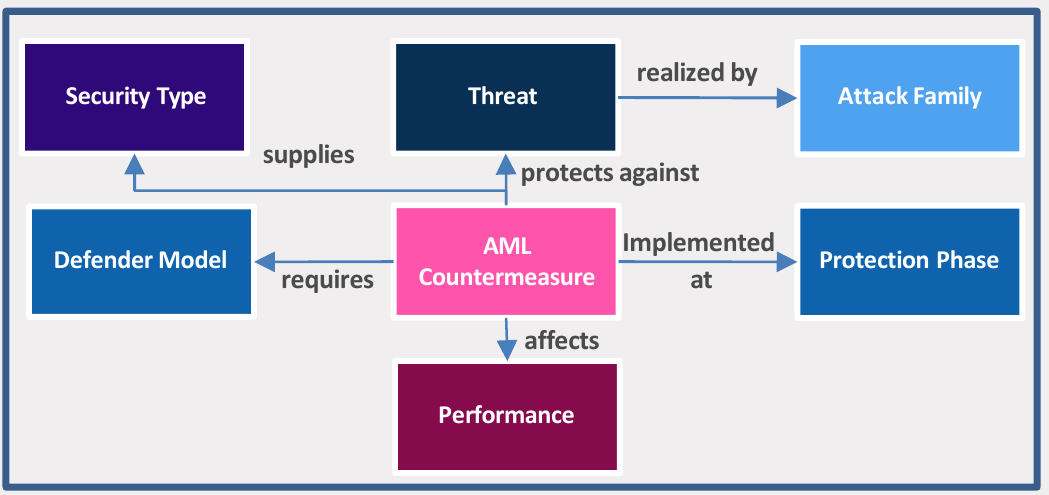}
    \caption{Countermeasure property ontology.}
    \label{fig:countermeasures_ontology}
\end{figure}

\begin{figure}[h]
    \centering
    \includegraphics[width=0.5\textwidth]{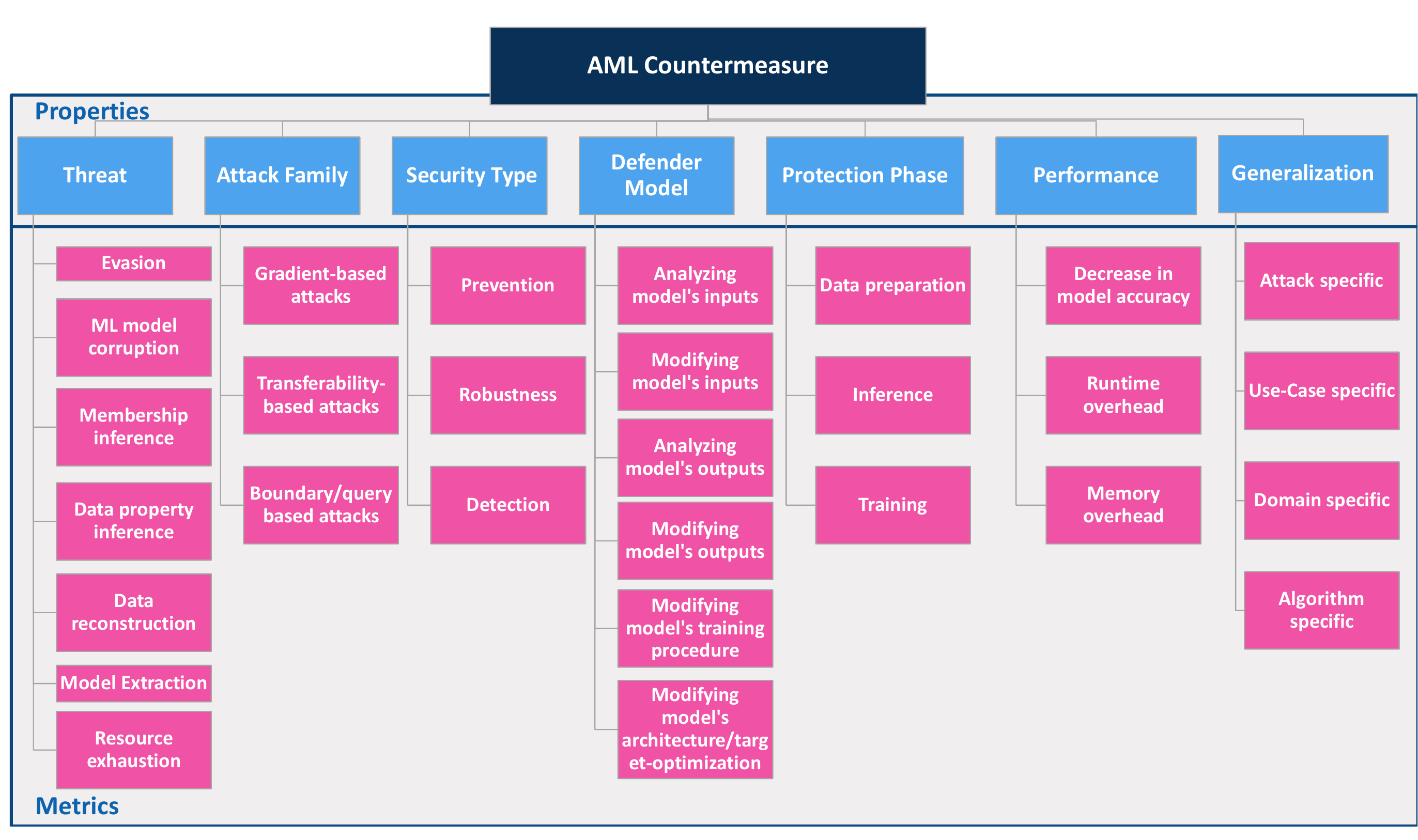}
    \caption{The attributes by which the countermeasures are measured.}
    \label{fig:countermeasures_taxonomy}
\end{figure}

In order to analyze the countermeasures and defense techniques, we defined an ontology (presented in Figure~\ref{fig:countermeasures_ontology}) to examine the properties that distinguish each method. The metrics that extends the properties are presented in Figure~\ref{fig:countermeasures_taxonomy}.
% \begin{figure}[h]
%     \centering
%     \includegraphics[width=0.48\textwidth]{sections/figures/countermeasures_ontology.pdf}
%     \caption{Countermeasure property ontology.}
%     \label{fig:countermeasures_ontology}
% \end{figure}

\noindent \textbf{[P1] Threat} This property pertains to the threats the countermeasures can deal with.
In this context the threats against a given system or ML model are as follows:
evasion, ML model corruption, membership inference, data property inference, data reconstruction, model extraction, and resource exhaustion. 
A description of the threats is provided in Section~\ref{sec:threat_category}.
It is important to note that some defense techniques can provide protection against more than one threat simultaneously.

\noindent \textbf{[P2] Attack Family} This property pertains to the attack families that the countermeasure addresses. 
The attack families derive from the attacks described in Section~\ref{sec:attack_tech}:

\noindent \textbf{Gradient-based attacks (white-box):} Attacks in which an adversary aims to use or impact the gradient of a model, perturbing the input with the gradient of the loss and gradually increasing the magnitude until the input is misclassified. 

\noindent \textbf{Transferability-based attacks (black-box):} Attacks in which an adversary aims  to acquire a specific ability of a given attack in the context of a particular ML model and use it effectively against another potentially unknown model. 

\noindent \textbf{Query-based attacks (black-box):} Attacks in which an adversary aims to use query access to a given victim ML model to exploit a threat vector and use it against the victim model.

\noindent \textbf{[P3] Security Type} 
This property pertains to the abstract defense mechanism of a specific defense method. 
We note three different types of defense mechanisms a countermeasure can serve.

\noindent \textbf{Prevention:} This defense mechanism aims to eliminate the effects (damage) of adversarial inputs, usually via modification of the input data. 

\noindent \textbf{Robustness:}  This defense mechanism aims to create a model that can withstand (and be minimally impacted) by adversarial attacks. 
This can be achieved by intervention in the model architecture or during training phase, in order to expose a given ML model to adversarial threats.

\noindent \textbf{Detection:} This defense mechanism aims to identify adversarial inputs as they are received, either during the training or inference phase, enabling the model to deal with detected adversarial samples and prevent harmful effects (e.g., gradient decline or accuracy loss). 

\noindent \textbf{[P4] Defender Model} 
This property pertains to the potential access requirements a given countermeasure requires. 
In the \oran environment, the defender model differentiates between the defender's privileges and access to the model they are required to protect.
\noindent \textbf{Analyzing the model's inputs:} Requires \textit{read-only} access to the model's inputs during the training or inference phase.

\noindent \textbf{Modifying the model's inputs:} Requires \textit{write} access to the model's inputs during the training or inference phase.

\noindent \textbf{Analyzing the model's outputs:} Requires \textit{read-only} access to the model's outputs during the training or inference phase.

\noindent \textbf{Modifying the model's outputs:} Requires \textit{write} access to the model's outputs during the training or inference phase.

\noindent \textbf{Modifying the model's training procedure:} Indicates that the countermeasure can change how the given ML model's training occurs.

\noindent \textbf{Modifying the model's architecture/target optimization:} Indicates that the countermeasure can change the model's structure and optimization.

\noindent \textbf{[P5] Pipeline Protection Phase} 
This property pertains to the stage of the ML model's pipeline (lifecycle) in which the countermeasure is implemented. 
The ML pipeline is described in Section~\ref{sec:ml_dep_oran} and consists of the following: data preparation phase, training phase, testing and inference phase.

\noindent \textbf{[P6] Performance} 
This property pertains to the characteristics of the original model that can be affected by the countermeasure implementation.
This property directly impacts the ability to implement it in different deployment scenarios; for example, we would like to avoid an impact on the duration of training in a deployment scenario where the training-host is in the Near-RT RIC.

\noindent \textbf{Decrease in model accuracy:} This metric indicates whether there is a decrease in the model's performance level, compared to normal conditions (without an attack), after implementing the countermeasure.

\noindent \textbf{Runtime overhead:} This parameter indicates whether the length of the inference time increases after implementing the countermeasure. This parameter is crucial in real-time systems where the model's response latency is essential.

\noindent \textbf{Memory overhead:} This parameter indicates whether additional memory resources are required to implement the countermeasure (above and beyond the memory that would have been required without its implementation).

\subsection{\label{subsec:counters}Countermeasure Analysis}
%\vspace{-5pt}
%     \item I have not read all the countermeasures that you review. But from the ones i have read, i noticed that there is a very loosely connection between the description of the countermeasure and the relevant taxonomy properties.
%     \item In my opinion, for each countermeasures that you are describing you must (a) map it to the relevant properties and metrics, (2) justify your mapping - i.e., **WHY** you selected this specific mapping and not other mapping (unless it is trivial) 
    
\begin{figure*}[tb]
    \centering
    \includegraphics[width=0.85\textwidth]{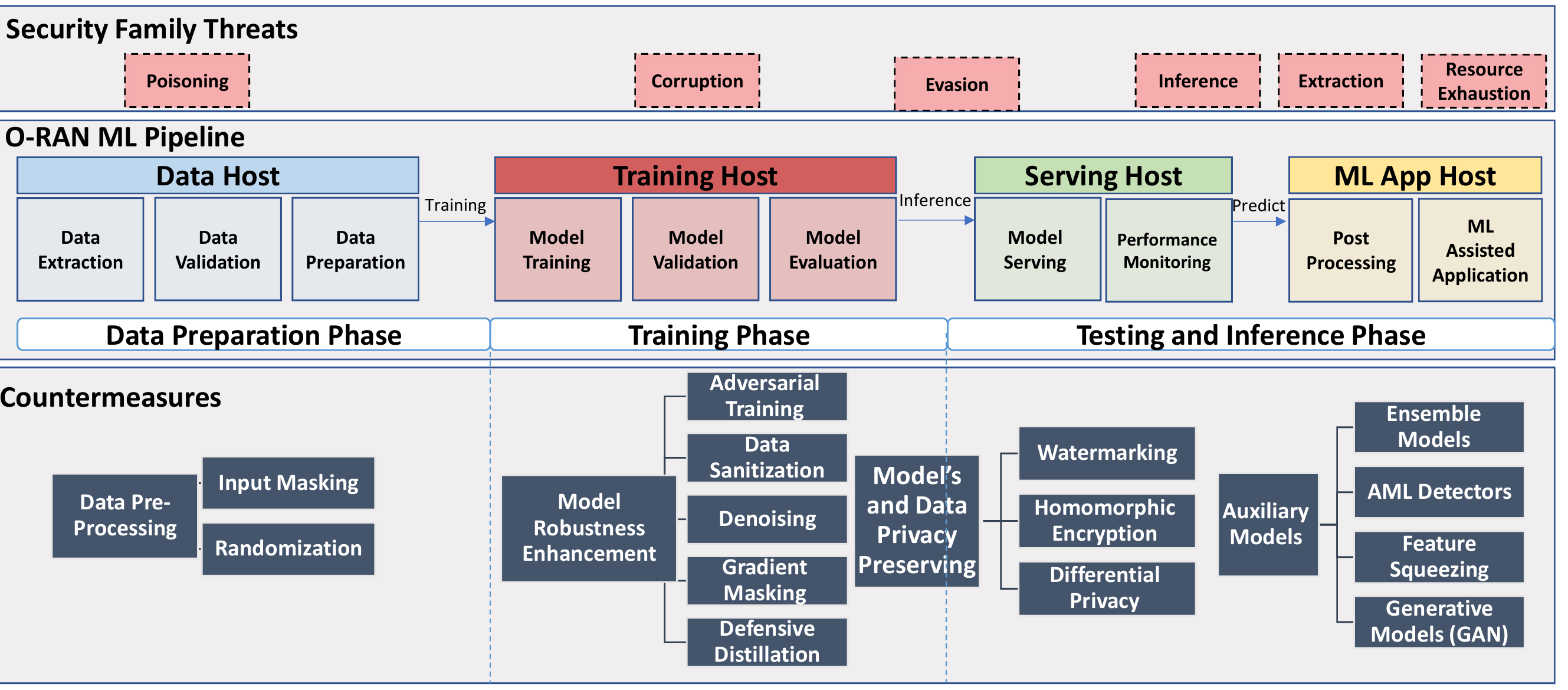}
    \caption{AML countermeasure mapping.}
    \label{fig:countermeasures_oran}
\end{figure*}

\begin{comment}
\begin{figure}[h]
    \centering
    \includegraphics[width=0.5\textwidth]{sections/figures/advcountermeasures-final.pdf}
    \caption{AML countermeasure mapping.}
    \label{fig:countermeasures_oran}
\end{figure}
\end{comment}
Below, we describe the various techniques in each of the countermeasure categories and the properties defined in Section~\ref{subsec:categories}.
The countermeasures are mapped based on their implementation phase in Figure~\ref{fig:countermeasures_oran}.

%%%%%%%%%%%%%%%%%%%%%%%%%%%%%%%%%%%%%%%%%
%\vspace{-8pt}
\noindent \textbf{Data Pre-processing Category:}
\newline
\noindent \textit{Randomization} is a technique that aims to prevent adversaries from constructing adversarial examples by causing the ML model to be non-deterministic. 
Randomization can be applied in the data preprocessing and inference phases by randomly nullifying features within samples or during inference~\cite{wang2017adversary}, making the model more robust to random perturbations.
Studies have shown that this technique can also be helpful at inference time by applying randomization operations to the input before it is fed into the model can also mitigate the effect of adversarial examples (e.g., Xie et al.~\cite{xie2017mitigating}).

\noindent \textit{Data augmentation} is a technique designed to improve the robustness of the model by learning a wide variety of samples.
Several proposed methods applied this technique to generate and generate augmented datasets that provide increased robustness to common corruptions, such as AutoAugment~\cite{cubuk2018autoaugment}, RandAugment~\cite{cubuk2020randaugment}, and DeepAugment~\cite{hendrycks2021many}.
Recent studies have shown how generative models trained solely on the original training set can be leveraged to artificially increase the size of the original training set and improve robustness~\cite{gowal2021improving}.

%%%%%%%%%%%%%%%%%%%%%%%%%%%%%%%%%%%%%%%
\noindent \textbf{Model Robustness Enhancement Category:}
\newline
\noindent \textit{Adversarial training} is a technique introduced by Goodfellow et al.~\cite{goodfellow2014explaining} for improving a model's resistance against the threat of evasion attacks at inference time, in which perturbed input results in the model outputting an incorrect label with high confidence. In this technique, adversarial samples are fed into the model during the training path and explicitly labeled as threatening thus creating a robust model for adversarial perturbation.
Adversarial training has been shown to be a generalized defense technique that is effective against a wide range of attacks, adding negligible runtime overhead during the training phase.
Over the past few years, various adversarial training improvements were proposed~\cite{bai2021recent} to improve the efficiency of the training phase and the model's accuracy in the testing phase.
% \begin{tcolorbox}[colback=gray!5!white,colframe=gray!75!black,fonttitle=\small,left=0pt,right=0pt,title=Countermeasure properties]
% \footnotesize
% \noindent \textbf{[P1] Threat} refers to the different types of adversarial attacks that a countermeasure can defend against, such as evasion, model corruption, and data property inference. \\
% \noindent \textbf{[P2] Attack Family} refers to the attack techniques that a countermeasure addresses, such as gradient-based attacks, transferability-based attacks, and query-based attacks.\\
% \noindent \textbf{[P3] Security Type}  refers to the general defense mechanism used by a countermeasure, such as prevention, robustness, and detection.\\
% \noindent \textbf{[P4] Defender Model}  refers to the access requirements a countermeasure has to the model it is protecting, such as the ability to analyze or modify the model's inputs or outputs, or modify the model's training procedure or architecture.\\
% \noindent \textbf{[P5] Pipeline Protection Phase}  refers to the stage of the machine learning pipeline at which a countermeasure is implemented, such as data preparation, training, testing, or inference as described in Section~\ref{sec:ml_dep_oran} .\\
% \noindent \textbf{[P6] Performance} refers to characteristics of the original model that can be affected by the countermeasure, such as a decrease in model accuracy, runtime overhead, and memory overhead.
% \end{tcolorbox}
\noindent \textit{Data sanitization} is a technique used to ensure the purity of the training data by detecting and separating adversarial samples from normal ones and removing the malicious samples~\cite{nelson2009misleading,chan2018data}. The main idea behind the data sanitization technique is to avoid training the model on samples that have a significant impact on the model; therefore,  this technique necessitates that the impact on the model's classification performance during the training phase be quantified; there is also the need to repeat this process several times, each time with a dataset from which highly influential samples were removed.
As a result,in an effort to produce a model trained on clean data without abnormal samples, data sanitization increases the training time. The study of \cite{chan2018data} shows that by applying this technique, the model's robustness increased by 10\%, the model's accuracy decreased by 1-3\%, and the training time was five times longer.

\noindent \textit{Feature denoising} is a technique used to remove noise from an example (e.g., signal, image) through architecture changes designed to increase the model's robustness to adversarial examples.
Xie et al.~\cite{xie2019feature} suggested an implementation of feature denoising for image classification models, which resulted in a robust model that achieved a high accuracy for white/black-box attacks. 

\noindent \textit{Gradient masking} is a technique that aims to produce a model that is less susceptible to gradient-based adversarial attacks by reducing the usefulness of the defended model’s gradients~\cite{papernot2017practical}. Countermeasures based on the gradient masking technique result in a very smooth model in specific directions of training points, making it harder for the adversary to find gradients indicating good candidate directions for perturbing an input in a way that harms the target model. It is important to note that while gradient masking provides robustness against gradient-based black-box attacks, the models are often still vulnerable to non-gradient-based attacks~\cite{athalye2018obfuscated}.

\noindent \textit{Defensive distillation} is a technique that aims to add flexibility to the model's algorithm and improve the model's robustness by making it smoother. The distillation concept proposed by Hinton et al.~\cite{hinton2015distilling} deals with transferring the knowledge received in the training phase (teacher model) to a second model (student model) that may be less susceptible to adversarial examples. Papernot et al.~\cite{papernot2016distillation} demonstrated how defensive distillation could reduce the effectiveness of adversarial sample creation to nearly zero.
This technique increases the training time, as it necessitates the creation of two models, but at the same time, the training phase produces a student model that may be smaller, less complex, and faster than the teacher model.

%%%%%%%%%%%%%%%%%%%%%%%%%%%%%%%%%%%%%%%
\noindent \textbf{Model and Data Privacy Preserving Category:}
\newline
\noindent\textit{Watermarking} is a technique that serves as the main component of existing methods for protecting a model's copyright and preventing model extraction attacks; in this case, the model is considered the intellectual property of the legitimate party it was trained by.
In this technique, secret information is embedded into the model's parameters to prove ownership.
There are several ways of implementing watermarking were originally proposed for implanting secrets in neural networks, e.g., modifying the training algorithm~\cite{chen2018deepmarks,gu2017badnets,le2020adversarial}, poisoning the training data~\cite{adi2018turning}, and writing the secret within the model's parameters after the training phase~\cite{song2017machine}.
It is important to note that inserting a watermark in a model does not prevent theft but enables legitimate owners to identify their model instances if they have been stolen.
Most watermarking methods either require minimal storage resources when hiding the secret in the model's parameters or increase the training time when assimilating the secret in the training process.

\noindent\textit{Homomorphic encryption} is a technique that focuses on preserving the training and testing data's privacy and preventing data property, membership inference, and data reconstruction attacks. The homomorphic encryption scheme~\cite{paillier1999public,gentry2009fully} allows computation on ciphertext, generating an encrypted result which, when decrypted, matches the result of the operations as if they had been performed on plaintext, thereby enabling evaluation on encrypted data. 
Existing homomorphic encryption schemes require custom work to adjust each ML model and a heavy computational load for the training phase~\cite{ryffel2018generic,takabi2016privacy, wood2020homomorphic}.

\noindent\textit{Differential privacy} is a technique proposed by Dwork et al.~\cite{dwork2008differential} that aims to produce an algorithm capable of learning statistical information about the population without disclosing information about individuals.
This technique is mainly used for quantifying the degree of privacy protection provided by an algorithm on the underlying dataset it operates on.
Therefore, ML algorithms and training mechanisms are designed to preserve privacy~\cite{ji2014differential}.
Abadi et al.~\cite{abadi2016deep} suggested an approach for training a deep neural network with differential privacy that can be adapted to many optimization methods, with a manageable cost in terms of the model's accuracy and training efficiency.

%%%%%%%%%%%%%%%%%%%%%%%%%%%%%%%%%%%%%%%%%%%%%%%%%%
\noindent \textbf{Auxiliary Model Category:}
\newline
\noindent\textit{Ensemble models} % combine several base models in order to produce a single optimal predictive model. 
%In the context of AML, 
multiple classifiers are trained together and combined to improve robustness. 
The ensemble technique requires significant storage resources and, at runtime, the ability to run the models in parallel, unless results in substantial damage in the testing and inferring phase.
Recently, several studies aimed at reducing the complexity and overhead inherent in this technique, e.g., Shen et al.~\cite{shen2019meal} presented a method for compressing large %, complex trained 
ensembles into a single network, where knowledge from a variety of trained deep neural networks (DNNs) is distilled and transferred to a single DNN.

\noindent\textit{Feature squeezing} is a technique used to reduce the noise of the input to highlight and reduce the effect of adversarial tampering. 
Although the technique was originally evaluated on image classification models, the feature-squeezing approach was used in many domains (e.g., toxic text detection~\cite{carlini2016hidden}, voice recognition~\cite{hosseini2017deceiving}, or malware detection~\cite{rosenberg2021sequence}).
Usually, feature squeezers are used as part of a detection framework~\cite{xu2017feature} (in AML detectors) to prevent evasion attacks. 
The use of squeezers as part of the predictive ML model as opposed to the detection framework is not widespread, as feature squeezers are known to degrade the model's accuracy significantly.

\noindent \textit{AML detectors} are based on auxiliary machine learning models whose purpose is to distinguish between benign and adversarial examples. The AML detector analyzes the model's input, output, or internal properties and disregards the sample or the prediction if a malicious sample is detected.
An AML detector requires storing and training an additional model, which affects the performance and usability of the base-model.

\noindent \textit{Generative models (GAN)} designed by Goodfellow~\cite{goodfellow2020generative}, are part of a class of ML models, %frameworks, 
which can be a component of defensive frameworks in order to reproduce the sample without adversarial perturbations.
Recent studies~\cite{samangouei2018defense, yang2021novel} have shown that GANs can reconstruct an input without adversarial perturbations. Therefore, the use of GANs in the target model as a protective layer before processing the input sample can serve as a prevention mechanism against adversarial attacks.

Table~\ref{tab:countermeasures} contains the mapping between the all the above-mentioned countermeasures according to the properties listed in Section~\ref{subsec:categories}.

\begin{comment}
\begin{table}[h]
    \centering
    \includegraphics[width=0.48\textwidth]{sections/figures/tab1_cropped.pdf}
    \caption{Mapping the countermeasures according to the different properties.}
    \label{tab:countermeasures}
\end{table}
\end{comment}
%\input{sections/Tables/countermeasures_table1.tex}

%\vspace{-15pt}
\subsection{Affect of Countermeasures on O-RAN}
%\vspace{-5pt}
In order to implement countermeasures within the \oran %environment
, it is necessary to consider the constraints and implications inherent in the \oran environment.
The models embedded in the O-RAN environment can be integrated into several areas of a system. 
In some deployment scenarios, where serving-host or application-host integrated in the Near-RT RIC, the model must maintain an almost immediate response time and consume few resources.
In addition, the models can derived from different sources, and accordingly, the permissions needed to access the model will be different, starting from a black-box model that is ready as plug-and-play to a  modifiable white-box model.
Several aspects can be affected by a countermeasure's implementation:

%\begin{tcolorbox}[colback=gray!5!white,colframe=gray!75!black,fonttitle=\small,left=0pt,right=0pt,title=Countermeasure requirements and constraints,enhanced,breakable,pad at break*=1mm,]
% \begin{tcolorbox}
% [colback=gray!5!white,colframe=gray!75!black,fonttitle=\small,left=2pt,right=2pt,title=Countermeasure requirements and constraints,boxrule=2pt,arc=0.6em,boxsep=0mm]
% \footnotesize
\noindent \textbf{Model Access:}
Implementation of the countermeasure may require architecture changes, modification in the code or the layers' structure, or changes in the loss functions or other hyperparameters, all of which require access to the influencing variables of the model (applied on \textit{white-box models}).
Some countermeasures do not require access to the influencing variables at all, and as a result, they can be implemented without revealing any information about its internal workings (applied on \textit{black-box models}).

\noindent \textbf{Data Access:}
Implementation of the countermeasure may require access to the data format and its structure and representation (\textit{feature data}) or the content of the data (\textit{training data}) used to train the model.

\noindent \textbf{Implementation Host:}
The implementation of the countermeasure may be suitable for different stages of the model's pipeline. Thus, it must be adapted for implementation in different O-RAN host environments:
\textit{data host, training host, serving host, application host}.

\noindent \textbf{\oran Component Constraints:}
The implementation of the diverse countermeasures may affect various characteristics and constraints of the model; in the O-RAN environment, it is of great importance that a countermeasure's implementation doesn't influence the model's response time in the inference phase (\textit{inference at Near RT-RIC}) as well as to maintain the model's simplicity in terms of its size and the resources it consumes (\textit{lightweight}).
Complex and heavy calculations are to be performed at Non-RT RIC.
% \end{tcolorbox}

The mapping between the various countermeasures and the necessary constraints for their implementation in the \oran environment is found in Table~\ref{tab:affectedaspects}.

\begin{table*}[h!]
%\tiny
\centering
%\resizebox{\linewidth}{!}{
%\begin{tabular}{|c||c|c||c|c||c|c||c|c|}
  % \resizebox{%
  %     \ifdim\width>\columnwidth
  %       \columnwidth
  %     \else
  %       \width
  %     \fi
  %   }{!}{%
\begin{tabular}{|c|c|cc|ccccc|ccc|cccc|}
%\begin{tabular}{|c|c|c|c|c|c|c|c|c|c|c|p{0.48\textwidth}|}
\hline
%\textbf{Threat} & \multicolumn{2}{c||}{\textbf{\#Events}} & \multicolumn{2}{c||}{\textbf{\#Machines}} & \multicolumn{2}{c||}{\textbf{Alerts}} & \multicolumn{2}{c|}{\textbf{Good Alerts}}\\ \hline

\multicolumn{2}{|c|}{\textbf{Categories}} & \multicolumn{2}{c|}{\makecell{\textbf{Data}\\\textbf{preprocessing}}} & \multicolumn{5}{c|}{\makecell{\textbf{Robustness}\\\textbf{enhancement}}} & \multicolumn{3}{c|}{\makecell{\textbf{Data and}\\\textbf{privacy}\\\textbf{preserving}}} & \multicolumn{4}{c|}{\makecell{\textbf{Auxilary}\\\textbf{model}}} \\ \hline

\multicolumn{2}{|c|}{AML countermeasures}  & \rotatebox{90}{\textbf{Data augmentation}} & \rotatebox{90}{\textbf{Randomization}} & \rotatebox{90}{\textbf{Adversarial training}} & \rotatebox{90}{\textbf{Data Sanitization}} & \rotatebox{90}{\textbf{Denoising}} & \rotatebox{90}{\textbf{Gradient Masking}} & \rotatebox{90}{\textbf{Defensive Distillation}} & \rotatebox{90}{\textbf{Watermarking}} & \rotatebox{90}{\textbf{Homomorphic Encryption}} & \rotatebox{90}{\textbf{Differential Privacy}} & \rotatebox{90}{\textbf{Ensemble}} & \rotatebox{90}{\textbf{AML Detector}} & \rotatebox{90}{\textbf{Feature Squeezers}} & \rotatebox{90}{\textbf{Generative Models}} \\ \hline

%\textbf{Threat} & {\textbf{Data reconstruction} & \multicolumn{2}{c||}{\textbf{\#Machines}} & \multicolumn{2}{c||}{\textbf{Alerts}} & \multicolumn{2}{c|}{\textbf{Good Alerts}}\\ \hline

\multirow{5}{*}{\textbf{Threat}} & \textbf{Data reconstruction} & $\circ$ & $\circ$ & $\circ$ & $\circ$ & $\circ$ & $\circ$ & $\circ$ & $\circ$ & $\bullet$ & $\bullet$ & $\circ$ & $\circ$ & $\circ$ & $\circ$ \\ 
 & \textbf{Data property inference} & $\circ$ & $\circ$ & $\circ$ & $\circ$ & $\circ$ & $\circ$ & $\circ$ & $\circ$ & $\bullet$ & $\bullet$ & $\circ$ & $\circ$ & $\circ$ &  $\circ$\\ 
 & \textbf{Model corruption} & $\circ$ & $\circ$ & $\circ$ & $\bullet$ & $\bullet$ & $\bullet$ & $\bullet$  &  $\circ$ &  $\circ$ &  $\circ$ &  $\circ$ & $\bullet$ & $\bullet$ & $\bullet$ \\ 
  & \textbf{Model extraction} & $\circ$  & $\circ$  & $\circ$  & $\circ$  & $\circ$  & $\circ$  & $\circ$  & $\bullet$ & $\bullet$ & $\circ$ & $\circ$ & $\circ$ & $\circ$ & $\circ$ \\
  & \textbf{Evasion} & $\bullet$ & $\bullet$ & $\bullet$ & $\bullet$ & $\bullet$ & $\bullet$ & $\bullet$ & $\circ$ & $\circ$ & $\circ$ & $\bullet$ & $\bullet$ & $\bullet$ & $\bullet$ \\ \hline

\multirow{3}{*}{\textbf{Attack Family}} & \textbf{Gradient based (white-box)} &  $\bullet$  &  $\bullet$  &  $\bullet$  &  $\bullet$  &  $\bullet$  &  $\bullet$  & $\circ$ & $\circ$ & $\bullet$ & $\bullet$ &  $\bullet$  &  $\bullet$  & $\circ$ &  $\bullet$  \\ 
 & \textbf{Transferability based (black-box)} & $\circ$ & $\circ$ & $\bullet$  & $\circ$ & $\bullet$ & $\circ$ & $\circ$ & $\circ$ & $\circ$ & $\circ$ & $\bullet$ & $\bullet$  & $\bullet$  & $\bullet$  \\ 
 & \textbf{Boundary/Query based (black-box) } & $\circ$ & $\circ$ & $\bullet$ & $\circ$ & $\bullet$ & $\circ$ & $\bullet$ & $\circ$  & $\bullet$ & $\bullet$ & $\bullet$ & $\bullet$ & $\bullet$ & $\bullet$ \\  \hline
 
\multirow{3}{*}{\textbf{Security Type}} & \textbf{Prevention} & $\circ$ & $\bullet$ & $\circ$ & $\bullet$ & $\circ$ & $\circ$ & $\circ$ & $\circ$ & $\bullet$ & $\bullet$ & $\circ$ & $\circ$ & $\circ$ & $\bullet$ \\ 
 & \textbf{Robustness} & $\bullet$ & $\bullet$ & $\bullet$ & $\circ$  & $\bullet$ & $\bullet$ & $\bullet$ & $\circ$  & $\circ$  & $\circ$  & $\bullet$ & $\circ$  & $\circ$  & $\circ$  \\ 
 & \textbf{Detection} & $\circ$  & $\circ$  & $\circ$  & $\bullet$ & $\circ$  & $\circ$  & $\circ$  & $\bullet$ & $\circ$  & $\circ$  & $\circ$  & $\bullet$ & $\bullet$ & $\bullet$ \\  \hline

\multirow{6}{*}{\textbf{Defender Model}} & \textbf{Analyzing model's inputs} & $\bullet$ & $\bullet$ & $\bullet$ & $\bullet$ & $\circ$ & $\circ$ & $\bullet$ & $\circ$  & $\bullet$ & $\bullet$ & $\circ$  & $\bullet$ & $\bullet$ & $\bullet$ \\ 
 & \textbf{Modifying model's inputs} &  $\bullet$ &  $\bullet$ &  $\bullet$ &  $\bullet$ & $\circ$  & $\circ$  &  $\bullet$ &  $\bullet$ & $\bullet$ & $\bullet$ & $\circ$ & $\circ$  &  $\bullet$ &  $\bullet$ \\ 
 & \textbf{Analyzing model's outputs} & $\circ$ & $\circ$  & $\circ$  & $\circ$  & $\circ$  & $\circ$  & $\circ$  & $\circ$  & $\bullet$ & $\bullet$ & $\circ$  & $\circ$  & $\circ$  & $\circ$  \\ 
 & \textbf{Modifying model's outputs} & $\circ$ & $\circ$  & $\circ$  & $\circ$  & $\circ$  & $\circ$  & $\circ$  & $\bullet$  & $\bullet$ & $\bullet$ & $\circ$  & $\circ$  & $\circ$  & $\circ$  \\ 
  & \textbf{Modifying training procedure} & $\circ$  & $\circ$  &  $\bullet$& $\bullet$  & $\circ$ & $\bullet$ & $\bullet$ & $\bullet$ & $\bullet$ & $\bullet$ & $\bullet$ & $\circ$   &  $\circ$ & $\bullet$  \\
  & \textbf{Modifying model's architecture} &  &  &  &  &  &  &  &  &  &  &  &  &  &  \\
 \hline

\multirow{2}{*}{\textbf{Pipeline Protection Phase}} & \textbf{Inference} & $\circ$ & $\circ$ & $\bullet$ & $\bullet$ & $\bullet$ & $\circ$ & $\bullet$ &  $\circ$& $\bullet$ & $\bullet$ & $\bullet$ & $\bullet$ & $\bullet$ & $\bullet$ \\ 
 & \textbf{Training } & $\bullet$  & $\bullet$  & $\circ$ & $\circ$ & $\circ$ & $\circ$ & $\circ$ & $\bullet$  & $\bullet$ & $\bullet$ & $\circ$ & $\circ$ & $\circ$ & $\circ$ \\  \hline
 \multirow{4}{*}{\textbf{Performance}} & \textbf{Accuracy decrease}& $\circ$  & $\circ$  & $\circ$ & $\circ$ & $\bullet$ & $\circ$ & $\bullet$ & $\circ$  & $\bullet$ & $\bullet$ & $\circ$ & $\circ$ & $\bullet$ & $\circ$ \\
 & \textbf{Runtime overhead (inference)} &$\circ$  &$\circ$  &$\circ$   &$\circ$   &$\circ$   &$\circ$   &$\circ$   &$\circ$   &$\bullet$   &$\bullet$   &$\bullet$   &$\bullet$   &$\bullet$   &$\circ$  \\ 
  & \textbf{Runtime overhead (training)} &$\bullet$  &$\bullet$  &$\bullet$   &$\bullet$   &$\bullet$   &$\bullet$   &$\bullet$   &$\bullet$   &$\bullet$   &$\bullet$   &$\bullet$   &$\circ$   &$\circ$   &$\circ$\\
 & \textbf{Memory overhead} &$\bullet$  &$\circ$  &$\bullet$   &$\circ$   &$\circ$   &$\circ$   &$\circ$   &$\bullet$   &$\bullet$   &$\bullet$   &$\bullet$   &$\bullet$   &$\bullet$   &$\bullet$  \\  \hline
  
\end{tabular}%
% }
\caption{Mapping the countermeasures according to the different properties.}
\label{tab:countermeasures}
\end{table*}

\begin{table*}[h!]
%\tiny
\centering
  % \resizebox{%
  %     \ifdim\width>\columnwidth
  %       \columnwidth
  %     \else
  %       \width
  %     \fi
  %   }{!}{%
\begin{tabular}{|c|c|cc|ccccc|ccc|cccc|}
\hline
\multicolumn{2}{|c|}{AML countermeasures}  & \rotatebox{90}{\textbf{Data augmentation}} & \rotatebox{90}{\textbf{Randomization}} & \rotatebox{90}{\textbf{Adversarial training}} & \rotatebox{90}{\textbf{Data Sanitization}} & \rotatebox{90}{\textbf{Denoising}} & \rotatebox{90}{\textbf{Gradient Masking}} & \rotatebox{90}{\textbf{Defensive Distillation}} & \rotatebox{90}{\textbf{Watermarking}} & \rotatebox{90}{\textbf{Homomorphic Encryption}} & \rotatebox{90}{\textbf{Differential Privacy}} & \rotatebox{90}{\textbf{Ensemble}} & \rotatebox{90}{\textbf{AML Detector}} & \rotatebox{90}{\textbf{Feature Squeezers}} & \rotatebox{90}{\textbf{Generative Models}} \\ \hline

\multirow{2}{*}{\textbf{Model Access}} & \textbf{Applied on Black-Box Model} & $\bullet$ & $\bullet$ & $\bullet$ & $\bullet$ & $\circ$ & $\circ$ & $\circ$ & $\circ$ & $\circ$ & $\circ$ & $\bullet$ & $\bullet$ & $\circ$ & $\bullet$ \\ 
 & \textbf{Applied on White-Box Model} & $\bullet$ & $\bullet$ & $\bullet$ & $\bullet$ & $\bullet$ & $\bullet$ & $\bullet$ & $\bullet$ & $\bullet$ & $\bullet$ & $\bullet$ & $\bullet$ & $\bullet$ &  $\bullet$\\ 
\hline
\multirow{3}{*}{\textbf{Data Access}} & \textbf{Requires Access to Training Data} & $\bullet$ & $\bullet$ & $\bullet$  & $\bullet$ & $\circ$ & $\circ$ & $\bullet$ & $\circ$ & $\circ$ & $\bullet$ & $\circ$ & $\bullet$  & $\circ$  & $\bullet$  \\ 
 & \textbf{Requires Access to Features Data} & $\circ$ & $\bullet$ & $\circ$  & $\circ$ & $\bullet$ & $\bullet$ & $\bullet$ & $\bullet$ & $\circ$ & $\bullet$ & $\circ$ & $\circ$  & $\bullet$  & $\circ$  \\ 
\hline
\multirow{4}{*}{\textbf{Applied at Implementation Host}} & \textbf{Data Host} & $\bullet$ & $\bullet$ & $\circ$ & $\circ$ & $\circ$ & $\circ$ & $\circ$ & $\circ$ & $\bullet$ & $\bullet$ & $\circ$ & $\circ$ & $\circ$ & $\circ$ \\ 
 & \textbf{Applied at Training Host} & $\circ$ & $\circ$ & $\bullet$ & $\bullet$  & $\bullet$ & $\bullet$ & $\bullet$ & $\bullet$  & $\circ$  & $\bullet$  & $\circ$ & $\circ$  & $\circ$  & $\bullet$  \\ 
 & \textbf{Applied at Serving Host} & $\circ$  & $\circ$  & $\circ$  & $\circ$ & $\circ$  & $\circ$  & $\bullet$  & $\bullet$ & $\circ$  & $\circ$  & $\bullet$  & $\bullet$ & $\bullet$ & $\bullet$ \\  
 & \textbf{Applied at Application Host} & $\circ$  & $\circ$  & $\circ$  & $\circ$ & $\circ$  & $\circ$  & $\circ$  & $\bullet$ & $\bullet$  & $\circ$  & $\circ$  & $\bullet$ & $\bullet$ & $\circ$ \\    \hline
\multirow{2}{*}{\textbf{\oran Component Constraints}} & \textbf{Applied at Near-RT RIC} &  $\circ$  &  $\bullet$  &  $\circ$  &  $\circ$  &  $\circ$  &  $\circ$  & $\circ$ & $\bullet$ & $\circ$ & $\circ$ &  $\bullet$  &  $\bullet$  & $\bullet$ &  $\bullet$  \\ 
 & \textbf{Applied at Non-RT RIC} & $\bullet$ & $\bullet$ & $\bullet$  & $\bullet$ & $\bullet$ & $\bullet$ & $\bullet$ & $\bullet$ & $\bullet$ & $\bullet$ & $\bullet$ & $\bullet$  & $\bullet$  & $\bullet$  \\ 
\hline
\end{tabular}%
% }
\caption{Mapping countermeasures according to the affected aspects and \oran requirements.}
\label{tab:affectedaspects}
\end{table*}

\begin{comment}
\begin{table}[h]
    \centering
    \includegraphics[width=0.48\textwidth]{sections/figures/tab2_cropped.pdf}
    \caption{Mapping the countermeasures according to the affected aspects and requirements of \oran environment.}
    \label{tab:affectedaspects}
\end{table}
\end{comment}

%\input{sections/Tables/countermeasures_table2.tex}
%\vspace{-10pt}
\section{\label{sec:secexamplegeneric}Risk Assessment Process}
%\vspace{-5pt}
In this section, we present a risk assessment process that can be applied to any ML use case in \oran.
Based on the threat mapping of a general ML usecase in \oran, presented in Sections~\ref{sec:ontology} - \ref{sec:attack_tech}, we define a procedure and a tool~\cite{ORANRISK} that can be utilized for conducting the risk assessment of a selected use case.
The risk assessment is conducted according to the following procedure (note that throughout the description of the risk assessment procedure, we provide an example of the application of the procedure to the traffic steering use case):
\begin{enumerate}%[leftmargin=*,noitemsep,topsep=0pt,itemindent=1.25em]
%\small
 \item \textbf{Provide title and description --} 
Provide the title and short description of the selected use case. 
An example of title and description for the traffic steering use case is presented in Figure~\ref{fig:tsriskassessment}.

\noindent \item \textbf{Select threat actor --} Select the deployment scenario (out of the five possible deployment scenarios presented in Section~\ref{sec:ml_dep_oran}) as well as the threat actor (out of the six threat actors defined in Section~\ref{subsec:threatactor}), for which the risk assessment is conducted.
We denote the threat actor $i$ as $TA_i$ ($i=1...6$). \\
For example, in Figure~\ref{fig:tsriskassessment}, we can see that for the traffic steering use case, the selected deployment scenario is $DS2$, and the risk assessment is conducted for the malicious UE(s) threat actor ($[A5]$).

 \item \textbf{Rank impact --} Rank the impact of each of the seven threat categories (defined in Section~\ref{sec:threat_category}) for the selected use case. \\
For example, for the traffic steering, we can see in Figure~\ref{fig:tsriskassessment} that the selected impact for $[T1]$, which refers to the an attack that causes the model to provide incorrect output/s for a \emph{specific} input/s selected by the attacker, is `Medium' since it will result in specific, targeted users that will receive higher/lower quality of service when they should receive lower/higher quality of service (i.e., the impact on the provider is limited). \\
We denote the impact of threat $T_j$ ($j=1...7$) by $Imp_{T_j}$.

 \item \textbf{Map capabilities and constraints to the analyzed use case --} Map the threat actor capabilities (access and knowledge) and additional constraints in the specific use case.
This is done by grading several questions (presented in Figure~\ref{fig:tsriskassessment}), where each question is mapped to a specific capability. 
Each grade is mapped to a numeric score (between 0 to 1), according to a predefined scale for each question/capability.
The higher the numeric score is, the stronger the capability of the attacker (i.e., more powerful attacker).
The score assigned to capability $k$ ($CAP_k$) is denoted by $Scr_{TA_i}(CAP_k)$. \\ 
For example, $Q1$ in Figure~\ref{fig:tsriskassessment} refers to the capability of an attacker to manipulate the sensor/raw data of the ML use case. 
In the traffic steering use case, it is `Easy' for a malicious UE threat actor to manipulate the sensor/raw data since the sensors and raw data include signal information from the UEs and cells. 
Since the UEs are not trusted, an attacker can use malicious UEs and manipulate the signals. 
The numeric score for grade `Easy' for $Q1$ is 0.8.

 \item \textbf{Map required capabilities and impacts to attack technique --} For each attack technique, denoted as $AT_n$, presented in Section~\ref{sec:attack_tech}, we map the capabilities that are required by the attacker, and the possible impacts. \\
$Req_{AT_n}(CAP_k)=1$ indicates that attack technique $n$ requires capability $k$ ($Req$ can be between 0 to 1, where 0 means that the capability is not required by the attack technique and 1 means that the attacker \emph{must} own the capability in order to use the attack technique). 
Similarly, $Imp_{T_j}({AT_n})=1$ indicates that attack technique $AT_n$ can result in threat $T_j$ being materialized. \\
For example, as can be seen in Table~\ref{tab:attack_techniques_with_actors}, the first attack technique, Gradient-based, white box evasion attack (such as C\&W, FGSM, PGD) requires full knowledge about the training data as well as the attacked model.
It also requires sensor data access in order to manipulate the adversarial samples. 
The possible impact of the attack in `Tempering', i.e., causing the model to provide incorrect output/s for a specific input/s. \\
In addition, we assign for each attack technique a score that reflect the effectiveness of the adversarial attack technique.
This score is derived based on the performance of the attack technique as reported in previous research.
$Ef_{AT_n}$ is a numeric value between (0 to 1) that indicates the effectiveness of adversarial attack technique $AT_n$.
In Figure~\ref{fig:attackmapping}, we can see that the first attack technique, Gradient-based, white box evasion attack (such as C\&W, FGSM, PGD), is highly effective with score equals to 1. \\
It is important to note, that the mapping performed in this step is done only once, and it is use case-agnostic.

 \item \textbf{Compute risk --} Based on the information provided in steps (1)-(5), we can derive the risk for each attack technique within the selected usecase. 

The risk of $AT_n$ is the multiplication of the ranked impact (step (3)), the likelihood of the attack, and the effectiveness of the attack technique.
Formally: 
\begin{equation}
\scriptsize
Risk(TA_i,AT_n,T_j) = Ef_{AT_n} \times (Imp_{T_j} \times Imp_{T_j}({AT_n})) \times LH(TA_i,AT_n)
\end{equation}
where, the likelihood of the attack technique $AT_n$ for the $TA_i$ is computed as follows (according to the mapping in step (5)): 
\begin{equation}
\scriptsize
LH(TA_i,AT_n) = AVERAGE_k(Req_{AT_n}(CAP_k) \times Scr_{TA_i}(CAP_k))
\end{equation}
An example of the resulted risk for partial list of attack techniques within the traffic steering usecase is presented in Figure~\ref{fig:tsriskassessmenttechnique}.
The likelihood of the attack is computed as the average of numeric scores of the attacker capabilities that are required by the attack technique (according to the mapping in step (5)).
\item \textbf{Prioritize threats --} Finally, based on the derived risk in step (6), we prioritize the threats for the selected usecase.
The prioritized list for the traffic steering usecase is presented in Figure~\ref{fig:attackprioritization}.
As can be seen, the threat with the highest risk score is a malicious UE threat actor executing attack technique $AT4.1$ -- Gradient-based, white box poisoning attacks.
This is with high risk score (7.90) since the threat actor has access to the model as well as the ability to poison the model.
In addition, the impact of this attack (model corruption) is high and results in wrong prediction for any input.

\item \textbf{Select countermeasures --} In this last step, in order to mitigate the high risk threats, we can utilize the countermeasure ontology (presented in Section~\ref{sec:countermeasures}) for selecting the proper countermeasures.
The best countermeasure(s) can be selected first by their ability to mitigate the threat, and then by considering the use case requirements (e.g., impact on model performance, run time overhead), and the ability to apply the countermeasures in the specific use case.
For example, for mitigating the threat with the highest risk in the traffic steering usecase, possible countermeasures are data sensitization, data denoising, or using auxiliary models such as an AML detector in order to detect the poisoned samples.
\end{enumerate}

\section{\label{sec:Tool}Tool for risk assessment}

%\begin{figure*}[h]
%\centering
%\begin{minipage}{\textwidth}
%\centering
\begin{figure*}[h]
\centering
\includegraphics[trim={1.5cm 1.5cm 2.5cm 0},clip, width=0.98\textwidth]{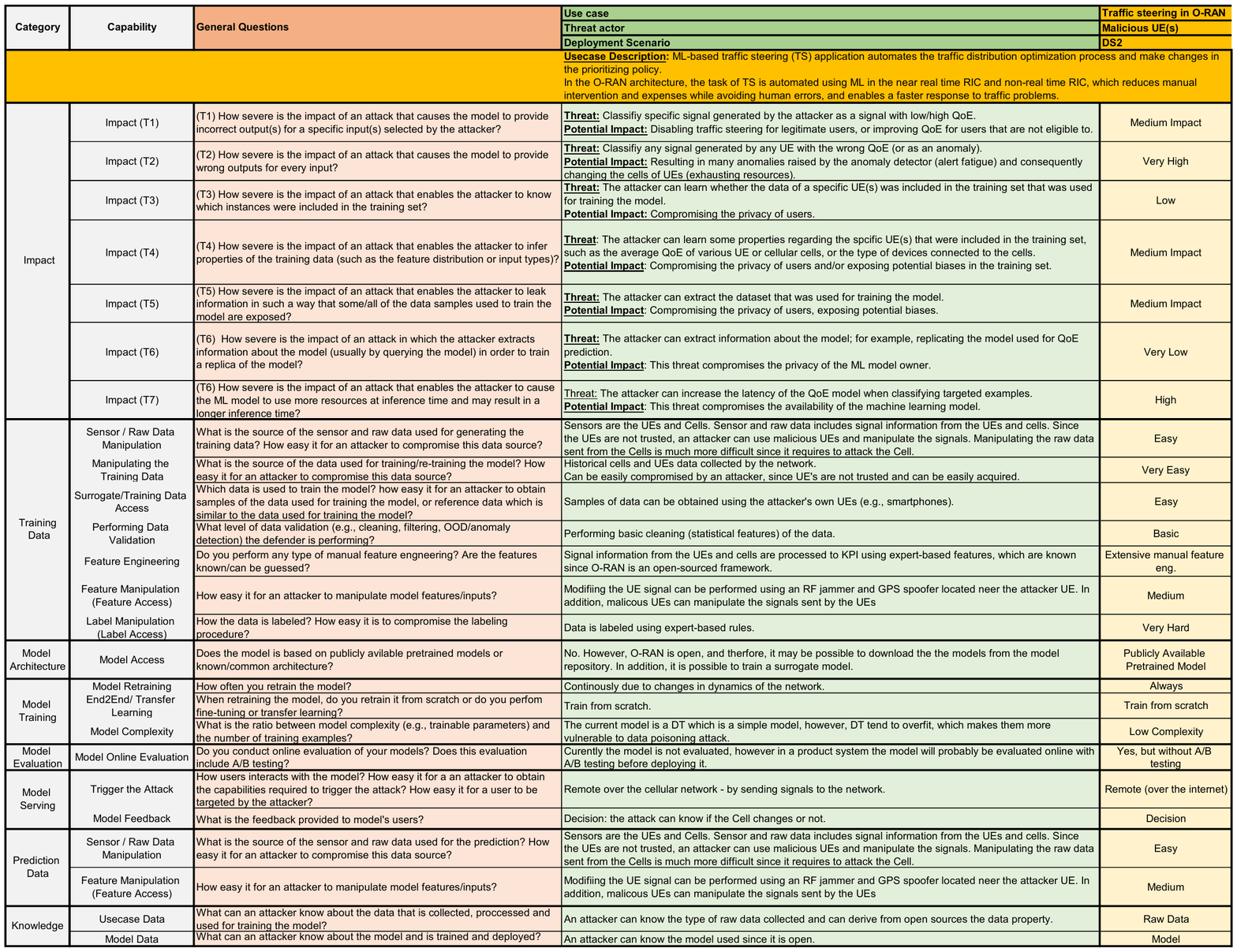}
\caption{Usecase Questions}
\label{fig:tsriskassessment}
\end{figure*}%
\begin{figure*}[h]%{0.98\linewidth}
\centering
\includegraphics[trim={1.8cm 11cm 1.8cm 0},clip, width=0.98\linewidth]{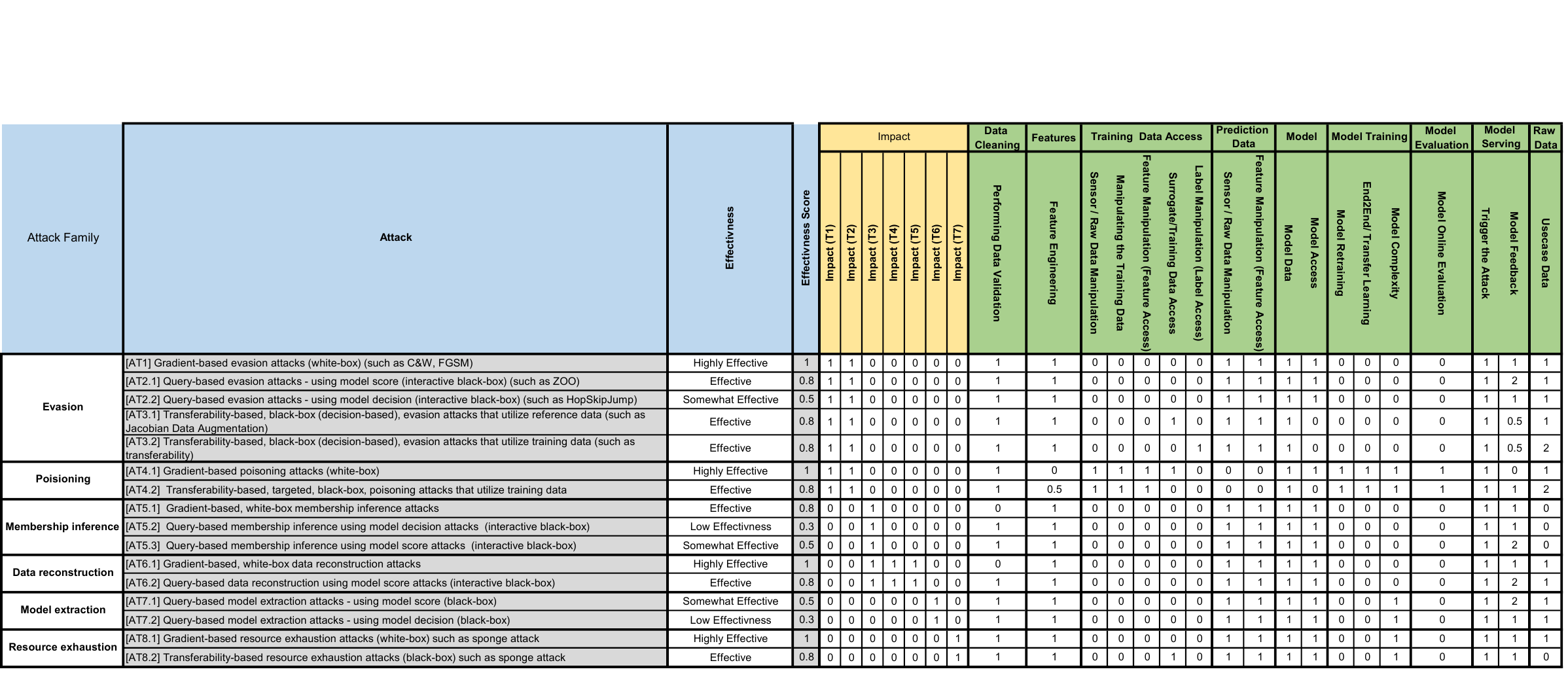}
\centering
\caption{Attack Mapping}
\label{fig:attackmapping}
\end{figure*}
\begin{figure*}[h]%{0.98\linewidth}
\centering
\includegraphics[trim={1.8cm 6.5cm 1.8cm 0},clip, width=0.98\textwidth]{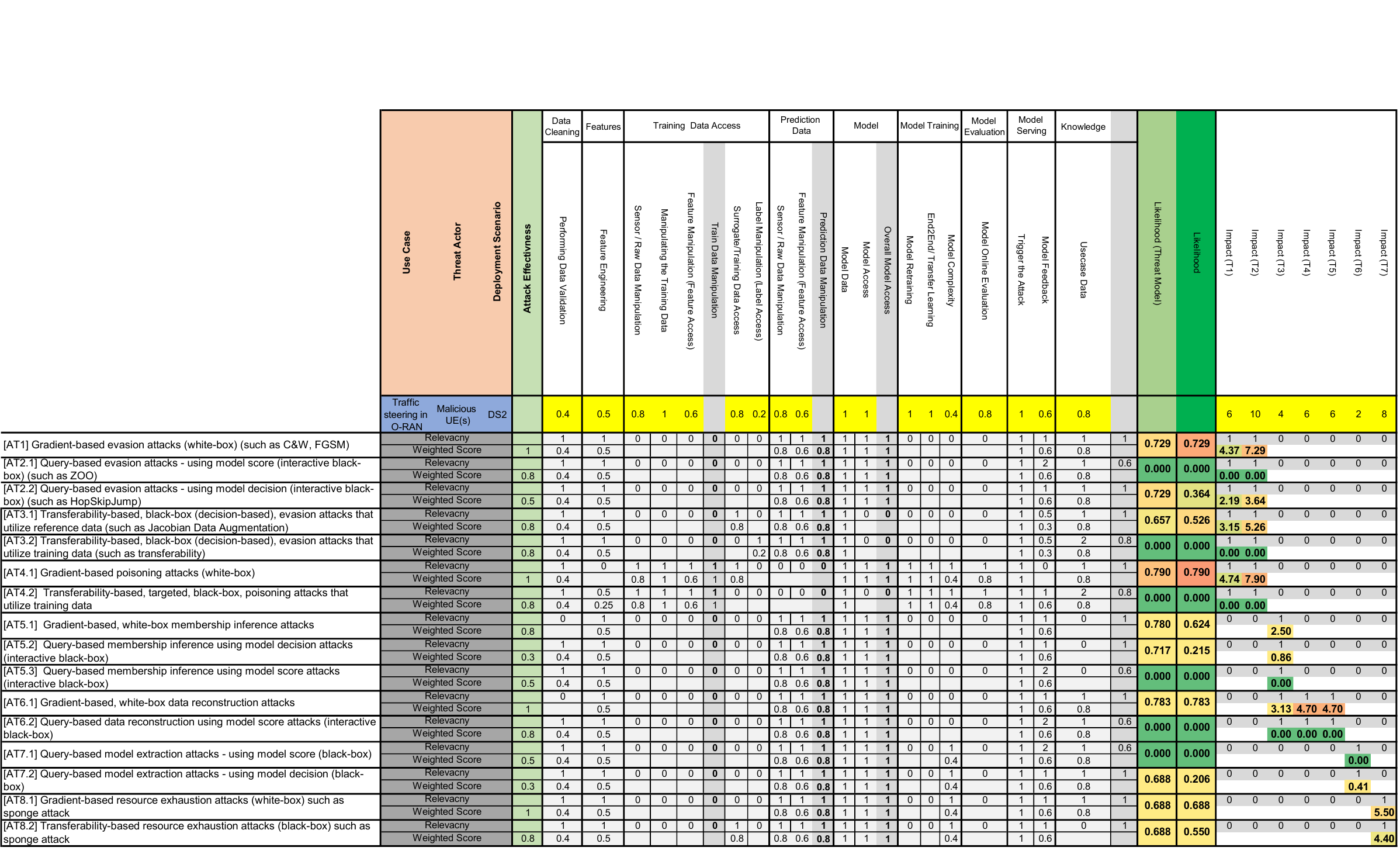}
\centering
\caption{Attack Mapping UC}
\label{fig:tsriskassessmenttechnique}
\end{figure*}%
\begin{figure*}[h]%{0.98\linewidth}
\centering
\includegraphics[trim={1.8cm 11.5cm 1.8cm 0},clip, width=0.98\textwidth]{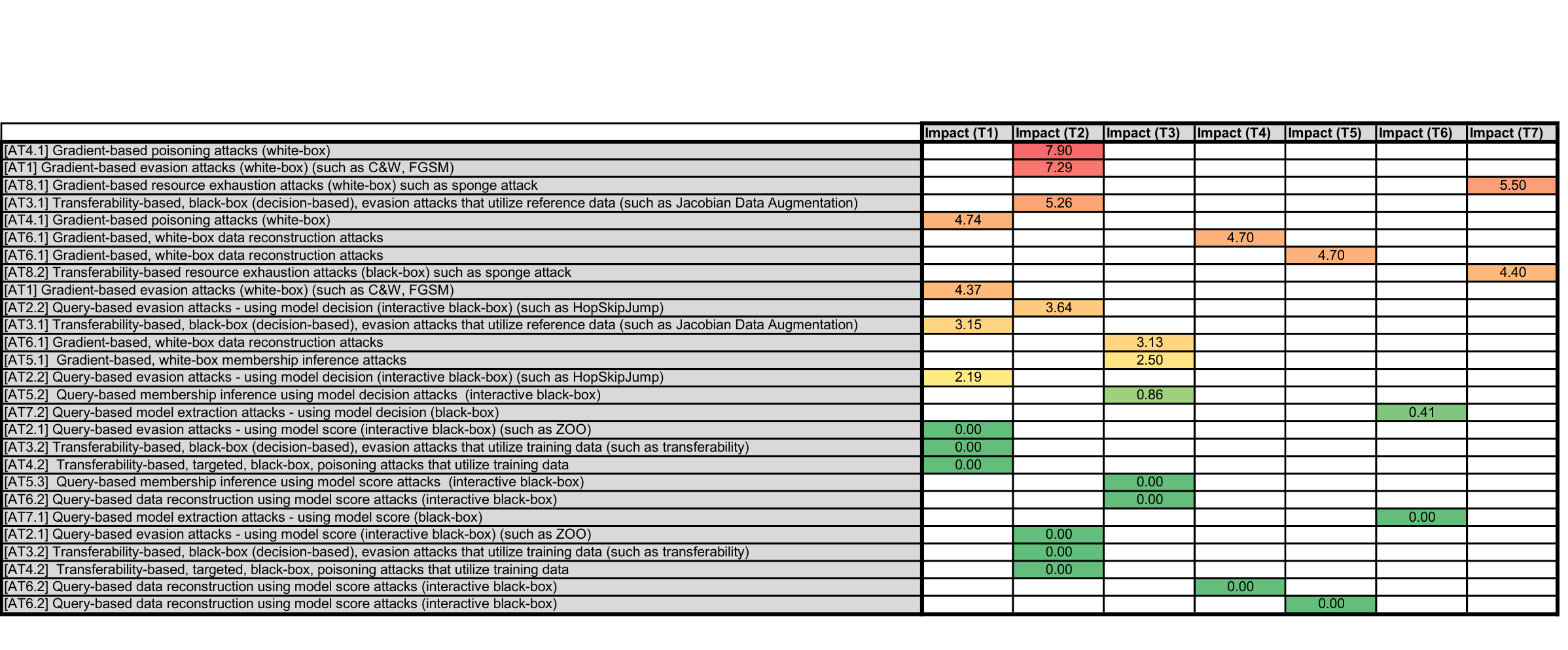}
\centering
\caption{Attack Prioritization}
\label{fig:attackprioritization}
\end{figure*}
%\centering
%\caption{AML Risk Tool Forms}
%\end{minipage}
%\end{figure*}

The illustration of the tool and its utilization of the traffic steering usecase can be found in~\cite{ORANRISK}.
The tool encompasses several forms, among which the \textit{usecase\_questions} (Fig.~\ref{fig:tsriskassessment}) form outlines the description of the use case and a compilation of questions utilized to determine the extent of the attack's likelihood and impact, as determined by the defined use case. The \textit{Attack Mapping} (Fig.~\ref{fig:attackmapping}) form features a matrix mapping the capabilities required for implementing the attack on a scale between 0 and 1, as well as the potential impact of the attack.
The \textit{Attack Mapping (UC)} form (Fig.~\ref{fig:tsriskassessmenttechnique})
presenting the derived risk scores for each
attack technique within the selected use case. The risk is the multiplication of the ranked impact, the likelihood of the attack and the effectiveness of the attack. These scores are consolidated in the \textit{AttackPrioritization} (Fig.~\ref{fig:attackprioritization}) form, where the attacks are sorted in accordance with their level of impact.

%\input{sections/D_sections/9_secassessexample.tex}
%\input{sections/D_sections/10_demonstration}
%\vspace{-10pt}
\section{Conclusions}
%\vspace{-5pt}
% \newText{One drawback is that the discussion is always at a quite high/conceptual level. It could be nice to provide a bit more details on the attacks such as the required effort and/or more concrete like the example given at the end.”
% Response 1. Examples of attack cases - specifically in ORAN at the threat level, 2. adding risk based on our previous paper and rank the attacks.
% }

% We present a systematic AML threat analysis of the O-RAN.
% We also demonstrate the applicability of an AML attack on the traffic steering use case implemented in the \oran~reference implementation.
% In future work, we intend to develop an extension to the MulVAL attack graph  framework~\cite{ou2005mulval} to incorporate the representation of cyberattacks for ML applications in the O-RAN. 
% We also plan to propose and evaluate a method for suggesting the optimal countermeasure deployment to address the identified risks.

We present a comprehensive threat assessment of ML use cases within the \oran using a common cybersecurity risk assessment ontology. We also conducted a systematic mapping of possible ML deployments, including identifying threat actors, their capabilities, and the attack techniques and families of AML specific to \oran.
Based on this threat assessment, a novel procedure and practical tool were introduced for the concrete identification and prioritization of AML threats, as well as the effective planning of countermeasures.
This approach was demonstrated through the application of the risk assessment procedure over the traffic steering use-case.

% ====== REFERENCE SECTION
\bibliographystyle{IEEEtran}
\bibliography{IEEEabrv,references}

% {\footnotesize \bibliographystyle{acm}
% \bibliography{references}}
% \bibliographystyle{ACM-Reference-Format}
% \bibliography{references}
% \input{sections/D_sections/12_appendix.tex}

\end{document}